\documentclass[12pt]{article}
\usepackage{amssymb}
\usepackage{graphicx}
\begin{document}
\newcommand{\beq}{\begin{equation}}
\newcommand{\eeq}{\end{equation}}
\newcommand{\beqa}{\begin{eqnarray}}
\newcommand{\eeqa}{\end{eqnarray}}
\newcommand{\beqar}{\begin{eqnarray*}}
\newcommand{\eeqar}{\end{eqnarray*}}
\newcommand{\al}{\alpha}
\newcommand{\be}{\beta}
\newcommand{\del}{\delta}
\newcommand{\D}{\Delta}
\newcommand{\eps}{\epsilon}
\newcommand{\ga}{\gamma}
\newcommand{\Ga}{\Gamma}
\newcommand{\ka}{\kappa}
\newcommand{\nn}{\nonumber}
\newcommand{\inn}{\!\cdot\!}
\newcommand{\h}{\eta}
\newcommand{\ii}{\iota}
\newcommand{\kk}{\varphi}
\newcommand\F{{}_3F_2}
\newcommand{\la}{\lambda}
\newcommand{\La}{\Lambda}
\newcommand{\na}{\prt}
\newcommand{\Om}{\Omega}
\newcommand{\om}{\omega}
\newcommand{\p}{\Phi}
\newcommand{\sig}{\sigma}
\renewcommand{\t}{\theta}
\newcommand{\z}{\zeta}
\newcommand{\ssc}{\scriptscriptstyle}
\newcommand{\eg}{{\it e.g.,}\ }
\newcommand{\ie}{{\it i.e.,}\ }
\newcommand{\labell}[1]{\label{#1}} 
\newcommand{\reef}[1]{(\ref{#1})}
\newcommand\prt{\partial}
\newcommand\veps{\varepsilon}
\newcommand{\pol}{\varepsilon}
\newcommand\vp{\varphi}
\newcommand\ls{\ell_s}
\newcommand\cF{{\cal F}}
\newcommand\cA{{\cal A}}
\newcommand\cS{{\cal S}}
\newcommand\cT{{\cal T}}
\newcommand\cV{{\cal V}}
\newcommand\cL{{\cal L}}
\newcommand\cM{{\cal M}}
\newcommand\cN{{\cal N}}
\newcommand\cG{{\cal G}}
\newcommand\cK{{\cal K}}
\newcommand\cH{{\cal H}}
\newcommand\cI{{\cal I}}
\newcommand\cJ{{\cal J}}
\newcommand\cl{{\iota}}
\newcommand\cP{{\cal P}}
\newcommand\cQ{{\cal Q}}
\newcommand\cg{{\tilde {{\cal G}}}}
\newcommand\cR{{\cal R}}
\newcommand\cB{{\cal B}}
\newcommand\cO{{\cal O}}
\newcommand\tcO{{\tilde {{\cal O}}}}
\newcommand\bz{\bar{z}}
\newcommand\bb{\bar{b}}
\newcommand\ba{\bar{a}}
\newcommand\bg{\bar{g}}
\newcommand\bc{\bar{c}}
\newcommand\bw{\bar{w}}
\newcommand\bX{\bar{X}}
\newcommand\bK{\bar{K}}
\newcommand\bA{\bar{A}}
\newcommand\bH{\bar{H}}
\newcommand\bF{\bar{F}}
\newcommand\bxi{\bar{\xi}}
\newcommand\bphi{\bar{\phi}}
\newcommand\bpsi{\bar{\psi}}
\newcommand\bprt{\bar{\prt}}
\newcommand\bet{\bar{\eta}}
\newcommand\btau{\bar{\tau}}
\newcommand\hF{\hat{F}}
\newcommand\hA{\hat{A}}
\newcommand\hT{\hat{T}}
\newcommand\htau{\hat{\tau}}
\newcommand\hD{\hat{D}}
\newcommand\hf{\hat{f}}
\newcommand\hK{\hat{K}}
\newcommand\hg{\hat{g}}
\newcommand\hp{\hat{\Phi}}
\newcommand\hi{\hat{i}}
\newcommand\ha{\hat{a}}
\newcommand\hb{\hat{b}}
\newcommand\hQ{\hat{Q}}
\newcommand\hP{\hat{\Phi}}
\newcommand\hS{\hat{S}}
\newcommand\hX{\hat{X}}
\newcommand\tL{\tilde{\cal L}}
\newcommand\hL{\hat{\cal L}}
\newcommand\tG{{\tilde G}}
\newcommand\tg{{\tilde g}}
\newcommand\tphi{{\widetilde \Phi}}
\newcommand\tPhi{{\widetilde \Phi}}
\newcommand\te{{\tilde e}}
\newcommand\tk{{\tilde k}}
\newcommand\tf{{\tilde f}}
\newcommand\tH{{\tilde H}}
\newcommand\ta{{\tilde a}}
\newcommand\tb{{\tilde b}}
\newcommand\tc{{\tilde c}}
\newcommand\td{{\tilde d}}
\newcommand\tm{{\tilde m}}
\newcommand\tmu{{\tilde \mu}}
\newcommand\tnu{{\tilde \nu}}
\newcommand\talpha{{\tilde \alpha}}
\newcommand\tbeta{{\tilde \beta}}
\newcommand\trho{{\tilde \rho}}
 \newcommand\tR{{\tilde R}}
\newcommand\teta{{\tilde \eta}}
\newcommand\tF{{\widetilde F}}
\newcommand\tK{{\tilde K}}
\newcommand\tE{{\widetilde E}}
\newcommand\tpsi{{\tilde \psi}}
\newcommand\tX{{\widetilde X}}
\newcommand\tD{{\widetilde D}}
\newcommand\tO{{\widetilde O}}
\newcommand\tS{{\tilde S}}
\newcommand\tB{{\tilde B}}
\newcommand\tA{{\widetilde A}}
\newcommand\tT{{\widetilde T}}
\newcommand\tC{{\widetilde C}}
\newcommand\tV{{\widetilde V}}
\newcommand\thF{{\widetilde {\hat {F}}}}
\newcommand\Tr{{\rm Tr}}
\newcommand\tr{{\rm tr}}
\newcommand\STr{{\rm STr}}
\newcommand\hR{\hat{R}}
\newcommand\M[2]{M^{#1}{}_{#2}}
\newcommand\MZ{\mathbb{Z}}
\newcommand\MR{\mathbb{R}}
\newcommand\bS{\textbf{ S}}
\newcommand\bI{\textbf{ I}}
\newcommand\bJ{\textbf{ J}}

\begin{titlepage}
\begin{center}

\vskip 0.5 cm
{\LARGE \bf  
Type IIA supergravity at one loop: $\alpha'^3$ terms \\  \vskip 0.25 cm in the metric-dilaton-RR one-form sector}\\
\vskip 1.25 cm
 Mohammad R. Garousi \footnote{garousi@um.ac.ir}

\vskip 1 cm
{{\it Department of Physics, Faculty of Science, Ferdowsi University of Mashhad\\}{\it P.O. Box 1436, Mashhad, Iran}\\}
\vskip .1 cm
 \end{center}

\begin{abstract}

The circle compactification of M-theory is dual to type IIA string theory, requiring that the dimensional reduction of the M-theory couplings \((t_8 t_8 - \frac{1}{4} \epsilon_8 \epsilon_8) R^4\) must reproduce the type IIA one-loop effective action at order \(\alpha'^3\), including contributions from the metric, dilaton, and RR one-form.  
Through compactification, we obtain 1,276 couplings involving Riemann, Ricci, and Ricci scalar tensors, along with first and second derivatives of the dilaton and RR one-form. By employing field redefinitions, we reduce these to a  basis of 359 independent couplings. Crucially, we observe that the dilaton cannot be entirely removed from the couplings via field redefinitions, even in the pure metric-dilaton sector. We validate our results by showing exact agreement between all four-field couplings and the corresponding string-theory S-matrix elements in the string frame.  

Further, upon compactifying on K3, we demonstrate that the resulting six-dimensional \(\alpha'\) couplings at one-loop level transform under S-duality into the tree-level \(\alpha'\) couplings of heterotic string theory on \(T^4\). This match necessitates carefully chosen field redefinitions for both the type IIA (on K3) and heterotic (on \(T^4\)) sectors, providing a stringent test of the duality.

\end{abstract}

\end{titlepage}

\section{Introduction}

It is well known that all ten-dimensional superstring theories \cite{Schwarz:1982jn,Gross:1986iv} are related to M-theory, the unique quantum theory of gravity in eleven dimensions \cite{Schwarz:1996bh}. In particular, M-theory compactified on a circle is equivalent to type IIA string theory. This correspondence is reflected in their low-energy effective actions: the dimensional reduction of eleven-dimensional supergravity on a circle yields type IIA supergravity \cite{Huq:1983im,Becker:2007zj}.
The first corrections to eleven-dimensional supergravity arise at the eight-derivative order. Upon circular reduction, the full eight-derivative couplings in M-theory should reproduce all one-loop corrections in type IIA theory. While the pure gravity sector of these corrections is known to take the form  \((t_8 t_8 - \frac{1}{4!} \epsilon_{11} \epsilon_{11}) R^4\) \cite{Vafa:1995fj,Green:1997di,Peeters:2000qj}, the couplings involving the 3-form field have only been determined at the four-field level—either through superparticle scattering methods \cite{Peeters:2005tb} or supersymmetry constraints \cite{Duff:1995wd,Antoniadis:1997eg,Russo:1997mk,Hyakutake:2025zqv}. Partial results are known beyond the four-field level \cite{Hyakutake:2007sm,Richards:2008jg,Richards:2008sa}. Additionally, supersymmetry requires the inclusion of the Chern-Simons coupling \(t_8 \epsilon_{11} A R^4\)  in M-theory \cite{Peeters:2000qj,Hyakutake:2006aq}.
In this paper, we focus on deriving the one-loop couplings in type IIA theory that correspond to the pure gravity sector of M-theory.

The circular reduction of the eleven-dimensional metric yields the metric, dilaton, and RR one-form in type IIA theory. Consequently, reducing the eleven-dimensional pure gravity couplings at the eight-derivative order should generate all one-loop couplings in type IIA theory involving the metric, dilaton, and RR one-form—also at the eight-derivative order in the string frame. These couplings span multiple field interactions, ranging from four-field up to eight-field terms.
Recently, some four-field couplings involving the RR one-form were derived in \cite{Aggarwal:2025lxf,Liu:2025uqu}, where their consistency with the corresponding type IIA S-matrix elements \cite{Policastro:2006vt} was also examined.

The circular reduction generates 1,276 couplings in type IIA theory. However, these are not all independent, as many are related through field redefinitions \cite{Metsaev:1987zx}, total derivative terms, and Bianchi identities. To streamline the analysis, we express them in terms of a minimal basis of metric-dilaton-RR couplings.
We find that this minimal basis consists of 377 independent couplings, which we present in a specific scheme. When reconstructing the original 1,276 couplings from this basis, we identify 359 non-zero couplings. Notably, many of these involve the dilaton—even in the gravity-dilaton sector.
We attempted to reformulate these 359 couplings in alternative schemes where the dilaton dependence might vanish, analogous to the tree-level NS-NS couplings \cite{Garousi:2020lof}. However, no such scheme exists: the dilaton persists in all cases, including the gravity-dilaton sector.

It is well known that, up to an overall factor, the sphere-level S-matrix element for four massless fields in type II superstring theory matches the torus-level S-matrix element \cite{Schwarz:1982jn}. Both consist solely of contact terms, which correspond to field-theory couplings in momentum space.  
The sphere-level coupling for four NS-NS fields, derived in \cite{Schwarz:1982jn,Gross:1986iv,Gross:1986mw}, takes the form \( t_8 t_8 \bar{R}^4 \), where \( \bar{R} \) is the linearized generalized Riemann curvature. While this coupling contains a dilaton dependence in the Einstein frame, the dilaton vanishes in the string-frame expression of \( \bar{R} \) \cite{Garousi:2012jp}. This aligns with the observation that there exists a scheme where all tree-level NS-NS couplings at the eight-derivative order in the string frame are dilaton-free \cite{Garousi:2020lof}, except for the overall factor \( e^{-2\Phi} \).  
Consequently, we expect the one-loop four-field couplings in the string frame to also lack dilaton terms when the RR one-form is zero. In this paper, we demonstrate that although the metric-dilaton couplings we derive are non-zero, their 4-point S-matrix elements vanish.

Several four-field couplings at the eight-derivative order involving RR fields in both type IIA and IIB theories were previously established in \cite{Peeters:2003pv,Garousi:2013lja,Bakhtiarizadeh:2013zia} by requiring consistency of the \( t_8 t_8 \bar{R}^4 \) coupling with S- and T-duality transformations. Notably, \cite{Bakhtiarizadeh:2013zia}  demonstrated the existence of multiple non-zero dilaton-RR couplings in the string frame. When the overall normalization factor is properly fixed, these couplings should naturally appear in the one-loop effective action.
Our present work confirms this expectation: the one-loop couplings we have derived precisely reproduce these previously identified dilaton-RR couplings.

It is well established that compactifying type IIA string theory on a K3 surface is equivalent to compactifying heterotic string theory on a four-torus $T^4$ \cite{Hull:1994ys,Becker:2007zj}. This equivalence manifests in their low-energy effective actions, where six-dimensional type IIA supergravity becomes identical to six-dimensional heterotic supergravity \cite{Liu:2013dna}. The field transformations relating these theories invert the sign of the dilaton, implementing an S-duality transformation that forms a $\MZ_2$ symmetry group.
Since the coupling constant inverts under S-duality, we can only directly compare effective action terms at derivative orders that receive no higher-loop corrections \cite{Tseytlin:1995fy}. The leading-order two-derivative terms, being tree-level exact, transform cleanly under S-duality: the supergravity description at weak coupling in one theory maps to the strong-coupling regime of the other.
The eight-derivative couplings in M-theory are quantum-exact, which implies their type IIA counterparts - including both the one-loop eight-derivative terms and their K3-reduced four-derivative versions in six dimensions - must similarly be exact. Conversely, the heterotic theory contains  four-derivative dilaton-gravity couplings of the form  $e^{-2\Phi}R_{\mu\nu\alpha\beta}R^{\mu\nu\alpha\beta}$ \cite{Metsaev:1987zx}, along with their $T^4$ compactification counterparts. In \cite{Ellis:1987dc,Ellis:1989fi}, it was shown that these couplings do not receive loop corrections and are therefore exact. These exact terms in both theories should map to each other under S-duality \cite{Antoniadis:1997eg}.
We demonstrate that with appropriate field redefinitions in both six-dimensional theories, these couplings do indeed transform into one another.

This paper is organized as follows: Section 2 reviews the circular reduction of the bosonic sector of 11-dimensional supergravity to obtain type IIA supergravity. Section 3 examines the reduction of M-theory's eight-derivative gravity couplings, beginning with a field redefinition of the \((t_8 t_8 - \frac{1}{4!} \epsilon_{11} \epsilon_{11}) R^4\) coupling into six Riemann curvature contractions, whose circular reduction yields 1,276 couplings involving Riemann, Ricci, and scalar curvatures along with first and second derivatives of the dilaton and RR one-form. Subsection 3.1 organizes these into a minimal basis of 377 eight-derivative gravity-dilaton-RR couplings, of which 359 are non-zero, while Subsection 3.2 demonstrates exact agreement between our four-field couplings and known string-frame results \cite{Garousi:2013lja,Bakhtiarizadeh:2013zia}. Section 4 explores K3 compactification and S-duality: Subsection 4.1 analyzes the reduction of type IIA supergravity (focusing on the gravity-dilaton-RR sector) and its S-dual correspondence with heterotic theory on \(T^4\). Subsection 4.2 demonstrates how eight-derivative terms in IIA reduce to four-derivative couplings that transform under S-duality into heterotic tree-level terms. A key distinction arises between derivative orders: while two-derivative S-duality holds without field redefinitions, the four-derivative (loop-level) case necessitates field redefinitions in both the six-dimensional theory. Section 5 summarizes our results, and the Appendix derives a minimal basis for eight-derivative couplings in a specific scheme.

\section{Two-derivative order reduction}

In this section, we establish our conventions by reviewing the derivation of the bosonic couplings in 10-dimensional type IIA supergravity via the circular reduction of 11-dimensional supergravity \cite{Cremmer:1979up,Huq:1983im,Becker:2007zj}. The bosonic sector of the 11-dimensional theory consists of the following couplings:
\beqa
\bS_0&=&-\frac{2}{\kappa_{11}^2}\Big[\int d^{11}x\sqrt{-g}(R-\frac{1}{2.4!}F_{abcd}F^{abcd})-\frac{1}{6}\int A^{(3)}\wedge F^{(4)}\wedge F^{(4)}\Big],\labell{11sugra}
\eeqa
where $\kappa_{11}^2=\frac{1}{\pi}(2\pi\ell_p)^9$.
The dimensional reduction of the 11-dimensional supergravity couplings on a circle of radius \( R_{11} = g_s^{2/3} \ell_p \) proceeds via the standard Kaluza-Klein (KK) ansatz for the metric:
\beqa
g_{ab}=e^{-2\Phi/3}\left(\matrix{G_{\mu\nu}+e^{2\Phi}C_\mu C_\nu & e^{2\Phi}C_\mu &\cr e^{2\Phi}C_\nu &e^{2\Phi}&}\right)\,\,;\,\,g^{ab}=e^{2\Phi/3}\left(\matrix{G^{\mu\nu} &  -C^\mu&\cr -C^\nu&e^{-2\Phi}+C_\alpha C^\alpha&}\right).\labell{reduc}\eeqa
Here, \( G^{\mu\nu} \) denotes the inverse of the 10-dimensional metric, which raises the index of the Ramond-Ramond (RR) vector field \( C_\mu \). The three-form field reduces as follows:
\beqa
A_{\mu\nu\alpha}=C_{\mu\nu\alpha}\,\,;\,\,
A_{\mu\nu y}=B_{\mu\nu}\,.\labell{CC}
\eeqa
Here, \( C^{(3)} \) denotes the RR three-form potential and \( B \) represents the NS-NS two-form field of type IIA superstring theory. Using these reduction rules, one can derive the 10-dimensional counterparts of various 11-dimensional couplings. For instance, the dimensional reduction of the overall factor \( \sqrt{-g} \) and the scalar curvature in \( \bf{S}_0 \) yields:
\beqa
\sqrt{-g}&=&e^{-8\Phi/3}\sqrt{-G}\,,\nn\\
R&=&e^{2\Phi/3}(R-\frac{16}{3}\nabla_\mu\Phi\nabla^\mu\Phi+\frac{14}{3}\nabla_\mu\nabla^\mu\Phi-\frac{1}{2.2!}e^{2\Phi}F_{\mu\nu}F^{\mu\nu})\,,
\eeqa
where \( F_{\mu\nu} = \partial_\mu C_\nu - \partial_\nu C_\mu \) is the field strength of the RR one-form potential. Up to a total derivative term, this yields the standard dimensional reduction, i.e.,
\beqa
\sqrt{-g}R&=&e^{-2\Phi}\sqrt{-G}(R+4\nabla_\mu\Phi\nabla^\mu\Phi-\frac{1}{2.2!}e^{2\Phi}F_{\mu\nu}F^{\mu\nu})\,.
\eeqa 
The dimensional reduction of the terms in the action \(\bf{S}_0\) involving the three-form field strength proceeds as follows:
\beqa
-\frac{1}{2.4!}\sqrt{-g}F_{abcd}F^{abcd}&=&e^{-2\Phi}\sqrt{-G}(-
\frac{1}{2.3!}H_{\mu\nu\alpha}H^{\mu\nu\alpha}-\frac{1}{2.4!}e^{2\Phi}\bar{F}_{\mu\nu\alpha\beta}\bar{F}^{\mu\nu\alpha\beta})\,,\nn\\
-\frac{1}{6} A^{(3)}\wedge F^{(4)}\wedge F^{(4)}&=&-\frac{1}{2}e^{-2\Phi}\Big[ B^{(2)}\wedge e^{\Phi} F^{(4)}\wedge e^{\Phi}F^{(4)}\Big]\,.
\eeqa
Here, the RR four-form field strength \(\bar{F}^{(4)}\) is given by
$
\bar{F}_{\mu\nu\alpha\beta} = F_{\mu\nu\alpha\beta} + 4C_{[\mu}H_{\nu\alpha\beta]},
$
where antisymmetrization is implicit in the indices. Note that a factor of \(e^\Phi\) can be absorbed into the RR fields, leaving the overall dilaton prefactor \(e^{-2\Phi}\). This confirms that the reduction of \(\bf{S_0}\) yields the sphere-level effective action of type IIA superstring theory.

Using the standard relations between the compactification radius, string coupling, and string length — \( R_{11} = g_s \sqrt{\alpha'} \) and \( \ell_p = g_s^{1/3} \sqrt{\alpha'} \) — the dimensional reduction of 11-dimensional supergravity  \reef{11sugra} yields the conventional type IIA supergravity action, which takes the form \cite{Huq:1983im,Becker:2007zj}:
\beqa
S_0&=&-\frac{2}{\kappa_{10}^2}\Big[\int d^{10}x \sqrt{-G}\left(e^{-2\Phi}(R+4\nabla_\mu\Phi\nabla^\mu\Phi-
\frac{1}{2}|H|^2)-\frac{1}{2}|F^{(2)}|^2-\frac{1}{2}|\bar{F}^{(4)}|^2\right)\nn\\
&&\qquad\qquad\qquad\qquad\qquad\qquad\qquad\qquad\qquad\qquad\qquad-\frac{1}{2}\int B^{(2)}\wedge F^{(4)}\wedge F^{(4)}\Big].\labell{S0}
\eeqa
Here, the 10-dimensional gravitational coupling is given by \(\kappa_{10}^2 = \kappa^2 g_s^2 = \frac{1}{\pi}(2\pi\ell_s)^8 g_s^2\), where \(\ell_s = \sqrt{\alpha'}\) defines the string length. Note that the dilaton \(\Phi\) is expressed as a fluctuation, excluding its constant background value. The sphere-level effective action of type IIA supergravity receives stringy corrections at \(\mathcal{O}(\alpha'^3)\) and higher, originating from non-zero KK modes in the compactification of 11-dimensional supergravity \cite{Green:1997as}. Furthermore, the dimensional reduction of higher-derivative corrections in 11-dimensional supergravity \reef{11sugra} generates loop-level higher-derivative couplings in type IIA theory, which are the primary focus of this work.

\section{Eight-derivative order reduction}

The first non-trivial higher-derivative correction to 11-dimensional supergravity appears at eight-derivative order (corresponding to \(\mathcal{O}(\ell_p^6)\)). While the complete structure of bosonic couplings at this order remains unknown, the purely gravitational sector has been determined and takes the form (see \eg  \cite{Hyakutake:2007vc}):
\beqa
\bS_6&=&-\frac{2}{\kappa_{11}^2}\frac{\pi^2\ell_p^6}{2^{11}.3^2}\int d^{11}x\sqrt{-g}\Big[(t_8t_8-\frac{1}{4!}\epsilon_{11}\epsilon_{11})R^4+\cdots\Big],\labell{ttee}
\eeqa
where the ellipsis denotes additional couplings involving the 3-form field \(A_{abc}\), which are not of interest here.
The tensor \(t_8\) is defined as \cite{Schwarz:1982jn}:
\beqa
t_8^{abcdefgh}&=&-2 g^{a f } g^{b e } g^{g d } 
g^{c h } + 8 g^{a d } g^{b e } 
g^{g f } g^{c h } + 8 g^{a h } g^{b 
e } g^{g d } g^{c f } \nn\\&&+ 8 
g^{a h } g^{b c } g^{g f } g^{d 
e } - 2 g^{a h } g^{b g } g^{c 
f } g^{d e } - 2 g^{a d } 
g^{b c } g^{g f } g^{e h }\,,
\eeqa
 and $\epsilon_{11}$ denotes the Levi-Civita tensor in eleven dimensions. Using these tensors, the action can be expressed through 27 distinct couplings  involving combinations of the Riemann tensor, Ricci tensor, and Ricci scalar. However, through appropriate field redefinitions, all terms involving the Ricci tensor or Ricci scalar can be eliminated, leaving only eight terms constructed purely from the Riemann curvature. After this simplification, one obtains:
\beqa
\bS_6&=&-\frac{2}{\kappa_{11}^2}\frac{\pi^2\ell_p^6}{2^{3}.3}\int d^{11}x\sqrt{-g}\Big[\frac{1}{4} R_{a b }{}^{d e 
} R^{a bc f } R_{c d }{}^{g h } R_{f e g h 
} -  R_{a b }{}^{d e } 
R^{a bc f } R_{c 
}{}^{g }{}_{d }{}^{h } R_{f g 
e h }\nn\\&&\qquad\qquad\qquad\qquad\qquad + \frac{1}{16} R_{a b }{}^{
d e } R^{a bc f } 
R_{c f }{}^{g h } R_{d 
e g h } + \frac{1}{2} R_{a 
}{}^{d }{}_{c }{}^{e } R^{a 
bc f } R_{b }{}^{g }{}_{f 
}{}^{h } R_{d g e h }\nn\\&&\qquad\qquad\qquad\qquad\qquad -  
R_{a bc }{}^{d } 
R^{a bc f } R_{f 
}{}^{e g h } R_{d g e 
h } + \frac{1}{32} R_{a bc f } 
R^{a bc f } R_{d 
e g h } R^{d e g h 
}\Big],\labell{ttee1}
\eeqa
where we have employed the cyclic symmetry of the Riemann tensor to express the original eight curvature terms in terms of six independent structures. 

Applying the dimensional reduction scheme outlined in  \reef{reduc}, we obtain the following couplings in the string frame of type IIA theory:
\beqa
S_6&=&-\frac{2}{\kappa^2}\frac{\pi^2\ell_s^6}{2^{3}.3}\int d^{10}x\sqrt{-G}\Big[\frac{1}{4} R_{\alpha \beta }{}^{\epsilon \varepsilon 
} R^{\alpha \beta \gamma \delta } R_{\gamma 
\epsilon }{}^{\mu \nu } R_{\delta \varepsilon \mu \nu 
} -  R_{\alpha \beta }{}^{\epsilon \varepsilon } 
R^{\alpha \beta \gamma \delta } R_{\gamma 
}{}^{\mu }{}_{\epsilon }{}^{\nu } R_{\delta \mu 
\varepsilon \nu }\labell{ttee1}\\&&\qquad\qquad\qquad\qquad\qquad + \frac{1}{16} R_{\alpha \beta }{}^{
\epsilon \varepsilon } R^{\alpha \beta \gamma \delta } 
R_{\gamma \delta }{}^{\mu \nu } R_{\epsilon 
\varepsilon \mu \nu } + \frac{1}{2} R_{\alpha 
}{}^{\epsilon }{}_{\gamma }{}^{\varepsilon } R^{\alpha 
\beta \gamma \delta } R_{\beta }{}^{\mu }{}_{\delta 
}{}^{\nu } R_{\epsilon \mu \varepsilon \nu }\nn\\&&\qquad\qquad\qquad\qquad\qquad -  
R_{\alpha \beta \gamma }{}^{\epsilon } 
R^{\alpha \beta \gamma \delta } R_{\delta 
}{}^{\varepsilon \mu \nu } R_{\epsilon \mu \varepsilon 
\nu } + \frac{1}{32} R_{\alpha \beta \gamma \delta } 
R^{\alpha \beta \gamma \delta } R_{\epsilon 
\varepsilon \mu \nu } R^{\epsilon \varepsilon \mu \nu 
}+\cdots\Big].\nn
\eeqa
The absence of an overall dilaton factor indicates that the reduced action \( S_6 \) corresponds to the torus-level effective action of type IIA theory. The ellipsis in the equation represents 1,270 couplings involving Riemann, Ricci, and Ricci scalar curvatures along with first and second derivatives of the dilaton and RR one-form. Unlike the 11-dimensional case, 10-dimensional field redefinitions eliminating Ricci terms, \( \nabla_\mu F^{\mu\nu} \), and \( \nabla_\mu\nabla^\mu\Phi \) are insufficient to minimize the couplings - even after removing these structures, 877 couplings remain, with additional field redefinitions possible. As shown in the Appendix, the minimal basis for eight-derivative dilaton-gravity-RR one-form couplings contains only 377 independent terms, implying the complete expression must be expressible in terms of at least this fundamental set, which we will construct in the following subsection.

\subsection{Eight-derivative couplings in a 10D minimal basis}

The couplings in \reef{ttee1} represent one-loop effective interactions in a specific scheme. While these can be transformed into alternative schemes through field redefinitions and integration by parts, constructing a representation with the minimal number of couplings presents a non-trivial challenge. Although we cannot determine the absolute minimum number of independent couplings, in this section we express them using a minimal basis of 377 eight-derivative dilaton-gravity-RR one-form interactions. As detailed in Appendix, this minimal basis admits multiple equivalent representations. Here we demonstrate that the couplings \reef{ttee1} can be expanded in this basis with 359 non-zero coefficients and 18 vanishing coefficients. The specific values of these coefficients depend on the chosen scheme for the minimal basis. A particularly interesting (but computationally intensive) problem would be to identify a scheme that maximizes the number of zero coefficients when matching the 10-dimensional couplings \reef{ttee1}. In this work, we determine the coefficients using the specific scheme selected in Appendix (see \reef{T55}), obtained by equating \reef{ttee1} with our minimal basis while accounting for field redefinitions, integration by parts
and Bianchi identities. That is 
 \beqa
S_6&\sim&-\frac{2}{\kappa^2}\frac{\pi^2\ell_s^6}{2^{3}.3}\int d^{10}x\sqrt{-G}\cL\,.\labell{ttee2}
\eeqa 
Here, $S_6$ represents the dimensionally reduced action from \reef{ttee1}, while $\mathcal{L}$ denotes the specific minimal basis of 377 couplings identified in Appendix (see \reef{T55}). The $\sim$ relation signifies equality modulo:
\begin{itemize}
\item Field redefinitions
\item Total derivative terms
\item Bianchi identities
\end{itemize}
The computational procedure mirrors our Appendix calculation for determining the minimal basis. From this matching, we uniquely fix all 377 coupling constants in \reef{T55} to the following values:
\beqa
&&c_{1}=2341/4608,c_{2}=-209747/921600,c_{3}=-1027/7200,c_{4}=-880957/1843200,\nn\\&&c_{5}=4625129/29491200,c_{6}=-66793/28800,c_{7}=-817/960,c_{8}=385121/230400,\nn\\&&c_{9}=-3433/46080,c_{10}=194693/368640,c_{11}=24653/19200,c_{12}=-979/1600,\nn\\&&c_{13}=4307/3840,c_{14}=2979/1600,c_{15}=-43241/9600,c_{16}=3351/3200,c_{17}=-1151/800,\nn\\&&c_{18}=-30283/4800,c_{19}=4,c_{20}=-7167/3200,c_{21}=-7/2,c_{22}=-3/2,c_{23}=1/2,\nn\\&&c_{24}=89527/115200,c_{25}=9721/115200,c_{26}=1,c_{27}=1027/640,c_{28}=-3849/3200,\nn\\&&c_{29}=-5873/3840,c_{30}=-75041/38400,c_{31}=17/128,c_{32}=16717/20480,c_{33}=3,\nn\\&&c_{34}=1/2,c_{35}=7167/6400,c_{36}=-13/2,c_{37}=-3/4,c_{38}=-3/8,c_{39}=0,c_{40}=-1,\nn\\&&c_{41}=-1,c_{42}=1/32,c_{43}=1119/6400,c_{44}=1693/3840,c_{45}=-4451/12800,\nn\\&&c_{46}=-647/3200,c_{47}=-4193/7680,c_{48}=-3/2,c_{49}=11/4,c_{50}=11/2,c_{51}=-7/8,\nn\\&&c_{52}=3/2,c_{53}=-5/4,c_{54}=1/16,c_{55}=1/4,c_{56}=0,c_{57}=3391/11520,\nn\\&&c_{58}=-173339/115200,c_{59}=3449/800,c_{60}=-5/2,c_{61}=73027/115200,\nn\\&&c_{62}=43031/57600,c_{63}=-267211/115200,c_{64}=-29399/230400,c_{65}=246391/921600,\nn\\&&c_{66}=-6197/3200,c_{67}=1559/9600,c_{68}=2159/9600,c_{69}=-8857/12800,c_{70}=3/4,\nn\\&&c_{71}=-1/4,c_{72}=5/4,c_{73}=14741/800,c_{74}=19219/1600,c_{75}=41/4,c_{76}=-6767/600,\nn\\&&c_{77}=6767/1200,c_{78}=-5497/400,c_{79}=-1217/400,c_{80}=14781/3200,c_{81}=-57/16,\nn\\&&c_{82}=-45/8,c_{83}=-186973/28800,c_{84}=144301/115200,c_{85}=58361/7680,\nn\\&&c_{86}=-1711343/921600,c_{87}=-16457/1600,c_{88}=23339/1600,c_{89}=-2259/1600,\nn\\&&c_{90}=12251/800,c_{91}=181/128,c_{92}=-5167/400,c_{93}=5,c_{94}=3567/400,c_{95}=3,c_{96}=2,\nn\\&&c_{97}=6367/200,c_{98}=14377/1600,c_{99}=-681/128,c_{100}=-1,c_{101}=1,c_{102}=-145/64,\nn\\&&c_{103}=-5471/640,c_{104}=-973/1600,c_{105}=-5667/400,c_{106}=3/2,c_{107}=6367/400,\nn\\&&c_{108}=-6167/3200,c_{109}=4367/400,c_{110}=-3/4,c_{111}=1,c_{112}=-6367/200,\nn\\&&c_{113}=141/100,c_{114}=-2761/800,c_{115}=-131/10,c_{116}=9/2,c_{117}=8/5,\nn\\&&c_{118}=-125353/230400,c_{119}=293867/1843200,c_{120}=-22253/19200,c_{121}=5/16,\nn\\&&c_{122}=37/1152,c_{123}=21661/57600,c_{124}=-5213/9600,c_{125}=-3827/1920,c_{126}=1,\nn\\&&c_{127}=1129/800,c_{128}=-10889/14400,c_{129}=-849/3200,c_{130}=1049/400,\nn\\&&c_{131}=-2879/800,c_{132}=2929/800,c_{133}=-7/2,c_{134}=-11/16,c_{135}=-3587/1920,\nn\\&&c_{136}=89/40,c_{137}=2,c_{138}=2081/800,c_{139}=-13/8,c_{140}=-1/5,c_{141}=-357/200,\nn\\&&c_{142}=0,c_{143}=-1/4,c_{144}=-5/16,c_{145}=-3/4,c_{146}=-1/2,c_{147}=-5,c_{148}=0,\nn\\&&c_{149}=3,c_{150}=1,c_{151}=0,c_{152}=3/2,c_{153}=3/4,c_{154}=0,c_{155}=1/4,c_{156}=0,c_{157}=0,\nn\\&&c_{158}=-14331/6400,c_{159}=-26579/800,c_{160}=5647/160,c_{161}=-1501/60,\nn\\&&c_{162}=7041/100,c_{163}=-563/60,c_{164}=-2927/200,c_{165}=125/4,c_{166}=2127/200,\nn\\&&c_{167}=-61/2,c_{168}=85/2,c_{169}=6927/100,c_{170}=2271/80,c_{171}=-1543/80,c_{172}=0,\nn\\&&c_{173}=-77/4,c_{174}=-1531/80,c_{175}=-1255/64,c_{176}=319/15,c_{177}=-3727/200,\nn\\&&c_{178}=-49/8,c_{179}=6927/200,c_{180}=-4527/1600,c_{181}=2427/200,c_{182}=61/8,c_{183}=2,\nn\\&&c_{184}=-6927/100,c_{185}=-552/5,c_{186}=0,c_{187}=65,c_{188}=4/5,c_{189}=-727/16,\nn\\&&c_{190}=-52291/3200,c_{191}=467/20,c_{192}=-3/10,c_{193}=-1/5,c_{194}=342/5,\nn\\&&c_{195}=-1/16,c_{196}=3/8,c_{197}=4,c_{198}=-1/4,c_{199}=-39/5,c_{200}=553/200,\nn\\&&c_{201}=-13/5,c_{202}=0,c_{203}=-27/16,c_{204}=-31/4,c_{205}=-5/2,c_{206}=-1779/25,\nn\\&&c_{207}=-26781/800,c_{208}=161/4,c_{209}=173/8,c_{210}=-4559/100,c_{211}=-147/2,\nn\\&&c_{212}=129/2,c_{213}=-29,c_{214}=-2599/50,c_{215}=353/40,c_{216}=-23/2,\nn\\&&c_{217}=-811/50,c_{218}=294/5,c_{219}=-979/50,c_{220}=-17383/200,c_{221}=-107,\nn\\&&c_{222}=157/2,c_{223}=8/5,c_{224}=53/10,c_{225}=-392/5,c_{226}=-499/100,c_{227}=499/100,\nn\\&&c_{228}=557/100,c_{229}=-657/400,c_{230}=1573/800,c_{231}=5/2,c_{232}=-5/2,\nn\\&&c_{233}=-121/5,c_{234}=-8/5,c_{235}=8/5,c_{236}=-1/2,c_{237}=13/4,c_{238}=4/5,c_{239}=133/5,\nn\\&&c_{240}=-197/80,c_{241}=2953/400,c_{242}=-512/5,c_{243}=106,c_{244}=168/5,c_{245}=-38,\nn\\&&c_{246}=19,c_{247}=84/5,c_{248}=-342/5,c_{249}=-128/5,c_{250}=5/4,c_{251}=63/20,c_{252}=15/4,\nn\\&&c_{253}=5/4,c_{254}=0,c_{255}=-5/2,c_{256}=71/40,c_{257}=-261/200,c_{258}=252/5,\nn\\&&c_{259}=-41/100,c_{260}=39/160,c_{261}=-8/5,c_{262}=-32/5,c_{263}=4/5,c_{264}=16,\nn\\&&c_{265}=1071/200,c_{266}=-10,c_{267}=807/20,c_{268}=617/10,c_{269}=-103/5,c_{270}=-7,\nn\\&&c_{271}=13,c_{272}=-316/5,c_{273}=-264/5,c_{274}=15029/14400,c_{275}=-6749/14400,\nn\\&&c_{276}=3389/9600,c_{277}=-1177/2304,c_{278}=967/3840,c_{279}=-1667/960,c_{280}=2387/960,\nn\\&&c_{281}=-389/320,c_{282}=1/4,c_{283}=-3/2,c_{284}=-3827/960,c_{285}=5/2,c_{286}=17/32,\nn\\&&c_{287}=3587/1920,c_{288}=-5/4,c_{289}=1/4,c_{290}=-119/40,c_{291}=3,c_{292}=0,\nn\\&&c_{293}=-11/10,c_{294}=-57/20,c_{295}=5,c_{296}=-31,c_{297}=-9/10,c_{298}=1/4,c_{299}=-4,\nn\\&&c_{300}=1403/1600,c_{301}=6449/28800,c_{302}=-5769/3200,c_{303}=343/128,c_{304}=-1327/1920,\nn\\&&c_{305}=1427/1920,c_{306}=1427/1920,c_{307}=-11/16,c_{308}=2987/960,c_{309}=-3827/960,\nn\\&&c_{310}=-3,c_{311}=3347/1920,c_{312}=-5/2,c_{313}=-1/3,c_{314}=2827/960,c_{315}=3667/960,\nn\\&&c_{316}=6,c_{317}=-8,c_{318}=-67/40,c_{319}=2,c_{320}=-3,c_{321}=-5/4,c_{322}=-283/5,\nn\\&&c_{323}=-1/128,c_{324}=-41/40,c_{325}=-13/40,c_{326}=235/4,c_{327}=85,c_{328}=112/5,\nn\\&&c_{329}=-2168/5,c_{330}=-3844/5,c_{331}=-502,c_{332}=-4,c_{333}=4/15,c_{334}=0,c_{335}=-32,\nn\\&&c_{336}=-333/400,c_{337}=0,c_{338}=84,c_{339}=-368/5,c_{340}=102,c_{341}=-112/5,c_{342}=0,\nn\\&&c_{343}=6,c_{344}=2,c_{345}=11/20,c_{346}=-2,c_{347}=2,c_{348}=2,c_{349}=-10,c_{350}=5,c_{351}=-1,\nn\\&&c_{352}=4,c_{353}=-2,c_{354}=0,c_{355}=-3/2,c_{356}=7/4,c_{357}=6,c_{358}=-7/2,c_{359}=-5/4,\nn\\&&c_{360}=-5/2,c_{361}=-5/2,c_{362}=5/4,c_{363}=7/2,c_{364}=-3/4,c_{365}=7,c_{366}=-9/2,\nn\\&&c_{367}=4,c_{368}=2,c_{369}=-9/2,c_{370}=-1/2,c_{371}=639/200,c_{372}=-4/5,c_{373}=-3,\nn\\&&c_{374}=4,c_{375}=1/16,c_{376}=3/32,c_{377}=0\labell{sol}
\eeqa
The minimal basis consists of 359 non-zero coupling constants and 18 vanishing couplings. Notably, the coefficient of coupling \reef{T53} (containing the Ricci tensor) vanishes. Additionally, the  structures $[F^3F'R\Phi']_1$, $[FF'R\Phi''\Phi']_2$, $[FF'\Phi''^2\Phi']_1$ and  $[FF'^3\Phi']_1$
have zero coefficients. 

The Lagrangian \reef{T55}, with the coupling constants set to the values above, represents our final result for the one-loop effective action of type IIA theory at the eight-derivative order in the metric-dilaton-RR one-form sector.
Moreover, additional derivative one-loop corrections appear at higher orders in $\ell_s$, stemming from non-zero KK modes in the compactification of the 11-dimensional couplings \reef{ttee1}. While these higher-order effects lie beyond the scope of our current analysis, they warrant further investigation in future work.

Our findings uncover a rich structure of non-trivial dilaton couplings. A particularly noteworthy example is the non-zero four-dilaton coupling in the string frame, which assumes the explicit form:
\beqa
[\Phi''^4]_2&=&-\frac{8}{5} \nabla_{\alpha  }\nabla^{\gamma  }\Phi  \nabla^{\alpha  }\nabla^{\beta  }\Phi  \nabla_{\beta  }\nabla^{\delta  }\Phi  \nabla_{\delta  }\nabla_{\gamma  }\Phi  +\frac{4}{5} \nabla^{\alpha  }\nabla^{\beta  }\Phi  \nabla_{\beta  }\nabla_{\alpha  }\Phi  \nabla^{\gamma  }\nabla^{\delta  }\Phi  \nabla_{\delta  }\nabla_{\gamma  }\Phi .
\eeqa
Expanding the covariant derivatives generates four-field couplings along with five-field and higher-order interactions. Our analysis reveals vanishing S-matrix elements for all four-field couplings, including specific cases in the structures $[R^3\Phi'']_3$, $[R^2\Phi''^2]_5$ and $[R\Phi''^3]_1$ which is consistent with the observations in \cite{Peeters:2005tb}. This null result is expected since one-loop four-field amplitudes are proportional to their classical-level counterparts, and no dilaton couplings exist at eight-derivative order in the string frame at classical level in the NS-NS sector in a specific scheme \cite{Garousi:2020lof}. Notably, the non-zero four-dilaton and two-dilaton-two-graviton S-matrix elements in the Einstein frame emerge from transforming the $[R^4]_6$ coupling to this frame, demonstrating the frame-dependence of these interactions while maintaining consistency with expected amplitudes.

The classical couplings involving two gravitons and two RR one-form fields were previously derived in \cite{Garousi:2013lja} through the application of S- and T-duality transformations to the four generalized Riemann curvature couplings (see eq.(21) in \cite{Garousi:2013lja}). At the one-loop level, the corresponding couplings maintain proportionality to these classical terms, specifically taking the form:
\beqa
S_{hhCC}&=&-\frac{2}{\kappa^2}\frac{\pi^2\ell_s^6}{2^{6}.3}\int d^{10}x\sqrt{-G}\Big[\frac{1}{4} R_{\mu \nu \rho \sigma } R^{\mu \nu \rho 
\sigma } \prt_{\gamma }F_{\alpha \beta } \prt^{\gamma 
}F^{\alpha \beta } + \frac{1}{2} R_{\tau \kappa }{}^{\alpha 
\beta } R^{\tau \kappa \mu \nu } \prt_{\gamma }F_{\alpha 
\beta } \prt^{\gamma }F_{\mu \nu } \nn\\&&-  R^{\mu \nu }{}_{\tau 
}{}^{\gamma } R^{\tau \kappa \alpha \beta } \prt_{\gamma 
}F_{\mu \nu } \prt_{\kappa }F_{\alpha \beta } + 2 R^{\beta 
\mu }{}_{\kappa \rho } R^{\tau \kappa }{}_{\alpha \beta } 
\prt^{\alpha }F^{\rho \gamma } \prt_{\mu }F_{\tau 
\gamma } + 2 R_{\beta \mu \kappa \rho } R^{\tau \kappa \alpha 
\beta } \prt_{\alpha }F_{\tau \gamma } \prt^{\mu 
}F^{\rho \gamma } \nn\\&&- 4 R_{\alpha }{}^{\nu \rho \gamma } 
R^{\tau \kappa \alpha \beta } \prt_{\beta }F_{\kappa \rho 
} \prt_{\nu }F_{\tau \gamma } + 2 R^{\alpha \beta \kappa 
\rho } R^{\tau }{}_{\kappa \alpha \beta } \prt_{\nu 
}F_{\tau \gamma } \prt^{\nu }F_{\rho }{}^{\gamma } + 
R^{\alpha \beta \rho \gamma } R^{\tau \kappa }{}_{\alpha 
\beta } \prt_{\nu }F_{\kappa \rho } \prt^{\nu }F_{\tau 
\gamma } \nn\\&&+ 2 R_{\beta }{}^{\mu }{}_{\kappa \rho } R^{\tau 
\kappa \alpha \beta } \prt^{\rho }F_{\mu }{}^{\nu } 
\prt_{\tau }F_{\alpha \nu } + R^{\mu \nu }{}_{\kappa 
}{}^{\rho } R^{\tau \kappa \alpha \beta } \prt_{\rho 
}F_{\alpha \beta } \prt_{\tau }F_{\mu \nu } + R^{\alpha 
\beta \kappa }{}_{\rho } R^{\tau }{}_{\kappa \alpha \beta } 
\prt^{\rho }F^{\mu \nu } \prt_{\tau }F_{\mu \nu } \nn\\&&+ 2 
R^{\beta \mu \kappa \rho } R_{\tau \kappa }{}^{\alpha 
}{}_{\beta } \prt_{\rho }F_{\alpha \nu } \prt^{\tau }F_{
\mu }{}^{\nu }\Big],\labell{RRFF}
\eeqa
where $R_{\mu\nu\alpha\beta}$ represents the linearized Riemann curvature and the graviton field $h_{\mu\nu}$ is expressed as $G_{\mu\nu} = \eta_{\mu\nu} + \kappa h_{\mu\nu}$. We have fixed the overall coefficient of these couplings to ensure consistency with the one-loop coupling $[R^2F'^2]_{13}$ derived in this work. Specifically, the one-loop coupling we have found  includes:
\beqa
S_{R^2F'^2}&=&-\frac{2}{\kappa^2}\frac{\pi^2\ell_s^6}{2^{3}.3}\int d^{10}x\sqrt{-G}[R^2F'^2]_{13}\,,
\eeqa
where the coupling $[R^2F'^2]_{13}$ is defined in equation \reef{RRF'F'}, incorporating the solution from \reef{sol}. This yields:
\beqa
[R^2F'^2]_{13}&=&- \frac{5}{2} e^{2 \Phi} R_{\gamma }{}^{\epsilon \varepsilon 
\mu } R_{\varepsilon \mu \delta \epsilon } \nabla_{\alpha 
}F_{\beta }{}^{\delta } \nabla^{\alpha }F^{\beta \gamma } + 
\frac{3}{4} e^{2 \Phi} R_{\varepsilon \mu \delta \epsilon } 
R^{\varepsilon \mu }{}_{\beta \gamma } \nabla_{\alpha 
}F^{\delta \epsilon } \nabla^{\alpha }F^{\beta \gamma }\nn\\&& -  
\frac{1}{4} e^{2 \Phi} R_{\varepsilon \mu \gamma \epsilon } 
R^{\varepsilon \mu }{}_{\beta \delta } \nabla_{\alpha 
}F^{\delta \epsilon } \nabla^{\alpha }F^{\beta \gamma } + 
\frac{5}{16} e^{2 \Phi} R^{\delta \epsilon \varepsilon \mu } 
R_{\varepsilon \mu \delta \epsilon } \nabla^{\alpha }F^{\beta 
\gamma } \nabla_{\beta }F_{\alpha \gamma } \nn\\&&+ e^{2 \Phi} 
R_{\gamma }{}^{\epsilon \varepsilon \mu } R_{\varepsilon \mu 
\delta \epsilon } \nabla^{\alpha }F^{\beta \gamma } 
\nabla_{\beta }F_{\alpha }{}^{\delta } -  \frac{7}{2} e^{2 
\Phi} R_{\delta }{}^{\varepsilon }{}_{\alpha }{}^{\mu } 
R_{\epsilon \mu \gamma \varepsilon } \nabla^{\alpha }F^{\beta 
\gamma } \nabla_{\beta }F^{\delta \epsilon } \nn\\&&-  \frac{3}{2} 
e^{2 \Phi} R_{\alpha }{}^{\varepsilon }{}_{\delta }{}^{\mu } R_{
\epsilon \mu \gamma \varepsilon } \nabla^{\alpha }F^{\beta 
\gamma } \nabla^{\delta }F_{\beta }{}^{\epsilon } + 
\frac{5}{2} e^{2 \Phi} R_{\gamma }{}^{\varepsilon }{}_{\alpha 
}{}^{\mu } R_{\epsilon \mu \delta \varepsilon } \nabla^{\alpha 
}F^{\beta \gamma } \nabla^{\delta }F_{\beta }{}^{\epsilon }\nn\\&& - 
 \frac{5}{4} e^{2 \Phi} R_{\varepsilon \mu \gamma \delta } R^{
\varepsilon \mu }{}_{\alpha \epsilon } \nabla^{\alpha 
}F^{\beta \gamma } \nabla^{\delta }F_{\beta }{}^{\epsilon } - 
3 e^{2 \Phi} R_{\beta }{}^{\mu }{}_{\alpha \gamma } R_{\epsilon 
\mu \delta \varepsilon } \nabla^{\alpha }F^{\beta \gamma } 
\nabla^{\delta }F^{\epsilon \varepsilon } \nn\\&&-  \frac{5}{2} e^{2 
\Phi} R_{\alpha }{}^{\mu }{}_{\delta \epsilon } R_{\varepsilon 
\mu \beta \gamma } \nabla^{\alpha }F^{\beta \gamma } 
\nabla^{\delta }F^{\epsilon \varepsilon } + 6 e^{2 \Phi} 
R_{\beta \epsilon \alpha }{}^{\mu } R_{\varepsilon \mu \gamma 
\delta } \nabla^{\alpha }F^{\beta \gamma } \nabla^{\delta 
}F^{\epsilon \varepsilon } \nn\\&&- 8 e^{2 \Phi} R_{\beta }{}^{\mu 
}{}_{\alpha \epsilon } R_{\varepsilon \mu \gamma \delta } 
\nabla^{\alpha }F^{\beta \gamma } \nabla^{\delta 
}F^{\epsilon \varepsilon }\labell{RRFF2}\,.
\eeqa
We find that the four-field truncation of these couplings reproduces precisely the same S-matrix elements as those derived from \reef{RRFF}. 

To see this explicitly, one should first remove the dilaton factor of \( e^{2\Phi} \) from the above couplings, replace the covariant derivative with the partial derivative, and substitute the Riemann curvature with its linearized form. Then, expressing the partial derivative of the RR one-form and the linearized Riemann curvature as
\beqa
\prt_\mu F_{\nu\alpha}&=&\kappa \left(\prt_\mu\prt_\nu c_\alpha-\prt_\mu\prt_\alpha c_\nu\right)\nn\\
R_{\mu\nu\alpha\beta}&=&\frac{\kappa}{2}\left(\prt_\beta\prt_\mu h_{\alpha\nu}-\prt_\beta\prt_\nu h_{\alpha\mu}-\prt_\alpha\prt_\mu h_{\beta\nu}+\prt_\alpha\prt_\nu h_{\beta\mu}\right)
\eeqa
one can proceed to momentum space.
Using the on-shell conditions \( k_1\cdot\zeta_1 = k_2\cdot\zeta_2 = 0 \) (where \( \zeta_1, \zeta_2 \) are the RR polarizations), \( k_3\cdot \epsilon_{3\mu} = k_4\cdot \epsilon_{4\mu} = 0 \) (where \( \epsilon_3, \epsilon_4 \) are the graviton polarization tensors), and employing momentum conservation to rewrite contractions of momenta in terms of the two independent Mandelstam variables \( k_1\cdot k_2 \) and \( k_1\cdot k_3 \), along with the identity \( k_3\cdot\epsilon_{4\mu} = -k_1\cdot\epsilon_{4\mu} - k_2\cdot\epsilon_{4\mu} \), one finds the couplings in \reef{RRFF} and \reef{RRFF2} produce identical S-matrix elements.

For the sphere-level amplitudes, reference \cite{Bakhtiarizadeh:2013zia} established the string-frame couplings involving two RR one-forms, one graviton, and one dilaton (see eq.(52) in \cite{Bakhtiarizadeh:2013zia}). The corresponding one-loop corrections take the form:
\beqa
S_{h\Phi CC}&=&-\frac{2}{\kappa^2}\frac{\pi^2\ell_s^6}{2^{6}.3}\int d^{10}x\sqrt{-G}\Big[-8 R_{\beta \mu \kappa \nu } \prt^{\beta }\prt^{\alpha }
\Phi \prt_{\gamma }F^{\mu \nu } \prt^{\kappa }F_{\alpha 
}{}^{\gamma } + 24 R_{\beta \mu \gamma \nu } \prt^{\beta 
}\prt^{\alpha }\Phi \prt_{\kappa }F^{\mu \nu } 
\prt^{\kappa }F_{\alpha }{}^{\gamma }\nn\\&& \qquad\qquad\qquad\qquad- 8 R_{\alpha \kappa 
\beta \nu } \prt^{\beta }\prt^{\alpha }\Phi 
\prt_{\gamma }F_{\mu }{}^{\nu } \prt^{\mu }F^{\gamma 
\kappa } + 8 R_{\beta \mu \gamma \nu } \prt^{\beta 
}\prt^{\alpha }\Phi \prt^{\kappa }F_{\alpha }{}^{\gamma } 
\prt^{\nu }F_{\kappa }{}^{\mu }\Big],\labell{FFRP}
\eeqa
where we have adopted the same normalization factor as in \reef{RRFF} for consistency. In contrast, the one-loop couplings derived in our work exhibit the following coupling:
\beqa
S_{RF'^2\Phi''}&=&-\frac{2}{\kappa^2}\frac{\pi^2\ell_s^6}{2^{3}.3}\int d^{10}x\sqrt{-G}[RF'^2\Phi'']_{9}\,,
\eeqa
where the coupling $[RF'^2\Phi'']_9$ is defined in equation \reef{RFFP}, incorporating the solution from \reef{sol}. This results in:
\beqa
[RF'^2\Phi'']_{9}&=&-10 e^{2 \Phi} R_{\epsilon \varepsilon \alpha \gamma } 
\nabla^{\alpha }F^{\beta \gamma } \nabla_{\beta }F^{\delta 
\epsilon } \nabla_{\delta }\nabla^{\varepsilon }\Phi + 3 e^{2 
\Phi} R_{\delta \varepsilon \gamma \epsilon } \nabla_{\alpha }
\nabla^{\varepsilon }\Phi \nabla^{\alpha }F^{\beta \gamma } 
\nabla^{\delta }F_{\beta }{}^{\epsilon } \nn\\&&+ 5 e^{2 \Phi} 
R_{\epsilon \varepsilon \alpha \delta } \nabla^{\alpha 
}F^{\beta \gamma } \nabla_{\gamma }\nabla^{\varepsilon }\Phi 
\nabla^{\delta }F_{\beta }{}^{\epsilon } + 2 e^{2 \Phi} 
R_{\epsilon \varepsilon \gamma \delta } \nabla^{\alpha 
}F^{\beta \gamma } \nabla_{\beta }\nabla_{\alpha }\Phi 
\nabla^{\delta }F^{\epsilon \varepsilon } \nn\\&&+ 2 e^{2 \Phi} 
R_{\delta \varepsilon \beta \gamma } \nabla^{\alpha }F^{\beta 
\gamma } \nabla^{\delta }F^{\epsilon \varepsilon } 
\nabla_{\epsilon }\nabla_{\alpha }\Phi - 2 e^{2 \Phi} 
R_{\gamma \varepsilon \alpha \delta } \nabla^{\alpha 
}F^{\beta \gamma } \nabla^{\delta }F^{\epsilon \varepsilon } 
\nabla_{\epsilon }\nabla_{\beta }\Phi\nn\\&& - 3 e^{2 \Phi} R_{\gamma 
\epsilon \delta \varepsilon } \nabla^{\alpha }F^{\beta \gamma 
} \nabla_{\beta }F_{\alpha }{}^{\delta } \nabla^{\epsilon 
}\nabla^{\varepsilon }\Phi + 4 e^{2 \Phi} R_{\delta \epsilon 
\alpha \varepsilon } \nabla^{\alpha }F^{\beta \gamma } 
\nabla_{\beta }F_{\gamma }{}^{\delta } \nabla^{\epsilon 
}\nabla^{\varepsilon }\Phi.
\eeqa
Note that one of the independent couplings in this structure is fixed to zero. Our analysis reveals that the four-field truncation of these couplings reproduces the identical S-matrix elements as those derived from equation \reef{FFRP}.

For the sphere-level interactions, previous work \cite{Bakhtiarizadeh:2013zia} has established the string-frame couplings involving two RR one-form fields and two dilatons (see eq.(66) in \cite{Bakhtiarizadeh:2013zia}). The corresponding one-loop corrections take the following form:
\beqa
S_{\Phi\Phi CC}&\!\!\!\!=\!\!\!\!&-\frac{2}{\kappa^2}\frac{\pi^2\ell_s^6}{2^{6}.3}\int d^{10}x\sqrt{-G}\Big[8 \prt^{\theta }\prt^{\eta }\Phi \prt^{\kappa 
}\prt_{\eta }\Phi \prt_{\nu }F_{\kappa \mu } 
\prt^{\nu }F_{\theta }{}^{\mu }\! - \!2 \prt_{\theta 
}\prt_{\eta }\Phi \prt^{\theta }\prt^{\eta }\Phi 
\prt_{\nu }F_{\kappa \mu } \prt^{\nu }F^{\kappa \mu }\Big].\labell{FFPP}
\eeqa
In contrast, the one-loop couplings derived in this work exhibit the following distinctive features: 
\beqa
S_{RF'^2\Phi''}&=&-\frac{2}{\kappa^2}\frac{\pi^2\ell_s^6}{2^{3}.3}\int d^{10}x\sqrt{-G}[F'^2\Phi''^2]_{6}\,.
\eeqa
where the coupling $[F'^2\Phi''^2]_6$ is defined in equation \reef{CCPP}, incorporating the solution from \reef{sol}. This yields:
\beqa
[F'^2\Phi''^2]_{6}&\!\!\!\!=\!\!\!\!&2 e^{2 \Phi} \nabla_{\alpha }F^{\delta \epsilon } 
\nabla^{\alpha }F^{\beta \gamma } \nabla_{\delta 
}\nabla_{\beta }\Phi \nabla_{\epsilon }\nabla_{\gamma }\Phi + 
2 e^{2 \Phi} \nabla^{\alpha }F^{\beta \gamma } \nabla_{\delta 
}\nabla_{\alpha }\Phi \nabla^{\delta }F_{\beta }{}^{\epsilon 
} \nabla_{\epsilon }\nabla_{\gamma }\Phi \nn\\&&+ 5 e^{2 \Phi} 
\nabla_{\alpha }F_{\beta }{}^{\delta } \nabla^{\alpha 
}F^{\beta \gamma } \nabla_{\gamma }\nabla^{\epsilon }\Phi 
\nabla_{\epsilon }\nabla_{\delta }\Phi -  e^{2 \Phi} 
\nabla^{\alpha }F^{\beta \gamma } \nabla_{\beta }F_{\alpha 
}{}^{\delta } \nabla_{\gamma }\nabla^{\epsilon }\Phi 
\nabla_{\epsilon }\nabla_{\delta }\Phi \nn\\&&- 2 e^{2 \Phi} \nabla^{
\alpha }F^{\beta \gamma } \nabla_{\gamma }\nabla_{\alpha 
}\Phi \nabla^{\delta }F_{\beta }{}^{\epsilon } 
\nabla_{\epsilon }\nabla_{\delta }\Phi -  \frac{3}{2} e^{2 
\Phi} \nabla^{\alpha }F^{\beta \gamma } \nabla_{\beta 
}F_{\alpha \gamma } \nabla^{\delta }\nabla^{\epsilon }\Phi 
\nabla_{\epsilon }\nabla_{\delta }\Phi.
\eeqa
We demonstrate that the four-field truncation of these couplings reproduces precisely the same S-matrix elements as those derived from equation \reef{FFPP}.

Given the non-vanishing S-matrix elements involving the RR one-form and dilaton in the string frame, it is fundamentally impossible to express the couplings in a scheme that eliminates dilaton derivatives. Our systematic investigation, following the approach of \cite{Garousi:2020lof}, confirms that  the 359 non-zero couplings can not be reformulated in a scheme that removes either: (i) all dilaton couplings or (ii) all gravity-dilaton interactions. This obstruction persists despite exhaustive attempts to find such representations.

\section{Reduction on K3 and S-duality in 6D}

Type IIA string theory compactified on a K3 surface is known to be S-dual to heterotic string theory on  $T^4$ (see \eg \cite{Becker:2007zj}), with the dilatons transforming as $\Phi\rightarrow -\Phi$; aside from 80 additional scalar moduli fields that we disregard, this duality requires the gauge fields to map between theories, with 24 gauge fields appearing in each case: for type IIA these consist of (1) the RR one-form, (2) twenty-two gauge fields from K3 harmonic expansion of the RR three-form, and (3) one gauge field from Hodge dualization of the RR four-form, while the heterotic theory has (1) four metric-derived gauge fields from  $T^4$ reduction, (2) sixteen from the Cartan subalgebra of $SO(32)$ or $E_8\times E_8$, and (3) four from $B$-field reduction on  $T^4$ - though we focus specifically on how the RR one-form in type IIA transforms into one particular metric-derived gauge field in heterotic theory under this S-duality.

\subsection{S-duality at lowest order in derivatives}

To establish our conventions, we first study the S-duality transformation at the two-derivative level. Working in the ansatz where the dilaton and RR one-form are K3-independent and the metric takes the block-diagonal form
\beqa
ds^2 = G_{\mu\nu}(x)dx^\mu dx^\nu + g_{ab}(y)dy^a dy^b,\label{K3red}
\eeqa
with $y^a$ denoting K3 coordinates, the dimensional reduction of the leading two-derivative action \reef{S0} in the metric-dilaton-RR sector yields
\beqa
{\bf S}^{(0)} = -\frac{2V}{\kappa_{10}^2}\int d^6 x \sqrt{-G} e^{-2\Phi} \left[ R + 4\nabla_\mu \Phi \nabla^\mu \Phi - \frac{1}{4} e^{2\Phi}F_{\mu\nu} F^{\mu\nu} \right], \label{IIA}
\eeqa
where $V \equiv \int_{K3} d^4 y \sqrt{g}$ is the K3 volume.

In contrast, the leading-order ten-dimensional effective action of heterotic string theory for the metric \( G' \) and dilaton \( \Phi' \) matches the corresponding terms in the type IIA action \reef{S0}. Under the assumption that the dilaton is independent of the \( T^4 \) coordinates and decomposing \( T^4 = S^1 \times T^3 \), the metric takes the form:
\beqa
ds'^2 = G'_{MN}(x)dx^M dx^N + g'_{ij}(z)dz^i dz^j,\labell{dsHet}
\eeqa
where  $z^i$ are the coordinates of $T^3$,  $x^M$ denotes both the circle coordinate of $S^{(1)}$ and the six-dimensional spacetime coordinates. Performing the KK reduction on the seven-dimensional metric using the ansatz:
\beqa  
G'_{MN} = \left(\matrix{G'_{\mu\nu} + R_1^{-2}C'_{\mu} C'_{\nu} &   C'_{\mu} \cr  C'_{\nu} & R_1^2 &}\!\!\!\!\!\right),  
  \labell{reduc1}  
\eeqa  
where $R_1$ represents the radius of the circle coordinate $y$, while indices  $\mu,\nu$ label the six-dimensional spacetime directions orthogonal to $y$. The dimensional reduction then yields the following effective action in six dimensions  ( see \eg \cite{Kaloper:1997ux}):
 \beqa
{\bf S'}^{(0)} = -\frac{2V'}{\kappa'^2_{10}}\int d^6 x \sqrt{-G'} e^{-2\Phi'} \left[ R' + 4\nabla_\mu \Phi' \nabla^\mu \Phi' - \frac{1}{4} F'_{\mu\nu} F'^{\mu\nu} \right], \label{Het}
\eeqa
where the ten-dimensional gravitational coupling is given by  $\kappa'^2_{10}=\frac{1}{\pi}(2\pi\ell'_s)^8g'^2_s$, with $\ell_s'$ being the heterotic string length scale, and $V'=\int_{S^{(1)}} R_1dy \int_{T^3} d^3 z \sqrt{g'} $ represents the $T^4$ volume.  In above action, the prime notation on the fields indicates that they belong to the heterotic theory.

The S-duality transformation identifies (i) the NS5-brane wrapped on K3 with the heterotic string, and (ii) the type IIA string with the NS5-brane wrapped on $T^4$ in the heterotic theory  \cite{Becker:2007zj}. Matching their tensions yields two key relations. First, the six-dimensional string couplings are related as \( g_{6H} = g_{6A}^{-1} \), where \( g_{6A}^2 = g_s^2/[V/(2\pi\ell_s)^4] \) and \( g_{6H}^2 = g_s'^2/[V'/(2\pi\ell_s')^4] \) define the respective couplings. Second, the six-dimensional gravitational coupling remains invariant:
\beqa
\kappa_{10}^2/V = \kappa_{10}'^2/V'\,.\label{VV'}
\eeqa
The six-dimensional field transformations
\beqa
G_{\mu\nu}' = e^{-2\Phi}G_{\mu\nu}\,,\quad \Phi' = -\Phi\,,\quad C'_\mu = C_\mu\,,\label{sduality}
\eeqa
then precisely map the tree-level heterotic action \reef{Het} to the type IIA tree-level action \reef{IIA}. The S-duality transformation exchanges the KK vector in heterotic theory with the RR one-form in type IIA theory, with these mappings satisfying the discrete group  $\MZ_2$.  

\subsection{S-duality at four-derivative order}

The K3 reduction of eight-derivative couplings via the ansatz \reef{K3red} generates both eight-derivative terms (which we disregard) and four-derivative couplings. Crucially, the non-flatness of K3 surfaces introduces non-vanishing curvature contributions. In particular, the integrated Riemann-squared term yields a topological invariant ( see \eg \cite{Liu:2019ses}):
\beqa
\frac{1}{32\pi^2}\int_{K3} d^4y\sqrt{g}R_{abcd}R^{abcd}&=&24\,.\labell{K3int}
\eeqa
This topological constraint plays a key role in producing four-derivative couplings when applied to eight-derivative couplings involving the Riemann-squared term. 

To determine the four-derivative couplings, we first restrict our analysis to the gravitational sector for simplicity. Using our earlier results, the K3 reduction of the ten-dimensional one-loop gravity couplings in \reef{ttee1} produces the following four-derivative term in six dimensions:
\beqa
S_{R^2}&=&-\frac{2}{\kappa^2}\frac{4\pi^4\ell_s^6}{3}\int d^{6}x\sqrt{-G}\Big[\frac{3}{2}R_{\mu\nu\alpha\beta}R^{\mu\nu\alpha\beta}\Big].\labell{SR2}
\eeqa
In contrast, the 10-dimensional heterotic theory at tree-level contains four-derivative gravity-dilaton couplings of the form \cite{Metsaev:1987zx}:
\beqa
S'_{R^2}&=&-\frac{2}{\kappa'^2_{10}}\frac{\ell'^2_s}{8}\int d^{10}x\sqrt{-G'}e^{-2\Phi'}R'_{\mu\nu\alpha\beta}R'^{\mu\nu\alpha\beta}\,.\labell{SHet}
\eeqa 
 The dimensional reduction of this term on $T^4$ yields the following six-dimensional gravity-dilaton coupling:
\beqa
S'_{R^2}&=&-\frac{2}{\kappa'^2_{10}}\frac{\ell'^2_sV'}{8}\int d^{6}x\sqrt{-G'}e^{-2\Phi'}R'_{\mu\nu\alpha\beta}R'^{\mu\nu\alpha\beta}\,.\labell{S'6}
\eeqa 
The NS5-brane of heterotic theory, when wrapped on $T^4$, transforms under S-duality into the fundamental string of type IIA theory. The equality of their tensions yields the relation
\beqa
\frac{2\pi V'}{(2\pi\ell_s')^6g_s'^2}&=&\frac{1}{2\pi\ell_s^2}\,.\labell{NS5F}
\eeqa
Together with the S-duality transformation \reef{sduality}, this relation maps the action \reef{S'6} to \reef{SR2} for a constant dilaton.


We now analyze the complete set of four-derivative couplings involving the gravitational sector, dilaton, and RR one-form in six dimensions. This includes both the curvature-squared term \reef{SR2} and additional structures involving the dilaton and RR field strengths. Applying the topological constraint \reef{K3int} from K3 compactification, we derive the following one-loop effective couplings in six dimensions:
\beqa
S_{6D}&\!\!\!\!\!=\!\!\!\!\!&-\frac{2}{\kappa^2}\frac{4\pi^4\ell_s^6}{3}\int d^{6}x\sqrt{-G}\Big[ 
\frac{3}{2} R_{\alpha \beta \gamma \delta } 
R^{\alpha \beta \gamma \delta }+\frac{51}{16} e^{4 \Phi} F_{\alpha }{}^{\gamma } F^{\alpha 
\beta } F_{\beta }{}^{\delta } F_{\gamma \delta } + 
\frac{50151}{2560} e^{4 \Phi} F_{\alpha \beta } F^{\alpha 
\beta } F_{\gamma \delta } F^{\gamma \delta }\nn\\&& - 3 e^{2 \Phi} 
F^{\alpha \beta } F^{\gamma \delta } R_{\alpha \beta 
\gamma \delta } - 3 e^{2 \Phi} F^{\alpha \beta } F^{\gamma 
\delta } R_{\alpha \gamma \beta \delta }  + \frac{44343}{400} 
e^{2 \Phi} F_{\beta \gamma } F^{\beta \gamma } 
\nabla_{\alpha }\Phi \nabla^{\alpha }\Phi \nn\\&&-  
\frac{13581}{200} e^{2 \Phi} F_{\alpha }{}^{\gamma } F_{\beta 
\gamma } \nabla^{\alpha }\Phi \nabla^{\beta }\Phi -  
\frac{36}{5} \nabla_{\alpha }\Phi \nabla^{\alpha }\Phi 
\nabla_{\beta }\Phi \nabla^{\beta }\Phi -  \frac{24}{5} 
\nabla^{\alpha }\Phi \nabla_{\beta }\nabla_{\alpha }\Phi 
\nabla^{\beta }\Phi\nn\\&& -  \frac{24}{5} \nabla_{\beta 
}\nabla_{\alpha }\Phi \nabla^{\beta }\nabla^{\alpha }\Phi + 
\frac{15}{2} e^{2 \Phi} F^{\beta \gamma } \nabla^{\alpha 
}\Phi \nabla_{\gamma }F_{\alpha \beta } -  \frac{18501}{400} 
e^{2 \Phi} F_{\alpha }{}^{\gamma } F^{\alpha \beta } \nabla_{
\gamma }\nabla_{\beta }\Phi \nn\\&&+ \frac{15}{2} e^{2 \Phi} 
\nabla_{\beta }F_{\alpha \gamma } \nabla^{\gamma }F^{\alpha 
\beta }\Big].\labell{SR21}
\eeqa
The above six-dimensional effective action is formulated in the specific minimal scheme developed in the Appendix.

As observed in \cite{Garousi:2025xqn}, S-duality remains uncorrected at higher-derivative orders. To study the action under S-duality, we must therefore extend it to the most general form—differing only through six-dimensional field redefinitions, integration by parts, and applications of the Bianchi identities.
To construct this generalization, we first derive a maximal basis in six dimensions that includes all independent terms up to total derivatives and Bianchi identities, following the procedure established for the ten-dimensional theory in the Appendix (but excluding field redefinition terms). This basis consists of 19 couplings. By equating it with the couplings in \reef{SR21} modulo field redefinitions, total derivatives, and Bianchi identities, we ultimately obtain the following action:
\beqa
S_{6D}&=&-\frac{2}{\kappa^2}\frac{4\pi^4\ell_s^6}{3}\int d^{6}x\sqrt{-G}\Big[\frac{3}{2} R_{\alpha \beta \gamma \delta 
} R^{\alpha \beta \gamma \delta } +a_{1} e^{4 \Phi} F_{\alpha }{}^{\gamma } F^{\alpha \beta } 
F_{\beta }{}^{\delta } F_{\gamma \delta } + a_{2} e^{4 \Phi} F_{
\alpha \beta } F^{\alpha \beta } F_{\gamma \delta } 
F^{\gamma \delta } \nn\\&& + (- \frac{9}{8} - 2 a_{1} + 2 a_{19} -  
\frac{1}{2} a_{3}) e^{2 \Phi} F_{\alpha }{}^{\gamma } F^{\alpha 
\beta } R_{\beta \gamma } + \frac{1}{4800} (-65403 + 
800 a_{1} + 1800 a_{10} \nn\\&&+ 900 a_{11} + 150 a_{12} - 825 a_{13} - 450 a_{14} + 450 a_{15} - 
525 a_{16} + 300 a_{17} + 1200 a_{19} \nn\\&&+ 3200 a_{2} - 1200 a_{3}) e^{2 \Phi} 
F_{\alpha \beta } F^{\alpha \beta } R + \frac{1}{16} 
(-12 + 4 a_{10} + 4 a_{11} - 2 a_{13} -  a_{14} + a_{15} \nn\\&&- 2 a_{16} - 8 a_{3}) 
R^2 + (- \frac{3}{4} -  a_{19}) e^{2 \Phi} F^{\alpha 
\beta } F^{\gamma \delta } R_{\alpha \beta \gamma 
\delta } + a_{3} R_{\alpha \beta } 
R^{\alpha \beta } + (\frac{20001}{400} 
+ \frac{2}{3} a_{1}\nn\\&& -  \frac{3}{8} a_{12} + \frac{5}{16} a_{13} + 
\frac{3}{8} a_{14} -  \frac{3}{16} a_{15} + \frac{1}{16} a_{16} -  
\frac{3}{4} a_{17} - 3 a_{19} + \frac{8}{3} a_{2}) e^{2 \Phi} F_{\beta 
\gamma } F^{\beta \gamma } \nabla_{\alpha }\Phi 
\nabla^{\alpha }\Phi\nn\\&& + a_{10} R \nabla_{\alpha }\Phi 
\nabla^{\alpha }\Phi + a_{11} R^{\alpha \beta } 
\nabla_{\beta }\nabla_{\alpha }\Phi + a_{12} e^{2 \Phi} 
F_{\alpha }{}^{\gamma } F_{\beta \gamma } \nabla^{\alpha 
}\Phi \nabla^{\beta }\Phi + a_{13} R_{\alpha \beta } 
\nabla^{\alpha }\Phi \nabla^{\beta }\Phi \nn\\&&+ a_{14} \nabla_{\alpha 
}\Phi \nabla^{\alpha }\Phi \nabla_{\beta }\Phi \nabla^{\beta 
}\Phi + a_{15} \nabla^{\alpha }\Phi \nabla_{\beta 
}\nabla_{\alpha }\Phi \nabla^{\beta }\Phi + a_{16} \nabla_{\beta 
}\nabla_{\alpha }\Phi \nabla^{\beta }\nabla^{\alpha }\Phi \nn\\&&+ 
a_{17} e^{2 \Phi} F^{\beta \gamma } \nabla^{\alpha }\Phi 
\nabla_{\gamma }F_{\alpha \beta } + \frac{1}{4} (-3 - 16 a_{1} + 
2 a_{12} + a_{13} -  a_{16}) e^{2 \Phi} F_{\alpha }{}^{\gamma } 
F^{\alpha \beta } \nabla_{\gamma }\nabla_{\beta }\Phi \nn\\&&+ a_{19} 
e^{2 \Phi} \nabla_{\gamma }F_{\alpha \beta } \nabla^{\gamma 
}F^{\alpha \beta }
\Big],\labell{SR22}
\eeqa
where ${a_{1},a_{2},a_{3},\cdots} $ represent twelve arbitrary scheme parameters. The actions \reef{SR22} and \reef{SR21} are physically equivalent, being related by allowed field redefinitions of the metric, dilaton and RR one-form fields.

In contrast, dimensional reduction of the tree-level heterotic effective action \reef{SHet} via the compactification schemes \reef{dsHet} and \reef{reduc1} yields the following six-dimensional couplings:
\beqa
S'_{6D}&=&-\frac{2}{\kappa'^2_{10}}\frac{\ell'^2_sV'}{8}\int d^{6}x\sqrt{-G'}e^{-2\Phi'}\Big[R'_{\alpha \beta \gamma \delta } R'^{\alpha 
\beta \gamma \delta }+\frac{5}{8} F'_{\alpha }{}^{\gamma } F'^{\alpha \beta } 
F'_{\beta }{}^{\delta } F'_{\gamma \delta } + \frac{3}{8} 
F'_{\alpha \beta } F'^{\alpha \beta } F'_{\gamma \delta } 
F'^{\gamma \delta }\nn\\&& -  F'^{\alpha \beta } F'^{\gamma \delta } 
R'_{\alpha \beta \gamma \delta } -  F'^{\alpha \beta } 
F'^{\gamma \delta } R'_{\alpha \gamma \beta \delta }  
 + \nabla_{\gamma }F'_{\alpha \beta } 
\nabla^{\gamma }F'^{\alpha \beta }\Big].\labell{SSHet}
\eeqa
This six-dimensional heterotic action inherits its scheme dependence from the original ten-dimensional action \reef{SHet}. To properly analyze its S-duality properties, we must generalize the action to an arbitrary scheme through field redefinitions. The complete form of the action in a general scheme is given by:
\beqa
S'_{6D}&=&-\frac{2}{\kappa^2}\frac{4\pi^4\ell^6_s}{3}\int d^{6}x\sqrt{-G'}e^{-2\Phi'}\Big[ \frac{3}{2} R'_{\alpha \beta \gamma 
\delta } R'^{\alpha \beta \gamma \delta }+b_{1} F'_{\alpha }{}^{\gamma } F'^{\alpha \beta } F'_{\beta 
}{}^{\delta } F'_{\gamma \delta } + b_{2} F'_{\alpha \beta } 
F'^{\alpha \beta } F'_{\gamma \delta } F'^{\gamma \delta }  \nn\\&&+ 
\frac{1}{4} (2 b_{11} + 2 b_{12} + b_{13} -  b_{16}) R'_{\alpha \beta 
} R'^{\alpha \beta } + \frac{1}{8} (-9 - 16 b_{1} - 2 b_{11} - 
2 b_{12} -  b_{13} + b_{16}  \nn\\&&+ 16 b_{19}) F'_{\alpha }{}^{\gamma } F'^{\alpha 
\beta } R'_{\beta \gamma } + \frac{1}{192} (162 + 32 b_{1} 
- 36 b_{10} - 12 b_{12} - 6 b_{13} + 9 b_{14} + 2 b_{16} + 8 b_{17}  \nn\\&&+ 8 b_{18} - 32 b_{19}- 
256 b_{2}) F'_{\alpha \beta } F'^{\alpha \beta } R + 
\frac{1}{16} (4 b_{10} -  b_{14}) R'^2 + (- \frac{3}{4} -  b_{19}) 
F'^{\alpha \beta } F'^{\gamma \delta } R'_{\alpha \beta 
\gamma \delta } +\nn\\&& + 
\frac{1}{48} (162 + 32 b_{1} - 12 b_{12} - 9 b_{14} - 3 b_{15} + 2 b_{16} + 8 b_{17} 
+ 8 b_{18} - 32 b_{19} \labell{S2Hetgen}\\&&- 256 b_{2}) F'_{\beta \gamma } F'^{\beta \gamma } 
\nabla_{\alpha }\Phi'\nabla^{\alpha }\Phi'+ b_{10} R' 
\nabla_{\alpha }\Phi'\nabla^{\alpha }\Phi'+ b_{11} 
R'^{\alpha \beta } \nabla_{\beta }\nabla_{\alpha }\Phi'
+ b_{12} F'_{\alpha }{}^{\gamma } F'_{\beta \gamma } 
\nabla^{\alpha }\Phi'\nabla^{\beta }\Phi'\nn\\&&+ b_{13} 
R'_{\alpha \beta } \nabla^{\alpha }\Phi'\nabla^{\beta 
}\Phi'+ b_{14} \nabla_{\alpha }\Phi'\nabla^{\alpha }\Phi'
\nabla_{\beta }\Phi'\nabla^{\beta }\Phi'+ b_{15} \nabla^{\alpha 
}\Phi'\nabla_{\beta }\nabla_{\alpha }\Phi'\nabla^{\beta 
}\Phi'\nn\\&&+ b_{16} \nabla_{\beta }\nabla_{\alpha }\Phi'\nabla^{\beta 
}\nabla^{\alpha }\Phi'\!+\! b_{17} F'^{\beta \gamma } \nabla^{\alpha 
}\Phi'\nabla_{\gamma }F'_{\alpha \beta } \!+ \!b_{18} F'_{\alpha }{}^{
\gamma } F'^{\alpha \beta } \nabla_{\gamma }\nabla_{\beta 
}\Phi'+ b_{19} \nabla_{\gamma }F'_{\alpha \beta } \nabla^{\gamma 
}F'^{\alpha \beta }\Big],\nn
\eeqa
where $b_{1},b_{2},\cdots$ represent twelve arbitrary scheme parameters, and we have incorporated the  relation \reef{S'6}.  The action \reef{S2Hetgen} is physically equivalent to \reef{SSHet}, being related by allowed field redefinitions of the metric, dilaton and KK vector fields.

Under the S-duality transformation \reef{sduality}, the action \reef{SR22} maps precisely to \reef{S2Hetgen} modulo total derivative terms and Bianchi identity applications. This correspondence establishes the following parameter relations between the two actions:
\beqa
&&a_{2} = 65403/6400 - 
  a_{1}/4 + (3 a_{13})/64 + (3 a_{14})/64 - (3 a_{15})/128, a_{3} = -(20901/200) + 
  a_{10} \nn\\&&+ a_{11}/2 - (3 a_{13})/4 - a_{14}/2 + (3 a_{15})/8 - a_{16}/4, b_{1} = (4 a_{1})/
  3, b_{10} = 
 13734/25 - 4 a_{10} \nn\\&&- (4 a_{11})/3 + 4 a_{13} + (8 a_{14})/3 - 
  2 a_{15} + (4 a_{16})/3, b_{11} = 
 27268/25 + (8 a_{10})/3 + (20 a_{11})/3 \nn\\&&+ (4 a_{13})/3 + 2 a_{14} - (2 a_{15})/3 - 
  4 a_{16}, b_{12} = 
 6767/25 - (8 a_{10})/3 - (4 a_{11})/3 + (4 a_{13})/3 \nn\\&&+ (4 a_{14})/3 - 
  a_{15} + (4 a_{16})/3, b_{13} = -(13834/25) + 12 a_{10} + (16 a_{11})/3 - 
  6 a_{13} - (13 a_{14})/3 \nn\\&&+ (11 a_{15})/3 - (14 a_{16})/3, b_{14} = 
 27468/25 - (32 a_{10})/3 - (16 a_{11})/3 + (32 a_{13})/3 + (20 a_{14})/
   3 \nn\\&&- (16 a_{15})/3 + (16 a_{16})/3, b_{15} = 
 26468/25 + (88 a_{10})/3 + (64 a_{11})/3 - (68 a_{13})/3 - (34 a_{14})/
   3 \nn\\&&+ (34 a_{15})/3 - 20 a_{16}, b_{16} = 
 13634/5 + (20 a_{10})/3 + (40 a_{11})/3 + (10 a_{13})/3 + 
  5 a_{14} - (5 a_{15})/3 \nn\\&&- (26 a_{16})/3, b_{17} = 
 130273/100 - (32 a_{1})/3 + 4 a_{10} + 
  2 a_{11} + (5 a_{12})/3 + (9 a_{13})/2 + (14 a_{14})/3\nn\\&& - (11 a_{15})/6 - (11 a_{16})/
   6 + 2 a_{17} + (8 a_{19})/3, b_{18} = 
 6742/25 - (16 a_{1})/3 - (8 a_{10})/3 - (4 a_{11})/3\nn\\&& - (2 a_{12})/3 + (5 a_{13})/
   3 + (4 a_{14})/3 - a_{15} + a_{16} + (32 a_{19})/3, b_{19} = (4 a_{19})/3, b_{2} = 
 21801/1600 \nn\\&&- a_{1}/3 + a_{13}/16 + a_{14}/16 - a_{15}/32\,.\labell{rel}
\eeqa
The parameter relations in \reef{rel}  exhibit an important subtlety: the S-duality transformation only produces exact equivalence between the generalized form of action \reef{SR21} and the heterotic action \reef{S2Hetgen} because the minimal scheme \reef{SR21} leads to two mutually incompatible parameter constraints that cannot be simultaneously satisfied. 

Since S-duality forms a $\mathbb{Z}_2$ group, the S-duality transformation of the classical six-dimensional heterotic action \reef{S2Hetgen} also yields the one-loop six-dimensional type IIA effective action \reef{SR22}, up to total derivative terms and applications of the Bianchi identities. Building on the known S-duality symmetry between $SO(32)$ heterotic and type I superstring theories \cite{Witten:1995ex}, analogous calculations were performed in \cite{Garousi:2025xqn} to derive the S-duality transformation of the classical ten-dimensional heterotic theory at four-derivative order. This transformation generates the corresponding disk-level and higher-genus couplings in the ten-dimensional type I theory.

The above calculations confirm that the two actions \reef{SR21} and \reef{SSHet} are indeed S-dual to each other when appropriate field redefinitions are included. This provides a nontrivial verification of S-duality between type IIA theory on K3 and heterotic theory on \( T^4 \).

\section{Conclusion}

In this work, we systematically investigate the dimensional reduction of pure gravity couplings at $\mathcal{O}(\ell_p^6)$, deriving the corresponding one-loop corrections in type IIA theory. These corrections are expressed in a minimal string-frame basis consisting of 359 non-zero couplings, whose four-field interactions exactly match known results obtained from linear S- and T-duality transformations of the standard $t_8t_8\bar{R}^4$ couplings.
Furthermore, by compactifying the type IIA couplings on K3, we demonstrate that the resulting one-loop four-derivative terms transform under S-duality into the tree-level $R^2$ couplings of heterotic theory on $T^4$. This correspondence requires appropriate field redefinitions in both six-dimensional theories.
The same S-duality relates the tree-level $\alpha'$ couplings of the six-dimensional type IIA theory to the one-loop couplings of the six-dimensional heterotic theory \cite{Antoniadis:1997eg}. While the latter couplings vanish \cite{Ellis:1987dc,Ellis:1989fi}, it has been shown in \cite{Liu:2019ses,Garousi:2025xqn} that the NS-NS couplings in the former case also vanish.

The four-field M-theory couplings $R^2F'^2$ and $F'^4$ were derived in \cite{Peeters:2005tb} using the superparticle method. When reduced with vanishing B-field, dilaton, and RR one-form, these couplings yield identical one-loop gravity-RR three-form couplings in type IIA theory. We have verified that these four-field couplings match those obtained in \cite{Garousi:2013lja} through S- and T-duality transformations of the standard $t_8t_8\bar{R}^4$ coupling.

\vskip .3 cm
{\bf Acknowledgments}:   I would like to thank Linus Wulff for useful conversations.

\vskip 0.5 cm
{\Large \bf Appendix: }{\large\bf Minimal coupling basis}
\vskip 0.5 cm

This appendix establishes the minimal basis of eight-derivative metric-dilaton-RR one-form couplings by first generating all covariant, RR-gauge invariant terms with even RR field strength counts and appropriate $e^{\Phi}$ dilaton factors, yielding 18,462 candidate couplings through xAct package \cite{Nutma:2013zea} as 
\beqa
\mathcal{L}'=& c'_1 R_{\beta}{}^{\delta \epsilon\nu} R_{\gamma\epsilon}{}^{\mu \zeta} \
R_{\delta\nu\mu\zeta}F_\alpha{}^\beta F^{\alpha\gamma}e^{2\Phi}+\cdots\,,\labell{L3}
\eeqa
where $c'_1,\cdots, c'_{18462}$ are some coupling constants.
However, these terms are not independent due to: (i) total derivative redundancies, (ii) field redefinition equivalences, and (iii) Bianchi identity constraints, necessitating systematic reduction to a minimal set.

To systematically eliminate redundant terms arising from total derivatives and field redefinition ambiguities in the Lagrangian $\mathcal{L}'$, we introduce the following compensating terms to the action:
\beqa
 \mathcal{J}&\!\!\!\!\!\equiv\!\!\!\!\!\!&  \nabla_\alpha ({\cal I}^\alpha)\,,\labell{J3}\nn\\
\mathcal{K}
&\!\!\!\!\!\equiv\!\!\!\!\!\!&e^{\Phi} \nabla_{\beta}F^{\beta\alpha }\delta C_{\alpha}-\frac{1}{8}e^{2\Phi}F_{\alpha\beta}F^{\alpha\beta}\delta G^{\mu}{}_\mu -(  R^{\alpha \beta}-\frac{1}{2} e^{2\Phi}F^{\alpha \gamma } F^{\beta}{}_{\gamma }+ 2 \nabla^{\beta}\nabla^{\alpha}\Phi)\delta G_{\alpha\beta}
\nn\\
&&-2( R+ 4 \nabla_{\alpha}\nabla^{\alpha}\Phi -4 \nabla_{\alpha}\Phi \nabla^{\alpha}\Phi)(\delta\Phi-\frac{1}{4}\delta G^{\mu}{}_\mu) .\labell{eq.13}
\eeqa
Here, the vector ${\cal I}^\alpha$ encompasses all possible covariant and gauge-invariant seven-derivative terms constructed from the metric, dilaton, and RR one-form fields. Our systematic classification identifies 10,349 such independent vectors, with corresponding coefficients  $J_1,\cdots, J_{10349}$. The field redefinition contributions arise from the following infinitesimal transformations of the fundamental fields:
\begin{eqnarray}
G_{\mu\nu}&\rightarrow &G_{\mu\nu}+\ell_s^6 \,e^{2\Phi}\delta G_{\mu\nu}\,,\nn\\
C_{\mu}&\rightarrow &C_{\mu}+ \ell_s^6\,e^{\Phi}\delta C_{\mu}\,,\nn\\
\Phi &\rightarrow &\Phi+ \ell_s^6\,e^{2\Phi}\delta\Phi\,.\labell{gbp}
\end{eqnarray}
The perturbations are introduced into the leading-order action \reef{S0} while retaining only linear terms, with integration by parts applied as shown in \reef{eq.13}, where $\delta C_{\mu}$ contains odd powers of the RR field strength while  $\delta G_{\mu\nu}$ and $\delta\Phi$ contain even powers - specifically manifesting as 3,265 metric perturbations (coefficients {$e_i$}), 1,621 RR-field perturbations {$g_i$}, and 656 dilaton perturbations {$f_i$}. By augmenting  $\mathcal{L}'$ with these field redefinitions and total derivative terms, we obtain an equivalent Lagrangian  ${\cal L}$ with transformed parameters ${c_i}$, yielding the fundamental relation:
\beqa
\Delta-{\cal J}-{\cal K}&=&0\,.\labell{DLK}
\eeqa
The difference $\Delta={ \cal L}-\mathcal{L}'$  preserves the same functional form as $\mathcal{L}'$ but with modified coefficients $\delta c_i= c_i-c'_i$ , representing the net effect of the field redefinitions and total derivative terms.

To systematically solve equation \reef{DLK}, we must first express it in terms of linearly independent couplings by enforcing the relevant Bianchi identities:
\beqa
 R_{\alpha[\beta\gamma\delta]}&=&0\,,\nn\\
 \nabla_{[\mu}R_{\alpha\beta]\gamma\delta}&=&0\,,\labell{bian}\\
\nabla_{[\mu}F_{\alpha\beta]}&=&0\,,\nn\\
{[}\nabla,\nabla{]}\mathcal{O}-R\mathcal{O}&=&0\,.\nn
\eeqa
To implement the Bianchi identities while working in a non-gauge-invariant formulation, we adopt a local inertial frame where covariant derivatives reduce to partial derivatives and first metric derivatives vanish, while simultaneously expressing all occurrences of $\prt F$ in \reef{DLK} through the fundamental relation $F = dC$ - this combined approach automatically satisfies all Bianchi identities for both the curvature tensor and RR field strength, as established in \cite{Garousi:2019cdn}.

Through this systematic procedure, all terms on the left-hand side of \reef{DLK} can be expressed in terms of linearly independent (though non-gauge-invariant) couplings, whose vanishing coefficients yield algebraic equations with two distinct classes of solutions: (i) 377 relations involving exclusively the $\delta c_i$ parameter variations, and (ii) additional equations mixing $\delta c_i$ with total derivative and field redefinition coefficients (which we disregard). The invariant count of 377 relations in the first class determines the dimension of the physically meaningful coupling space in ${\cal L}'$, as this number remains unchanged under scheme transformations - while particular schemes may nullify certain coefficients in ${\cal L}'$, substituting these back into \reef{DLK} preserves exactly 377 constraints among the  $\delta c_i$ parameters.

To systematically eliminate redundant couplings while preserving the 377 fundamental relations between the $\delta c_i$ parameters, we impose a specific scheme choice by nullifying all coefficients in ${\cal L}'$ containing the following structures: $R$,  $\nabla_\mu F^{\mu\alpha}$,  $\nabla_\mu\nabla^\mu\Phi$, $\nabla_\mu\nabla_\nu F_{\alpha\beta}$, $\nabla_\mu\nabla_\nu \nabla_{\alpha}\Phi$, $\nabla_\mu R_{\nu\alpha\beta\gamma}$, $R_{\mu\nu}R_{\alpha\beta}$, $R_{\mu\nu\alpha\beta}R_{\gamma\lambda}$ or $F_{\alpha\beta}R_{\mu\nu}$ - crucially, this elimination preserves exactly 377 constraints among the remaining $\delta c_i$ variations, demonstrating that these terms represent non-essential redundancies in the effective action.

We next eliminate all couplings in ${\cal L}'$ containing $R_{\mu\nu}$ by setting their coefficients to zero. Solving \reef{DLK} under this constraint yields 376 relations among the $\delta c_i$ parameters, demonstrating that at least one independent coupling must contain Ricci curvature terms. To identify this essential coupling, we implemented a binary search strategy: we divided the Ricci-containing terms into subsets, nullified one subset's coefficients, and verified whether \reef{DLK} generated 377 relations among the remaining $\delta c_i$'s. If not, we retained the complementary subset. Through iterative application of this method, we uniquely determined the independent coupling:
\beqa
[R^2 R'']_1 &=& c_{56} R^{\alpha\beta\gamma\delta}R^{\mu\nu}{}_{\gamma\delta}\nabla_{\alpha}\nabla_{\mu}R_{\beta\nu}\,.\labell{T53}
\eeqa

While numerous coefficient choices satisfy the 377 relations $\delta c_i = 0$, our selected scheme organizes these couplings into 52 distinct structures. These fundamental structures form the basis and are explicitly listed below:
\beqa
{\cal L}&\!\!\!\!\!=\!\!\!\!\! &[R^4]_7+[F'^4]_5+[R\Phi'^2F'^2]_{11}+[RF'^2\Phi'']_9+[R^2F'^2]_{13}+[F^2R^3]_{16}+[R^3\Phi'']_3+[R^3\Phi'^2]_5\nn\\&&+[F^2R\Phi''\Phi'^2]_8+[R^2\Phi''^2]_5+[R^2\Phi''\Phi'^2]_4+[R^2\Phi'^4]_3+[F^3F'\Phi'^3]_{9}+[F^4R^2]_{20}+[F^2R^2\Phi'']_{14}\nn\\&&+[F^2R^2\Phi'^2]_{19}+[F^4R\Phi'^2]_{12}
+[F^2RF'^2]_{33}+[F^6R]_7+[F^6\Phi'']_4+[F^6\Phi'^2]_{1}+[F^4\Phi'^2\Phi'']_{2}\nn\\&&+[FR^2F'\Phi']_{15}+[F^4\Phi'^4]_{5}+[F^2R\Phi'^4]_{4}+[FR\Phi'^3F']_{10}
+[F^4R\Phi'']_{9}+[F^2\Phi'^6]_2+[F^4\Phi''^2]_8\nn\\&&+[F^2R\Phi''^2]_9+[F^2F'^2\Phi'']_{21}
+[F^3R\Phi'F']_{1}+[FR\Phi''\Phi'F']_{2}+[F^2\Phi''^2\Phi'^2]_4+[F^2\Phi'^2F'^2]_{17}\nn\\&&+[FF'\Phi'\Phi''^2]_1+[FF'\Phi'^3\Phi'']_5+[R\Phi'^2\Phi''^2]_2+[\Phi'^4F'^2]_4+[\Phi''^4]_2+[FF'\Phi'^5]_2+[F^2\Phi'^4\Phi'']_2\nn\\&&+[R\Phi''^3]_1+[R\Phi'^4\Phi'']_1+[F'^2\Phi''^2]_6+[F'^2\Phi'^2\Phi'']_4+[FF'^3\Phi']_1+[\Phi'^4\Phi''^2]_2+[\Phi'^6\Phi'']_1\nn\\&&+[F^8]_5+[F^4F'^2]_{20}+[R^2R'']_1\,.\labell{T55}
\eeqa
The notation $[X]_n$ denotes that structure $X$ admits $n$ distinct contractions, each with an independent coupling constant. A prime symbol  indicates covariant differentiation of the associated field. Among these structures, the coupling $[R^2R'']_1$ is explicitly given in \reef{T53}, while all others are the following:
\beqa
[R^4]_7&=&c_{23} R^{\alpha  \beta  \gamma  \delta  } R_{\gamma  }{}^{\epsilon  }{}_{\alpha  }{}^{\varepsilon  } R_{\delta  }{}^{\mu  }{}_{\beta  }{}^{\nu  } R_{\varepsilon  \nu  \epsilon  \mu  } +c_{39} R_{\alpha  }{}^{\epsilon  \varepsilon  \mu  } R^{\alpha  \beta  \gamma  \delta  } R_{\gamma  }{}^{\nu  }{}_{\beta  \varepsilon  } R_{\mu  \nu  \delta  \epsilon  } +c_{40} R_{\alpha  }{}^{\epsilon  }{}_{\gamma  }{}^{\varepsilon  } R^{\alpha  \beta  \gamma  \delta  } R_{\beta  \epsilon  }{}^{\mu  \nu  } R_{\mu  \nu  \delta  \varepsilon  } \nn\\&&+c_{41} R_{\alpha  \gamma  \beta  }{}^{\epsilon  } R^{\alpha  \beta  \gamma  \delta  } R_{\delta  }{}^{\varepsilon  \mu  \nu  } R_{\mu  \nu  \epsilon  \varepsilon  } +c_{42} R^{\alpha  \beta  \gamma  \delta  } R_{\gamma  \delta  \alpha  \beta  } R^{\epsilon  \varepsilon  \mu  \nu  } R_{\mu  \nu  \epsilon  \varepsilon  } +c_{54} R_{\alpha  \beta  }{}^{\epsilon  \varepsilon  } R^{\alpha  \beta  \gamma  \delta  } R_{\mu  \nu  \epsilon  \varepsilon  } R^{\mu  \nu  }{}_{\gamma  \delta  } \nn\\&&+c_{55} R_{\alpha  \beta  }{}^{\epsilon  \varepsilon  } R^{\alpha  \beta  \gamma  \delta  } R_{\mu  \nu  \delta  \varepsilon  } R^{\mu  \nu  }{}_{\gamma  \epsilon  } \,,
\eeqa
\beqa
[F'^4]_5&=&e^{4 \Phi }c_{254} \nabla^{\alpha  }F^{\beta  \gamma  } \nabla_{\beta  }F_{\alpha  }{}^{\delta  } \nabla_{\gamma  }F^{\epsilon  \varepsilon  } \nabla_{\delta  }F_{\epsilon  \varepsilon  } + e^{4 \Phi }c_{323} \nabla_{\alpha  }F_{\beta  \gamma  } \nabla^{\alpha  }F^{\beta  \gamma  } \nabla_{\delta  }F_{\epsilon  \varepsilon  } \nabla^{\delta  }F^{\epsilon  \varepsilon  } \nn\\&&+ e^{4 \Phi }c_{342} \nabla^{\alpha  }F^{\beta  \gamma  } \nabla_{\beta  }F_{\gamma  }{}^{\delta  } \nabla_{\delta  }F^{\epsilon  \varepsilon  } \nabla_{\epsilon  }F_{\alpha  \varepsilon  } + e^{4 \Phi }c_{375} \nabla^{\alpha  }F^{\beta  \gamma  } \nabla_{\beta  }F_{\alpha  }{}^{\delta  } \nabla^{\epsilon  }F_{\gamma  }{}^{\varepsilon  } \nabla_{\varepsilon  }F_{\delta  \epsilon  } \nn\\&&+ e^{4 \Phi }c_{376} \nabla_{\alpha  }F^{\delta  \epsilon  } \nabla^{\alpha  }F^{\beta  \gamma  } \nabla_{\varepsilon  }F_{\gamma  \epsilon  } \nabla^{\varepsilon  }F_{\beta  \delta  }\,,
\eeqa
\beqa
[R\Phi'^2F'^2]_{11}&=&e^{2 \Phi }c_{198} R_{\epsilon  \varepsilon  \beta  \delta  } \nabla_{\alpha  }\Phi  \nabla^{\alpha  }\Phi  \nabla^{\beta  }F^{\gamma  \delta  } \nabla_{\gamma  }F^{\epsilon  \varepsilon  } + e^{2 \Phi }c_{253} R_{\alpha  \epsilon  \beta  \varepsilon  } \nabla^{\alpha  }\Phi  \nabla^{\beta  }\Phi  \nabla^{\gamma  }F^{\delta  \epsilon  } \nabla_{\delta  }F_{\gamma  }{}^{\varepsilon  } \nn\\&&+ e^{2 \Phi }c_{255} R_{\alpha  \varepsilon  \beta  \gamma  } \nabla^{\alpha  }\Phi  \nabla^{\beta  }\Phi  \nabla^{\gamma  }F^{\delta  \epsilon  } \nabla_{\delta  }F_{\epsilon  }{}^{\varepsilon  } + e^{2 \Phi }c_{321} R_{\epsilon  \varepsilon  \gamma  \delta  } \nabla_{\alpha  }F_{\beta  }{}^{\gamma  } \nabla^{\alpha  }\Phi  \nabla^{\beta  }\Phi  \nabla^{\delta  }F^{\epsilon  \varepsilon  } \nn\\&&+ e^{2 \Phi }c_{359} R_{\epsilon  \varepsilon  \gamma  \delta  } \nabla_{\alpha  }F^{\gamma  \delta  } \nabla^{\alpha  }\Phi  \nabla^{\beta  }\Phi  \nabla^{\epsilon  }F_{\beta  }{}^{\varepsilon  } + e^{2 \Phi }c_{362} R_{\delta  \varepsilon  \gamma  \epsilon  } \nabla^{\alpha  }\Phi  \nabla^{\beta  }\Phi  \nabla^{\gamma  }F_{\alpha  }{}^{\delta  } \nabla^{\epsilon  }F_{\beta  }{}^{\varepsilon  }\nn\\&& + e^{2 \Phi }c_{364} R_{\delta  \varepsilon  \beta  \epsilon  } \nabla_{\alpha  }\Phi  \nabla^{\alpha  }\Phi  \nabla^{\beta  }F^{\gamma  \delta  } \nabla^{\epsilon  }F_{\gamma  }{}^{\varepsilon  } + e^{2 \Phi }c_{365} R_{\delta  \varepsilon  \beta  \epsilon  } \nabla_{\alpha  }F^{\gamma  \delta  } \nabla^{\alpha  }\Phi  \nabla^{\beta  }\Phi  \nabla^{\epsilon  }F_{\gamma  }{}^{\varepsilon  } \nn\\&&+ e^{2 \Phi }c_{366} R_{\epsilon  \varepsilon  \beta  \delta  } \nabla_{\alpha  }F^{\gamma  \delta  } \nabla^{\alpha  }\Phi  \nabla^{\beta  }\Phi  \nabla^{\epsilon  }F_{\gamma  }{}^{\varepsilon  } + e^{2 \Phi }c_{369} R_{\delta  \varepsilon  \beta  \epsilon  } \nabla^{\alpha  }\Phi  \nabla^{\beta  }\Phi  \nabla^{\gamma  }F_{\alpha  }{}^{\delta  } \nabla^{\epsilon  }F_{\gamma  }{}^{\varepsilon  } \nn\\&&+ e^{2 \Phi }c_{370} R_{\epsilon  \varepsilon  \beta  \delta  } \nabla^{\alpha  }\Phi  \nabla^{\beta  }\Phi  \nabla^{\gamma  }F_{\alpha  }{}^{\delta  } \nabla^{\epsilon  }F_{\gamma  }{}^{\varepsilon  }\,,
\eeqa
\beqa
[RF'^2\Phi'']_{9}&=&e^{2 \Phi }c_{266} R_{\epsilon  \varepsilon  \alpha  \gamma  } \nabla^{\alpha  }F^{\beta  \gamma  } \nabla_{\beta  }F^{\delta  \epsilon  } \nabla_{\delta  }\nabla^{\varepsilon  }\Phi  + e^{2 \Phi }c_{291} R_{\delta  \varepsilon  \gamma  \epsilon  } \nabla_{\alpha  }\nabla^{\varepsilon  }\Phi  \nabla^{\alpha  }F^{\beta  \gamma  } \nabla^{\delta  }F_{\beta  }{}^{\epsilon  } \nn\\&& + e^{2 \Phi }c_{292} R_{\epsilon  \varepsilon  \gamma  \delta  } \nabla_{\alpha  }\nabla^{\varepsilon  }\Phi  \nabla^{\alpha  }F^{\beta  \gamma  } \nabla^{\delta  }F_{\beta  }{}^{\epsilon  } + e^{2 \Phi }c_{295} R_{\epsilon  \varepsilon  \alpha  \delta  } \nabla^{\alpha  }F^{\beta  \gamma  } \nabla_{\gamma  }\nabla^{\varepsilon  }\Phi  \nabla^{\delta  }F_{\beta  }{}^{\epsilon  }  \nn\\&&+ e^{2 \Phi }c_{319} R_{\epsilon  \varepsilon  \gamma  \delta  } \nabla^{\alpha  }F^{\beta  \gamma  } \nabla_{\beta  }\nabla_{\alpha  }\Phi  \nabla^{\delta  }F^{\epsilon  \varepsilon  } + e^{2 \Phi }c_{344} R_{\delta  \varepsilon  \beta  \gamma  } \nabla^{\alpha  }F^{\beta  \gamma  } \nabla^{\delta  }F^{\epsilon  \varepsilon  } \nabla_{\epsilon  }\nabla_{\alpha  }\Phi   \nn\\&&+ e^{2 \Phi }c_{346} R_{\gamma  \varepsilon  \alpha  \delta  } \nabla^{\alpha  }F^{\beta  \gamma  } \nabla^{\delta  }F^{\epsilon  \varepsilon  } \nabla_{\epsilon  }\nabla_{\beta  }\Phi  + e^{2 \Phi }c_{373} R_{\gamma  \epsilon  \delta  \varepsilon  } \nabla^{\alpha  }F^{\beta  \gamma  } \nabla_{\beta  }F_{\alpha  }{}^{\delta  } \nabla^{\epsilon  }\nabla^{\varepsilon  }\Phi  \nn\\&& + e^{2 \Phi }c_{374} R_{\delta  \epsilon  \alpha  \varepsilon  } \nabla^{\alpha  }F^{\beta  \gamma  } \nabla_{\beta  }F_{\gamma  }{}^{\delta  } \nabla^{\epsilon  }\nabla^{\varepsilon  }\Phi\,,\labell{RFFP}
\eeqa
\beqa
[R^2F'^2]_{13}&=&e^{2 \Phi }c_{60} R_{\gamma  }{}^{\epsilon  \varepsilon  \mu  } R_{\varepsilon  \mu  \delta  \epsilon  } \nabla_{\alpha  }F_{\beta  }{}^{\delta  } \nabla^{\alpha  }F^{\beta  \gamma  } + e^{2 \Phi }c_{70} R_{\varepsilon  \mu  \delta  \epsilon  } R^{\varepsilon  \mu  }{}_{\beta  \gamma  } \nabla_{\alpha  }F^{\delta  \epsilon  } \nabla^{\alpha  }F^{\beta  \gamma  }  \nn\\&&+ e^{2 \Phi }c_{71} R_{\varepsilon  \mu  \gamma  \epsilon  } R^{\varepsilon  \mu  }{}_{\beta  \delta  } \nabla_{\alpha  }F^{\delta  \epsilon  } \nabla^{\alpha  }F^{\beta  \gamma  } + e^{2 \Phi }c_{121} R^{\delta  \epsilon  \varepsilon  \mu  } R_{\varepsilon  \mu  \delta  \epsilon  } \nabla^{\alpha  }F^{\beta  \gamma  } \nabla_{\beta  }F_{\alpha  \gamma  } \nn\\&& + e^{2 \Phi }c_{126} R_{\gamma  }{}^{\epsilon  \varepsilon  \mu  } R_{\varepsilon  \mu  \delta  \epsilon  } \nabla^{\alpha  }F^{\beta  \gamma  } \nabla_{\beta  }F_{\alpha  }{}^{\delta  } + e^{2 \Phi }c_{133} R_{\delta  }{}^{\varepsilon  }{}_{\alpha  }{}^{\mu  } R_{\epsilon  \mu  \gamma  \varepsilon  } \nabla^{\alpha  }F^{\beta  \gamma  } \nabla_{\beta  }F^{\delta  \epsilon  }  \nn\\&&+ e^{2 \Phi }c_{283} R_{\alpha  }{}^{\varepsilon  }{}_{\delta  }{}^{\mu  } R_{\epsilon  \mu  \gamma  \varepsilon  } \nabla^{\alpha  }F^{\beta  \gamma  } \nabla^{\delta  }F_{\beta  }{}^{\epsilon  } + e^{2 \Phi }c_{285} R_{\gamma  }{}^{\varepsilon  }{}_{\alpha  }{}^{\mu  } R_{\epsilon  \mu  \delta  \varepsilon  } \nabla^{\alpha  }F^{\beta  \gamma  } \nabla^{\delta  }F_{\beta  }{}^{\epsilon  } \nn\\&& + e^{2 \Phi }c_{288} R_{\varepsilon  \mu  \gamma  \delta  } R^{\varepsilon  \mu  }{}_{\alpha  \epsilon  } \nabla^{\alpha  }F^{\beta  \gamma  } \nabla^{\delta  }F_{\beta  }{}^{\epsilon  } + e^{2 \Phi }c_{310} R_{\beta  }{}^{\mu  }{}_{\alpha  \gamma  } R_{\epsilon  \mu  \delta  \varepsilon  } \nabla^{\alpha  }F^{\beta  \gamma  } \nabla^{\delta  }F^{\epsilon  \varepsilon  } \nn\\&& + e^{2 \Phi }c_{312} R_{\alpha  }{}^{\mu  }{}_{\delta  \epsilon  } R_{\varepsilon  \mu  \beta  \gamma  } \nabla^{\alpha  }F^{\beta  \gamma  } \nabla^{\delta  }F^{\epsilon  \varepsilon  } + e^{2 \Phi }c_{316} R_{\beta  \epsilon  \alpha  }{}^{\mu  } R_{\varepsilon  \mu  \gamma  \delta  } \nabla^{\alpha  }F^{\beta  \gamma  } \nabla^{\delta  }F^{\epsilon  \varepsilon  }  \nn\\&&+ e^{2 \Phi }c_{317} R_{\beta  }{}^{\mu  }{}_{\alpha  \epsilon  } R_{\varepsilon  \mu  \gamma  \delta  } \nabla^{\alpha  }F^{\beta  \gamma  } \nabla^{\delta  }F^{\epsilon  \varepsilon  }\,,\labell{RRF'F'}
\eeqa
\beqa
[F^2R^3]_{16}&=&e^{2 \Phi }c_{19} F^{\alpha  \beta  } F^{\gamma  \delta  } R_{\alpha  }{}^{\epsilon  }{}_{\gamma  }{}^{\varepsilon  } R_{\epsilon  }{}^{\mu  }{}_{\beta  }{}^{\nu  } R_{\varepsilon  \nu  \delta  \mu  } + e^{2 \Phi }c_{20} F_{\alpha  \beta  } F^{\alpha  \beta  } R^{\gamma  \delta  \epsilon  \varepsilon  } R_{\epsilon  }{}^{\mu  }{}_{\gamma  }{}^{\nu  } R_{\varepsilon  \nu  \delta  \mu  }  \nn\\&&+ e^{2 \Phi }c_{21} F^{\alpha  \beta  } F^{\gamma  \delta  } R_{\alpha  }{}^{\epsilon  }{}_{\gamma  }{}^{\varepsilon  } R_{\delta  }{}^{\mu  }{}_{\beta  }{}^{\nu  } R_{\varepsilon  \nu  \epsilon  \mu  } + e^{2 \Phi }c_{22} F_{\alpha  }{}^{\gamma  } F^{\alpha  \beta  } R_{\delta  }{}^{\mu  }{}_{\gamma  }{}^{\nu  } R^{\delta  \epsilon  }{}_{\beta  }{}^{\varepsilon  } R_{\varepsilon  \nu  \epsilon  \mu  } \nn\\&& + e^{2 \Phi }c_{33} F^{\alpha  \beta  } F^{\gamma  \delta  } R_{\alpha  }{}^{\epsilon  \varepsilon  \mu  } R_{\varepsilon  }{}^{\nu  }{}_{\beta  \epsilon  } R_{\mu  \nu  \gamma  \delta  } + e^{2 \Phi }c_{34} F^{\alpha  \beta  } F^{\gamma  \delta  } R_{\alpha  }{}^{\epsilon  \varepsilon  \mu  } R_{\varepsilon  }{}^{\nu  }{}_{\beta  \gamma  } R_{\mu  \nu  \delta  \epsilon  }  \nn\\&&+ e^{2 \Phi }c_{35} F_{\alpha  \beta  } F^{\alpha  \beta  } R_{\gamma  \epsilon  }{}^{\mu  \nu  } R^{\gamma  \delta  \epsilon  \varepsilon  } R_{\mu  \nu  \delta  \varepsilon  } + e^{2 \Phi }c_{36} F^{\alpha  \beta  } F^{\gamma  \delta  } R_{\alpha  }{}^{\epsilon  }{}_{\beta  \gamma  } R_{\delta  }{}^{\varepsilon  \mu  \nu  } R_{\mu  \nu  \epsilon  \varepsilon  }  \nn\\&&+ e^{2 \Phi }c_{37} F_{\alpha  }{}^{\gamma  } F^{\alpha  \beta  } R_{\beta  }{}^{\delta  }{}_{\gamma  }{}^{\epsilon  } R_{\delta  }{}^{\varepsilon  \mu  \nu  } R_{\mu  \nu  \epsilon  \varepsilon  } + e^{2 \Phi }c_{38} F^{\alpha  \beta  } F^{\gamma  \delta  } R_{\alpha  \gamma  \beta  \delta  } R^{\epsilon  \varepsilon  \mu  \nu  } R_{\mu  \nu  \epsilon  \varepsilon  }  \nn\\&&+ e^{2 \Phi }c_{48} F^{\alpha  \beta  } F^{\gamma  \delta  } R_{\alpha  }{}^{\epsilon  }{}_{\gamma  }{}^{\varepsilon  } R_{\mu  \nu  \epsilon  \varepsilon  } R^{\mu  \nu  }{}_{\beta  \delta  } + e^{2 \Phi }c_{49} F^{\alpha  \beta  } F^{\gamma  \delta  } R^{\epsilon  \varepsilon  }{}_{\alpha  \gamma  } R_{\mu  \nu  \delta  \varepsilon  } R^{\mu  \nu  }{}_{\beta  \epsilon  }  \nn\\&&+ e^{2 \Phi }c_{50} F^{\alpha  \beta  } F^{\gamma  \delta  } R_{\alpha  }{}^{\epsilon  }{}_{\gamma  }{}^{\varepsilon  } R_{\mu  \nu  \delta  \epsilon  } R^{\mu  \nu  }{}_{\beta  \varepsilon  } + e^{2 \Phi }c_{51} F^{\alpha  \beta  } F^{\gamma  \delta  } R_{\alpha  }{}^{\epsilon  }{}_{\beta  }{}^{\varepsilon  } R_{\mu  \nu  \epsilon  \varepsilon  } R^{\mu  \nu  }{}_{\gamma  \delta  } \nn\\&& + e^{2 \Phi }c_{52} F_{\alpha  }{}^{\gamma  } F^{\alpha  \beta  } R_{\beta  }{}^{\delta  \epsilon  \varepsilon  } R_{\mu  \nu  \delta  \varepsilon  } R^{\mu  \nu  }{}_{\gamma  \epsilon  } + e^{2 \Phi }c_{53} F^{\alpha  \beta  } F^{\gamma  \delta  } R^{\epsilon  \varepsilon  }{}_{\alpha  \beta  } R_{\mu  \nu  \delta  \varepsilon  } R^{\mu  \nu  }{}_{\gamma  \epsilon  }\,,
\eeqa
\beqa
[R^3\Phi'']_3&=&c_{97} R_{\gamma  }{}^{\varepsilon  }{}_{\beta  }{}^{\mu  } R^{\gamma  \delta  }{}_{\alpha  }{}^{\epsilon  } R_{\epsilon  \mu  \delta  \varepsilon  } \nabla^{\alpha  }\nabla^{\beta  }\Phi  +c_{107} R_{\alpha  }{}^{\gamma  }{}_{\beta  }{}^{\delta  } R_{\gamma  }{}^{\epsilon  \varepsilon  \mu  } R_{\varepsilon  \mu  \delta  \epsilon  } \nabla^{\alpha  }\nabla^{\beta  }\Phi\nn\\&&  +c_{112} R_{\alpha  }{}^{\gamma  \delta  \epsilon  } R_{\varepsilon  \mu  \gamma  \epsilon  } R^{\varepsilon  \mu  }{}_{\beta  \delta  } \nabla^{\alpha  }\nabla^{\beta  }\Phi \,,
\eeqa
\beqa
[R^3\Phi'^2]_5&=&c_{76} R^{\beta  \gamma  \delta  \epsilon  } R_{\delta  }{}^{\varepsilon  }{}_{\beta  }{}^{\mu  } R_{\epsilon  \mu  \gamma  \varepsilon  } \nabla_{\alpha  }\Phi  \nabla^{\alpha  }\Phi  +c_{77} R_{\beta  \delta  }{}^{\varepsilon  \mu  } R^{\beta  \gamma  \delta  \epsilon  } R_{\varepsilon  \mu  \gamma  \epsilon  } \nabla_{\alpha  }\Phi  \nabla^{\alpha  }\Phi  \nn\\&& +c_{169} R_{\gamma  }{}^{\varepsilon  }{}_{\beta  }{}^{\mu  } R^{\gamma  \delta  }{}_{\alpha  }{}^{\epsilon  } R_{\epsilon  \mu  \delta  \varepsilon  } \nabla^{\alpha  }\Phi  \nabla^{\beta  }\Phi  +c_{179} R_{\alpha  }{}^{\gamma  }{}_{\beta  }{}^{\delta  } R_{\gamma  }{}^{\epsilon  \varepsilon  \mu  } R_{\varepsilon  \mu  \delta  \epsilon  } \nabla^{\alpha  }\Phi  \nabla^{\beta  }\Phi \nn\\&& +c_{184} R_{\alpha  }{}^{\gamma  \delta  \epsilon  } R_{\varepsilon  \mu  \gamma  \epsilon  } R^{\varepsilon  \mu  }{}_{\beta  \delta  } \nabla^{\alpha  }\Phi  \nabla^{\beta  }\Phi\,,
\eeqa
\beqa
[F^2R\Phi''\Phi'^2]_8&\!\!\!\!\!=\!\!\!\!\!&e^{2 \Phi }c_{185} F_{\delta  }{}^{\varepsilon  } F^{\delta  \epsilon  } R_{\gamma  \epsilon  \beta  \varepsilon  } \nabla_{\alpha  }\nabla^{\gamma  }\Phi  \nabla^{\alpha  }\Phi  \nabla^{\beta  }\Phi  + e^{2 \Phi }c_{186} F_{\gamma  }{}^{\delta  } F^{\epsilon  \varepsilon  } R_{\epsilon  \varepsilon  \beta  \delta  } \nabla_{\alpha  }\nabla^{\gamma  }\Phi  \nabla^{\alpha  }\Phi  \nabla^{\beta  }\Phi \nn\\&& + e^{2 \Phi }c_{187} F_{\beta  }{}^{\delta  } F^{\epsilon  \varepsilon  } R_{\epsilon  \varepsilon  \gamma  \delta  } \nabla_{\alpha  }\nabla^{\gamma  }\Phi  \nabla^{\alpha  }\Phi  \nabla^{\beta  }\Phi  + e^{2 \Phi }c_{241} F_{\epsilon  \varepsilon  } F^{\epsilon  \varepsilon  } R_{\alpha  \gamma  \beta  \delta  } \nabla^{\alpha  }\Phi  \nabla^{\beta  }\Phi  \nabla^{\gamma  }\nabla^{\delta  }\Phi \nn\\&& + e^{2 \Phi }c_{242} F_{\alpha  }{}^{\epsilon  } F_{\epsilon  }{}^{\varepsilon  } R_{\gamma  \varepsilon  \beta  \delta  } \nabla^{\alpha  }\Phi  \nabla^{\beta  }\Phi  \nabla^{\gamma  }\nabla^{\delta  }\Phi  + e^{2 \Phi }c_{243} F_{\alpha  }{}^{\epsilon  } F_{\gamma  }{}^{\varepsilon  } R_{\delta  \varepsilon  \beta  \epsilon  } \nabla^{\alpha  }\Phi  \nabla^{\beta  }\Phi  \nabla^{\gamma  }\nabla^{\delta  }\Phi  \nn\\&&+ e^{2 \Phi }c_{245} F_{\alpha  }{}^{\epsilon  } F_{\gamma  }{}^{\varepsilon  } R_{\epsilon  \varepsilon  \beta  \delta  } \nabla^{\alpha  }\Phi  \nabla^{\beta  }\Phi  \nabla^{\gamma  }\nabla^{\delta  }\Phi \nn\\&& + e^{2 \Phi }c_{246} F_{\alpha  \gamma  } F^{\epsilon  \varepsilon  } R_{\epsilon  \varepsilon  \beta  \delta  } \nabla^{\alpha  }\Phi  \nabla^{\beta  }\Phi  \nabla^{\gamma  }\nabla^{\delta  }\Phi,
\eeqa
\beqa
[R^2\Phi''^2]_5&=&c_{117} R_{\beta  }{}^{\delta  \epsilon  \varepsilon  } R_{\epsilon  \varepsilon  \gamma  \delta  } \nabla_{\alpha  }\nabla^{\gamma  }\Phi  \nabla^{\alpha  }\nabla^{\beta  }\Phi  +c_{140} R^{\gamma  \delta  \epsilon  \varepsilon  } R_{\epsilon  \varepsilon  \gamma  \delta  } \nabla^{\alpha  }\nabla^{\beta  }\Phi  \nabla_{\beta  }\nabla_{\alpha  }\Phi  \nn\\&&+c_{234} R_{\alpha  }{}^{\epsilon  }{}_{\beta  }{}^{\varepsilon  } R_{\gamma  \varepsilon  \delta  \epsilon  } \nabla^{\alpha  }\nabla^{\beta  }\Phi  \nabla^{\gamma  }\nabla^{\delta  }\Phi  +c_{235} R_{\alpha  }{}^{\epsilon  }{}_{\gamma  }{}^{\varepsilon  } R_{\delta  \varepsilon  \beta  \epsilon  } \nabla^{\alpha  }\nabla^{\beta  }\Phi  \nabla^{\gamma  }\nabla^{\delta  }\Phi  \nn\\&&+c_{238} R_{\epsilon  \varepsilon  \beta  \delta  } R^{\epsilon  \varepsilon  }{}_{\alpha  \gamma  } \nabla^{\alpha  }\nabla^{\beta  }\Phi  \nabla^{\gamma  }\nabla^{\delta  }\Phi \,,
\eeqa
\beqa
[R^2\Phi''\Phi'^2]_4&=&c_{188} R_{\beta  }{}^{\delta  \epsilon  \varepsilon  } R_{\epsilon  \varepsilon  \gamma  \delta  } \nabla_{\alpha  }\nabla^{\gamma  }\Phi  \nabla^{\alpha  }\Phi  \nabla^{\beta  }\Phi  +c_{193} R^{\gamma  \delta  \epsilon  \varepsilon  } R_{\epsilon  \varepsilon  \gamma  \delta  } \nabla^{\alpha  }\Phi  \nabla_{\beta  }\nabla_{\alpha  }\Phi  \nabla^{\beta  }\Phi  \nn\\&& +c_{244} R_{\gamma  }{}^{\epsilon  }{}_{\alpha  }{}^{\varepsilon  } R_{\delta  \varepsilon  \beta  \epsilon  } \nabla^{\alpha  }\Phi  \nabla^{\beta  }\Phi  \nabla^{\gamma  }\nabla^{\delta  }\Phi  +c_{247} R_{\epsilon  \varepsilon  \beta  \delta  } R^{\epsilon  \varepsilon  }{}_{\alpha  \gamma  } \nabla^{\alpha  }\Phi  \nabla^{\beta  }\Phi  \nabla^{\gamma  }\nabla^{\delta  }\Phi ,
\eeqa
\beqa
[R^2\Phi'^4]_3&=&c_{192} R^{\gamma  \delta  \epsilon  \varepsilon  } R_{\epsilon  \varepsilon  \gamma  \delta  } \nabla_{\alpha  }\Phi  \nabla^{\alpha  }\Phi  \nabla_{\beta  }\Phi  \nabla^{\beta  }\Phi  +c_{223} R_{\beta  }{}^{\delta  \epsilon  \varepsilon  } R_{\epsilon  \varepsilon  \gamma  \delta  } \nabla_{\alpha  }\Phi  \nabla^{\alpha  }\Phi  \nabla^{\beta  }\Phi  \nabla^{\gamma  }\Phi  \nn\\&&+c_{328} R_{\alpha  }{}^{\epsilon  }{}_{\beta  }{}^{\varepsilon  } R_{\gamma  \varepsilon  \delta  \epsilon  } \nabla^{\alpha  }\Phi  \nabla^{\beta  }\Phi  \nabla^{\gamma  }\Phi  \nabla^{\delta  }\Phi\,,
\eeqa
\beqa
[F^3F'\Phi'^3]_9&\!\!=\!\!&e^{4 \Phi }c_{206} F_{\gamma  }{}^{\epsilon  } F_{\delta  }{}^{\varepsilon  } F_{\epsilon  \varepsilon  } \nabla_{\alpha  }\Phi  \nabla^{\alpha  }\Phi  \nabla^{\beta  }\Phi  \nabla^{\gamma  }F_{\beta  }{}^{\delta  } + e^{4 \Phi }c_{207} F_{\gamma  \delta  } F_{\epsilon  \varepsilon  } F^{\epsilon  \varepsilon  } \nabla_{\alpha  }\Phi  \nabla^{\alpha  }\Phi  \nabla^{\beta  }\Phi  \nabla^{\gamma  }F_{\beta  }{}^{\delta  }  \nn\\&&+ e^{4 \Phi }c_{209} F_{\beta  }{}^{\varepsilon  } F_{\gamma  \delta  } F_{\epsilon  \varepsilon  } \nabla_{\alpha  }\Phi  \nabla^{\alpha  }\Phi  \nabla^{\beta  }\Phi  \nabla^{\gamma  }F^{\delta  \epsilon  } + e^{4 \Phi }c_{210} F_{\beta  \delta  } F_{\gamma  }{}^{\varepsilon  } F_{\epsilon  \varepsilon  } \nabla_{\alpha  }\Phi  \nabla^{\alpha  }\Phi  \nabla^{\beta  }\Phi  \nabla^{\gamma  }F^{\delta  \epsilon  }  \nn\\&&+ e^{4 \Phi }c_{214} F_{\gamma  }{}^{\epsilon  } F_{\delta  }{}^{\varepsilon  } F_{\epsilon  \varepsilon  } \nabla_{\alpha  }F_{\beta  }{}^{\delta  } \nabla^{\alpha  }\Phi  \nabla^{\beta  }\Phi  \nabla^{\gamma  }\Phi  + e^{4 \Phi }c_{215} F_{\gamma  \delta  } F_{\epsilon  \varepsilon  } F^{\epsilon  \varepsilon  } \nabla_{\alpha  }F_{\beta  }{}^{\delta  } \nabla^{\alpha  }\Phi  \nabla^{\beta  }\Phi  \nabla^{\gamma  }\Phi  \nn\\&& + e^{4 \Phi }c_{217} F_{\beta  \delta  } F_{\gamma  }{}^{\varepsilon  } F_{\epsilon  \varepsilon  } \nabla_{\alpha  }F^{\delta  \epsilon  } \nabla^{\alpha  }\Phi  \nabla^{\beta  }\Phi  \nabla^{\gamma  }\Phi  + e^{4 \Phi }c_{267} F_{\beta  }{}^{\varepsilon  } F_{\gamma  \varepsilon  } F_{\delta  \epsilon  } \nabla^{\alpha  }\Phi  \nabla^{\beta  }\Phi  \nabla^{\gamma  }\Phi  \nabla^{\delta  }F_{\alpha  }{}^{\epsilon  }  \nn\\&&+ e^{4 \Phi }c_{268} F_{\beta  \delta  } F_{\gamma  }{}^{\varepsilon  } F_{\epsilon  \varepsilon  } \nabla^{\alpha  }\Phi  \nabla^{\beta  }\Phi  \nabla^{\gamma  }\Phi  \nabla^{\delta  }F_{\alpha  }{}^{\epsilon  }\,,
\eeqa
\beqa
[F^4R^2]_{20}&=&e^{4 \Phi }c_{11} F^{\alpha  \beta  } F^{\gamma  \delta  } F^{\epsilon  \varepsilon  } F^{\mu  \nu  } R_{\alpha  \gamma  \beta  \delta  } R_{\epsilon  \mu  \varepsilon  \nu  } + e^{4 \Phi }c_{12} F_{\alpha  }{}^{\gamma  } F^{\alpha  \beta  } F_{\delta  }{}^{\varepsilon  } F^{\delta  \epsilon  } R_{\beta  }{}^{\mu  }{}_{\gamma  }{}^{\nu  } R_{\epsilon  \nu  \varepsilon  \mu  }  \nn\\&&+ e^{4 \Phi }c_{13} F^{\alpha  \beta  } F^{\gamma  \delta  } F^{\epsilon  \varepsilon  } F^{\mu  \nu  } R_{\alpha  \gamma  \epsilon  \mu  } R_{\varepsilon  \nu  \beta  \delta  } + e^{4 \Phi }c_{14} F_{\alpha  }{}^{\gamma  } F^{\alpha  \beta  } F_{\delta  }{}^{\varepsilon  } F^{\delta  \epsilon  } R_{\beta  }{}^{\mu  }{}_{\epsilon  }{}^{\nu  } R_{\varepsilon  \nu  \gamma  \mu  }  \nn\\&&+ e^{4 \Phi }c_{15} F_{\alpha  }{}^{\gamma  } F^{\alpha  \beta  } F^{\delta  \epsilon  } F^{\varepsilon  \mu  } R_{\delta  }{}^{\nu  }{}_{\beta  \epsilon  } R_{\varepsilon  \nu  \gamma  \mu  } + e^{4 \Phi }c_{16} F_{\alpha  \beta  } F^{\alpha  \beta  } F^{\gamma  \delta  } F^{\epsilon  \varepsilon  } R_{\gamma  }{}^{\mu  }{}_{\epsilon  }{}^{\nu  } R_{\varepsilon  \nu  \delta  \mu  }  \nn\\&&+ e^{4 \Phi }c_{17} F_{\alpha  }{}^{\gamma  } F^{\alpha  \beta  } F_{\beta  }{}^{\delta  } F^{\epsilon  \varepsilon  } R_{\epsilon  }{}^{\mu  }{}_{\gamma  }{}^{\nu  } R_{\varepsilon  \nu  \delta  \mu  } + e^{4 \Phi }c_{18} F_{\alpha  }{}^{\gamma  } F^{\alpha  \beta  } F^{\delta  \epsilon  } F^{\varepsilon  \mu  } R_{\beta  }{}^{\nu  }{}_{\gamma  \delta  } R_{\varepsilon  \nu  \epsilon  \mu  }  \nn\\&&+ e^{4 \Phi }c_{26} F_{\alpha  }{}^{\gamma  } F^{\alpha  \beta  } F^{\delta  \epsilon  } F^{\varepsilon  \mu  } R_{\beta  \delta  \varepsilon  }{}^{\nu  } R_{\mu  \nu  \gamma  \epsilon  } + e^{4 \Phi }c_{27} F_{\alpha  }{}^{\gamma  } F^{\alpha  \beta  } F^{\delta  \epsilon  } F^{\varepsilon  \mu  } R_{\delta  }{}^{\nu  }{}_{\beta  \varepsilon  } R_{\mu  \nu  \gamma  \epsilon  }  \nn\\&&+ e^{4 \Phi }c_{28} F^{\alpha  \beta  } F^{\gamma  \delta  } F^{\epsilon  \varepsilon  } F^{\mu  \nu  } R_{\alpha  \beta  \gamma  \epsilon  } R_{\mu  \nu  \delta  \varepsilon  } + e^{4 \Phi }c_{29} F_{\alpha  }{}^{\gamma  } F^{\alpha  \beta  } F_{\beta  }{}^{\delta  } F_{\gamma  }{}^{\epsilon  } R_{\delta  }{}^{\varepsilon  \mu  \nu  } R_{\mu  \nu  \epsilon  \varepsilon  } \nn\\&& + e^{4 \Phi }c_{30} F_{\alpha  \beta  } F^{\alpha  \beta  } F_{\gamma  }{}^{\epsilon  } F^{\gamma  \delta  } R_{\delta  }{}^{\varepsilon  \mu  \nu  } R_{\mu  \nu  \epsilon  \varepsilon  } + e^{4 \Phi }c_{31} F_{\alpha  }{}^{\gamma  } F^{\alpha  \beta  } F_{\beta  }{}^{\delta  } F_{\gamma  \delta  } R^{\epsilon  \varepsilon  \mu  \nu  } R_{\mu  \nu  \epsilon  \varepsilon  }  \nn\\&&+ e^{4 \Phi }c_{32} F_{\alpha  \beta  } F^{\alpha  \beta  } F_{\gamma  \delta  } F^{\gamma  \delta  } R^{\epsilon  \varepsilon  \mu  \nu  } R_{\mu  \nu  \epsilon  \varepsilon  } + e^{4 \Phi }c_{43} F_{\alpha  }{}^{\gamma  } F^{\alpha  \beta  } F_{\delta  }{}^{\varepsilon  } F^{\delta  \epsilon  } R_{\mu  \nu  \gamma  \varepsilon  } R^{\mu  \nu  }{}_{\beta  \epsilon  }  \nn\\&&+ e^{4 \Phi }c_{44} F_{\alpha  }{}^{\gamma  } F^{\alpha  \beta  } F_{\beta  }{}^{\delta  } F^{\epsilon  \varepsilon  } R_{\mu  \nu  \epsilon  \varepsilon  } R^{\mu  \nu  }{}_{\gamma  \delta  } + e^{4 \Phi }c_{45} F_{\alpha  \beta  } F^{\alpha  \beta  } F^{\gamma  \delta  } F^{\epsilon  \varepsilon  } R_{\mu  \nu  \epsilon  \varepsilon  } R^{\mu  \nu  }{}_{\gamma  \delta  } \nn\\&& + e^{4 \Phi }c_{46} F_{\alpha  }{}^{\gamma  } F^{\alpha  \beta  } F_{\beta  }{}^{\delta  } F^{\epsilon  \varepsilon  } R_{\mu  \nu  \delta  \varepsilon  } R^{\mu  \nu  }{}_{\gamma  \epsilon  } \nn\\&&+ e^{4 \Phi }c_{47} F_{\alpha  \beta  } F^{\alpha  \beta  } F^{\gamma  \delta  } F^{\epsilon  \varepsilon  } R_{\mu  \nu  \delta  \varepsilon  } R^{\mu  \nu  }{}_{\gamma  \epsilon  }\,,
\eeqa
\beqa
[F^2R^2\Phi'']_{14}&=&e^{2 \Phi }c_{92} F_{\gamma  }{}^{\epsilon  } F^{\gamma  \delta  } R_{\alpha  }{}^{\varepsilon  }{}_{\beta  }{}^{\mu  } R_{\delta  \mu  \epsilon  \varepsilon  } \nabla^{\alpha  }\nabla^{\beta  }\Phi  + e^{2 \Phi }c_{93} F^{\gamma  \delta  } F^{\epsilon  \varepsilon  } R_{\gamma  }{}^{\mu  }{}_{\alpha  \delta  } R_{\epsilon  \mu  \beta  \varepsilon  } \nabla^{\alpha  }\nabla^{\beta  }\Phi  \nn\\&&+ e^{2 \Phi }c_{94} F_{\gamma  }{}^{\epsilon  } F^{\gamma  \delta  } R_{\delta  }{}^{\varepsilon  }{}_{\alpha  }{}^{\mu  } R_{\epsilon  \mu  \beta  \varepsilon  } \nabla^{\alpha  }\nabla^{\beta  }\Phi  + e^{2 \Phi }c_{95} F_{\alpha  }{}^{\gamma  } F^{\delta  \epsilon  } R_{\delta  }{}^{\varepsilon  }{}_{\beta  }{}^{\mu  } R_{\epsilon  \mu  \gamma  \varepsilon  } \nabla^{\alpha  }\nabla^{\beta  }\Phi \nn\\&& + e^{2 \Phi }c_{96} F^{\gamma  \delta  } F^{\epsilon  \varepsilon  } R_{\alpha  }{}^{\mu  }{}_{\beta  \gamma  } R_{\epsilon  \mu  \delta  \varepsilon  } \nabla^{\alpha  }\nabla^{\beta  }\Phi  + e^{2 \Phi }c_{100} F^{\gamma  \delta  } F^{\epsilon  \varepsilon  } R_{\alpha  \gamma  \epsilon  }{}^{\mu  } R_{\varepsilon  \mu  \beta  \delta  } \nabla^{\alpha  }\nabla^{\beta  }\Phi  \nn\\&&+ e^{2 \Phi }c_{101} F^{\gamma  \delta  } F^{\epsilon  \varepsilon  } R_{\gamma  }{}^{\mu  }{}_{\alpha  \epsilon  } R_{\varepsilon  \mu  \beta  \delta  } \nabla^{\alpha  }\nabla^{\beta  }\Phi  + e^{2 \Phi }c_{103} F_{\gamma  \delta  } F^{\gamma  \delta  } R_{\alpha  }{}^{\epsilon  \varepsilon  \mu  } R_{\varepsilon  \mu  \beta  \epsilon  } \nabla^{\alpha  }\nabla^{\beta  }\Phi \nn\\&& + e^{2 \Phi }c_{105} F_{\alpha  }{}^{\gamma  } F_{\gamma  }{}^{\delta  } R_{\beta  }{}^{\epsilon  \varepsilon  \mu  } R_{\varepsilon  \mu  \delta  \epsilon  } \nabla^{\alpha  }\nabla^{\beta  }\Phi  + e^{2 \Phi }c_{106} F_{\alpha  }{}^{\gamma  } F_{\beta  }{}^{\delta  } R_{\gamma  }{}^{\epsilon  \varepsilon  \mu  } R_{\varepsilon  \mu  \delta  \epsilon  } \nabla^{\alpha  }\nabla^{\beta  }\Phi \nn\\&& + e^{2 \Phi }c_{108} F_{\alpha  }{}^{\gamma  } F_{\beta  \gamma  } R^{\delta  \epsilon  \varepsilon  \mu  } R_{\varepsilon  \mu  \delta  \epsilon  } \nabla^{\alpha  }\nabla^{\beta  }\Phi  + e^{2 \Phi }c_{109} F_{\gamma  }{}^{\epsilon  } F^{\gamma  \delta  } R_{\varepsilon  \mu  \beta  \epsilon  } R^{\varepsilon  \mu  }{}_{\alpha  \delta  } \nabla^{\alpha  }\nabla^{\beta  }\Phi  \nn\\&&+ e^{2 \Phi }c_{110} F_{\alpha  }{}^{\gamma  } F^{\delta  \epsilon  } R_{\varepsilon  \mu  \delta  \epsilon  } R^{\varepsilon  \mu  }{}_{\beta  \gamma  } \nabla^{\alpha  }\nabla^{\beta  }\Phi  \nn\\&&+ e^{2 \Phi }c_{111} F_{\alpha  }{}^{\gamma  } F^{\delta  \epsilon  } R_{\varepsilon  \mu  \gamma  \epsilon  } R^{\varepsilon  \mu  }{}_{\beta  \delta  } \nabla^{\alpha  }\nabla^{\beta  }\Phi,
\eeqa
\beqa
[F^2R^2\Phi'^2]_{19}&=&e^{2 \Phi }c_{75} F^{\beta  \gamma  } F^{\delta  \epsilon  } R_{\beta  }{}^{\varepsilon  }{}_{\delta  }{}^{\mu  } R_{\epsilon  \mu  \gamma  \varepsilon  } \nabla_{\alpha  }\Phi  \nabla^{\alpha  }\Phi  + e^{2 \Phi }c_{79} F_{\beta  }{}^{\delta  } F^{\beta  \gamma  } R_{\gamma  }{}^{\epsilon  \varepsilon  \mu  } R_{\varepsilon  \mu  \delta  \epsilon  } \nabla_{\alpha  }\Phi  \nabla^{\alpha  }\Phi  \nn\\&& + e^{2 \Phi }c_{80} F_{\beta  \gamma  } F^{\beta  \gamma  } R^{\delta  \epsilon  \varepsilon  \mu  } R_{\varepsilon  \mu  \delta  \epsilon  } \nabla_{\alpha  }\Phi  \nabla^{\alpha  }\Phi  + e^{2 \Phi }c_{81} F^{\beta  \gamma  } F^{\delta  \epsilon  } R_{\varepsilon  \mu  \delta  \epsilon  } R^{\varepsilon  \mu  }{}_{\beta  \gamma  } \nabla_{\alpha  }\Phi  \nabla^{\alpha  }\Phi   \nn\\&&+ e^{2 \Phi }c_{82} F^{\beta  \gamma  } F^{\delta  \epsilon  } R_{\varepsilon  \mu  \gamma  \epsilon  } R^{\varepsilon  \mu  }{}_{\beta  \delta  } \nabla_{\alpha  }\Phi  \nabla^{\alpha  }\Phi  + e^{2 \Phi }c_{164} F_{\gamma  }{}^{\epsilon  } F^{\gamma  \delta  } R_{\alpha  }{}^{\varepsilon  }{}_{\beta  }{}^{\mu  } R_{\delta  \mu  \epsilon  \varepsilon  } \nabla^{\alpha  }\Phi  \nabla^{\beta  }\Phi  \nn\\&& + e^{2 \Phi }c_{165} F^{\gamma  \delta  } F^{\epsilon  \varepsilon  } R_{\gamma  }{}^{\mu  }{}_{\alpha  \delta  } R_{\epsilon  \mu  \beta  \varepsilon  } \nabla^{\alpha  }\Phi  \nabla^{\beta  }\Phi  + e^{2 \Phi }c_{166} F_{\gamma  }{}^{\epsilon  } F^{\gamma  \delta  } R_{\delta  }{}^{\varepsilon  }{}_{\alpha  }{}^{\mu  } R_{\epsilon  \mu  \beta  \varepsilon  } \nabla^{\alpha  }\Phi  \nabla^{\beta  }\Phi   \nn\\&&+ e^{2 \Phi }c_{167} F_{\alpha  }{}^{\gamma  } F^{\delta  \epsilon  } R_{\delta  }{}^{\varepsilon  }{}_{\beta  }{}^{\mu  } R_{\epsilon  \mu  \gamma  \varepsilon  } \nabla^{\alpha  }\Phi  \nabla^{\beta  }\Phi  + e^{2 \Phi }c_{168} F^{\gamma  \delta  } F^{\epsilon  \varepsilon  } R_{\alpha  }{}^{\mu  }{}_{\beta  \gamma  } R_{\epsilon  \mu  \delta  \varepsilon  } \nabla^{\alpha  }\Phi  \nabla^{\beta  }\Phi \nn\\&&  + e^{2 \Phi }c_{172} F^{\gamma  \delta  } F^{\epsilon  \varepsilon  } R_{\alpha  \gamma  \epsilon  }{}^{\mu  } R_{\varepsilon  \mu  \beta  \delta  } \nabla^{\alpha  }\Phi  \nabla^{\beta  }\Phi  + e^{2 \Phi }c_{173} F^{\gamma  \delta  } F^{\epsilon  \varepsilon  } R_{\gamma  }{}^{\mu  }{}_{\alpha  \epsilon  } R_{\varepsilon  \mu  \beta  \delta  } \nabla^{\alpha  }\Phi  \nabla^{\beta  }\Phi   \nn\\&&+ e^{2 \Phi }c_{175} F_{\gamma  \delta  } F^{\gamma  \delta  } R_{\alpha  }{}^{\epsilon  \varepsilon  \mu  } R_{\varepsilon  \mu  \beta  \epsilon  } \nabla^{\alpha  }\Phi  \nabla^{\beta  }\Phi  + e^{2 \Phi }c_{177} F_{\alpha  }{}^{\gamma  } F_{\gamma  }{}^{\delta  } R_{\beta  }{}^{\epsilon  \varepsilon  \mu  } R_{\varepsilon  \mu  \delta  \epsilon  } \nabla^{\alpha  }\Phi  \nabla^{\beta  }\Phi   \nn\\&&+ e^{2 \Phi }c_{178} F_{\alpha  }{}^{\gamma  } F_{\beta  }{}^{\delta  } R_{\gamma  }{}^{\epsilon  \varepsilon  \mu  } R_{\varepsilon  \mu  \delta  \epsilon  } \nabla^{\alpha  }\Phi  \nabla^{\beta  }\Phi  + e^{2 \Phi }c_{180} F_{\alpha  }{}^{\gamma  } F_{\beta  \gamma  } R^{\delta  \epsilon  \varepsilon  \mu  } R_{\varepsilon  \mu  \delta  \epsilon  } \nabla^{\alpha  }\Phi  \nabla^{\beta  }\Phi  \nn\\&& + e^{2 \Phi }c_{181} F_{\gamma  }{}^{\epsilon  } F^{\gamma  \delta  } R_{\varepsilon  \mu  \beta  \epsilon  } R^{\varepsilon  \mu  }{}_{\alpha  \delta  } \nabla^{\alpha  }\Phi  \nabla^{\beta  }\Phi  + e^{2 \Phi }c_{182} F_{\alpha  }{}^{\gamma  } F^{\delta  \epsilon  } R_{\varepsilon  \mu  \delta  \epsilon  } R^{\varepsilon  \mu  }{}_{\beta  \gamma  } \nabla^{\alpha  }\Phi  \nabla^{\beta  }\Phi   \nn\\&&+ e^{2 \Phi }c_{183} F_{\alpha  }{}^{\gamma  } F^{\delta  \epsilon  } R_{\varepsilon  \mu  \gamma  \epsilon  } R^{\varepsilon  \mu  }{}_{\beta  \delta  } \nabla^{\alpha  }\Phi  \nabla^{\beta  }\Phi \,,
\eeqa
\beqa
[F^4R\Phi'^2]_{12}&=&e^{4 \Phi }c_{73} F_{\beta  }{}^{\delta  } F^{\beta  \gamma  } F_{\epsilon  }{}^{\mu  } F^{\epsilon  \varepsilon  } R_{\gamma  \varepsilon  \delta  \mu  } \nabla_{\alpha  }\Phi  \nabla^{\alpha  }\Phi  + e^{4 \Phi }c_{74} F_{\beta  \gamma  } F^{\beta  \gamma  } F^{\delta  \epsilon  } F^{\varepsilon  \mu  } R_{\delta  \varepsilon  \epsilon  \mu  } \nabla_{\alpha  }\Phi  \nabla^{\alpha  }\Phi  \nn\\&& + e^{4 \Phi }c_{78} F_{\beta  }{}^{\delta  } F^{\beta  \gamma  } F_{\gamma  }{}^{\epsilon  } F^{\varepsilon  \mu  } R_{\varepsilon  \mu  \delta  \epsilon  } \nabla_{\alpha  }\Phi  \nabla^{\alpha  }\Phi  + e^{4 \Phi }c_{159} F_{\gamma  }{}^{\epsilon  } F^{\gamma  \delta  } F_{\delta  }{}^{\varepsilon  } F_{\epsilon  }{}^{\mu  } R_{\alpha  \varepsilon  \beta  \mu  } \nabla^{\alpha  }\Phi  \nabla^{\beta  }\Phi  \nn\\&& + e^{4 \Phi }c_{160} F_{\gamma  \delta  } F^{\gamma  \delta  } F_{\epsilon  }{}^{\mu  } F^{\epsilon  \varepsilon  } R_{\alpha  \varepsilon  \beta  \mu  } \nabla^{\alpha  }\Phi  \nabla^{\beta  }\Phi  + e^{4 \Phi }c_{161} F_{\alpha  }{}^{\gamma  } F_{\beta  }{}^{\delta  } F_{\epsilon  }{}^{\mu  } F^{\epsilon  \varepsilon  } R_{\gamma  \varepsilon  \delta  \mu  } \nabla^{\alpha  }\Phi  \nabla^{\beta  }\Phi  \nn\\&& + e^{4 \Phi }c_{162} F_{\alpha  }{}^{\gamma  } F_{\gamma  }{}^{\delta  } F_{\epsilon  }{}^{\mu  } F^{\epsilon  \varepsilon  } R_{\delta  \varepsilon  \beta  \mu  } \nabla^{\alpha  }\Phi  \nabla^{\beta  }\Phi  + e^{4 \Phi }c_{163} F_{\alpha  }{}^{\gamma  } F_{\beta  \gamma  } F^{\delta  \epsilon  } F^{\varepsilon  \mu  } R_{\delta  \varepsilon  \epsilon  \mu  } \nabla^{\alpha  }\Phi  \nabla^{\beta  }\Phi   \nn\\&&+ e^{4 \Phi }c_{170} F_{\alpha  }{}^{\gamma  } F_{\delta  }{}^{\varepsilon  } F^{\delta  \epsilon  } F_{\epsilon  }{}^{\mu  } R_{\varepsilon  \mu  \beta  \gamma  } \nabla^{\alpha  }\Phi  \nabla^{\beta  }\Phi  + e^{4 \Phi }c_{171} F_{\alpha  }{}^{\gamma  } F_{\delta  \epsilon  } F^{\delta  \epsilon  } F^{\varepsilon  \mu  } R_{\varepsilon  \mu  \beta  \gamma  } \nabla^{\alpha  }\Phi  \nabla^{\beta  }\Phi   \nn\\&&+ e^{4 \Phi }c_{174} F_{\alpha  }{}^{\gamma  } F_{\gamma  }{}^{\delta  } F_{\delta  }{}^{\epsilon  } F^{\varepsilon  \mu  } R_{\varepsilon  \mu  \beta  \epsilon  } \nabla^{\alpha  }\Phi  \nabla^{\beta  }\Phi \nn\\&& + e^{4 \Phi }c_{176} F_{\alpha  }{}^{\gamma  } F_{\beta  }{}^{\delta  } F_{\gamma  }{}^{\epsilon  } F^{\varepsilon  \mu  } R_{\varepsilon  \mu  \delta  \epsilon  } \nabla^{\alpha  }\Phi  \nabla^{\beta  }\Phi,
\eeqa
\beqa
[F^2RF'^2]_{33}&=&e^{4 \Phi }c_{59} F_{\gamma  }{}^{\epsilon  } F^{\varepsilon  \mu  } R_{\varepsilon  \mu  \delta  \epsilon  } \nabla_{\alpha  }F_{\beta  }{}^{\delta  } \nabla^{\alpha  }F^{\beta  \gamma  } + e^{4 \Phi }c_{66} F_{\beta  }{}^{\varepsilon  } F_{\delta  }{}^{\mu  } R_{\varepsilon  \mu  \gamma  \epsilon  } \nabla_{\alpha  }F^{\delta  \epsilon  } \nabla^{\alpha  }F^{\beta  \gamma  } \nn\\&& + e^{4 \Phi }c_{67} F_{\beta  \delta  } F^{\varepsilon  \mu  } R_{\varepsilon  \mu  \gamma  \epsilon  } \nabla_{\alpha  }F^{\delta  \epsilon  } \nabla^{\alpha  }F^{\beta  \gamma  } + e^{4 \Phi }c_{68} F_{\beta  }{}^{\varepsilon  } F_{\gamma  }{}^{\mu  } R_{\varepsilon  \mu  \delta  \epsilon  } \nabla_{\alpha  }F^{\delta  \epsilon  } \nabla^{\alpha  }F^{\beta  \gamma  }  \nn\\&&+ e^{4 \Phi }c_{69} F_{\beta  \gamma  } F^{\varepsilon  \mu  } R_{\varepsilon  \mu  \delta  \epsilon  } \nabla_{\alpha  }F^{\delta  \epsilon  } \nabla^{\alpha  }F^{\beta  \gamma  } + e^{4 \Phi }c_{120} F^{\delta  \epsilon  } F^{\varepsilon  \mu  } R_{\delta  \varepsilon  \epsilon  \mu  } \nabla^{\alpha  }F^{\beta  \gamma  } \nabla_{\beta  }F_{\alpha  \gamma  }  \nn\\&&+ e^{4 \Phi }c_{124} F_{\epsilon  }{}^{\mu  } F^{\epsilon  \varepsilon  } R_{\gamma  \varepsilon  \delta  \mu  } \nabla^{\alpha  }F^{\beta  \gamma  } \nabla_{\beta  }F_{\alpha  }{}^{\delta  } + e^{4 \Phi }c_{125} F_{\gamma  }{}^{\epsilon  } F^{\varepsilon  \mu  } R_{\varepsilon  \mu  \delta  \epsilon  } \nabla^{\alpha  }F^{\beta  \gamma  } \nabla_{\beta  }F_{\alpha  }{}^{\delta  }  \nn\\&&+ e^{4 \Phi }c_{127} F_{\epsilon  }{}^{\mu  } F^{\epsilon  \varepsilon  } R_{\delta  \varepsilon  \alpha  \mu  } \nabla^{\alpha  }F^{\beta  \gamma  } \nabla_{\beta  }F_{\gamma  }{}^{\delta  } + e^{4 \Phi }c_{129} F_{\varepsilon  \mu  } F^{\varepsilon  \mu  } R_{\delta  \epsilon  \alpha  \gamma  } \nabla^{\alpha  }F^{\beta  \gamma  } \nabla_{\beta  }F^{\delta  \epsilon  }  \nn\\&&+ e^{4 \Phi }c_{130} F_{\delta  }{}^{\varepsilon  } F_{\varepsilon  }{}^{\mu  } R_{\epsilon  \mu  \alpha  \gamma  } \nabla^{\alpha  }F^{\beta  \gamma  } \nabla_{\beta  }F^{\delta  \epsilon  } + e^{4 \Phi }c_{131} F_{\gamma  }{}^{\varepsilon  } F_{\delta  }{}^{\mu  } R_{\epsilon  \mu  \alpha  \varepsilon  } \nabla^{\alpha  }F^{\beta  \gamma  } \nabla_{\beta  }F^{\delta  \epsilon  }  \nn\\&&+ e^{4 \Phi }c_{132} F_{\alpha  }{}^{\varepsilon  } F_{\delta  }{}^{\mu  } R_{\epsilon  \mu  \gamma  \varepsilon  } \nabla^{\alpha  }F^{\beta  \gamma  } \nabla_{\beta  }F^{\delta  \epsilon  } + e^{4 \Phi }c_{134} F_{\alpha  }{}^{\varepsilon  } F_{\delta  }{}^{\mu  } R_{\varepsilon  \mu  \gamma  \epsilon  } \nabla^{\alpha  }F^{\beta  \gamma  } \nabla_{\beta  }F^{\delta  \epsilon  }  \nn\\&&+ e^{4 \Phi }c_{135} F_{\alpha  \delta  } F^{\varepsilon  \mu  } R_{\varepsilon  \mu  \gamma  \epsilon  } \nabla^{\alpha  }F^{\beta  \gamma  } \nabla_{\beta  }F^{\delta  \epsilon  } + e^{4 \Phi }c_{278} F_{\varepsilon  \mu  } F^{\varepsilon  \mu  } R_{\gamma  \epsilon  \alpha  \delta  } \nabla^{\alpha  }F^{\beta  \gamma  } \nabla^{\delta  }F_{\beta  }{}^{\epsilon  } \nn\\&& + e^{4 \Phi }c_{279} F_{\alpha  }{}^{\varepsilon  } F_{\varepsilon  }{}^{\mu  } R_{\delta  \mu  \gamma  \epsilon  } \nabla^{\alpha  }F^{\beta  \gamma  } \nabla^{\delta  }F_{\beta  }{}^{\epsilon  } + e^{4 \Phi }c_{280} F_{\alpha  }{}^{\varepsilon  } F_{\epsilon  }{}^{\mu  } R_{\delta  \mu  \gamma  \varepsilon  } \nabla^{\alpha  }F^{\beta  \gamma  } \nabla^{\delta  }F_{\beta  }{}^{\epsilon  }  \nn\\&&+ e^{4 \Phi }c_{281} F_{\gamma  }{}^{\varepsilon  } F_{\varepsilon  }{}^{\mu  } R_{\epsilon  \mu  \alpha  \delta  } \nabla^{\alpha  }F^{\beta  \gamma  } \nabla^{\delta  }F_{\beta  }{}^{\epsilon  } + e^{4 \Phi }c_{282} F_{\alpha  }{}^{\varepsilon  } F_{\varepsilon  }{}^{\mu  } R_{\epsilon  \mu  \gamma  \delta  } \nabla^{\alpha  }F^{\beta  \gamma  } \nabla^{\delta  }F_{\beta  }{}^{\epsilon  } \nn\\&& + e^{4 \Phi }c_{284} F_{\alpha  }{}^{\varepsilon  } F_{\gamma  }{}^{\mu  } R_{\epsilon  \mu  \delta  \varepsilon  } \nabla^{\alpha  }F^{\beta  \gamma  } \nabla^{\delta  }F_{\beta  }{}^{\epsilon  } + e^{4 \Phi }c_{286} F_{\alpha  }{}^{\varepsilon  } F_{\delta  }{}^{\mu  } R_{\varepsilon  \mu  \gamma  \epsilon  } \nabla^{\alpha  }F^{\beta  \gamma  } \nabla^{\delta  }F_{\beta  }{}^{\epsilon  }  \nn\\&&+ e^{4 \Phi }c_{287} F_{\alpha  \delta  } F^{\varepsilon  \mu  } R_{\varepsilon  \mu  \gamma  \epsilon  } \nabla^{\alpha  }F^{\beta  \gamma  } \nabla^{\delta  }F_{\beta  }{}^{\epsilon  } + e^{4 \Phi }c_{304} F_{\beta  }{}^{\mu  } F_{\epsilon  \mu  } R_{\gamma  \varepsilon  \alpha  \delta  } \nabla^{\alpha  }F^{\beta  \gamma  } \nabla^{\delta  }F^{\epsilon  \varepsilon  } \nn\\&& + e^{4 \Phi }c_{305} F_{\alpha  }{}^{\mu  } F_{\epsilon  \mu  } R_{\delta  \varepsilon  \beta  \gamma  } \nabla^{\alpha  }F^{\beta  \gamma  } \nabla^{\delta  }F^{\epsilon  \varepsilon  } + e^{4 \Phi }c_{306} F_{\alpha  }{}^{\mu  } F_{\beta  \mu  } R_{\epsilon  \varepsilon  \gamma  \delta  } \nabla^{\alpha  }F^{\beta  \gamma  } \nabla^{\delta  }F^{\epsilon  \varepsilon  }  \nn\\&&+ e^{4 \Phi }c_{307} F_{\alpha  \delta  } F_{\beta  }{}^{\mu  } R_{\epsilon  \mu  \gamma  \varepsilon  } \nabla^{\alpha  }F^{\beta  \gamma  } \nabla^{\delta  }F^{\epsilon  \varepsilon  } + e^{4 \Phi }c_{308} F_{\alpha  \beta  } F_{\delta  }{}^{\mu  } R_{\epsilon  \mu  \gamma  \varepsilon  } \nabla^{\alpha  }F^{\beta  \gamma  } \nabla^{\delta  }F^{\epsilon  \varepsilon  }  \nn\\&&+ e^{4 \Phi }c_{309} F_{\alpha  \beta  } F_{\gamma  }{}^{\mu  } R_{\epsilon  \mu  \delta  \varepsilon  } \nabla^{\alpha  }F^{\beta  \gamma  } \nabla^{\delta  }F^{\epsilon  \varepsilon  } + e^{4 \Phi }c_{311} F_{\alpha  \epsilon  } F_{\delta  }{}^{\mu  } R_{\varepsilon  \mu  \beta  \gamma  } \nabla^{\alpha  }F^{\beta  \gamma  } \nabla^{\delta  }F^{\epsilon  \varepsilon  } \nn\\&& + e^{4 \Phi }c_{313} F_{\alpha  }{}^{\mu  } F_{\beta  \epsilon  } R_{\varepsilon  \mu  \gamma  \delta  } \nabla^{\alpha  }F^{\beta  \gamma  } \nabla^{\delta  }F^{\epsilon  \varepsilon  } + e^{4 \Phi }c_{314} F_{\alpha  \epsilon  } F_{\beta  }{}^{\mu  } R_{\varepsilon  \mu  \gamma  \delta  } \nabla^{\alpha  }F^{\beta  \gamma  } \nabla^{\delta  }F^{\epsilon  \varepsilon  }  \nn\\&&+ e^{4 \Phi }c_{315} F_{\alpha  \beta  } F_{\epsilon  }{}^{\mu  } R_{\varepsilon  \mu  \gamma  \delta  } \nabla^{\alpha  }F^{\beta  \gamma  } \nabla^{\delta  }F^{\epsilon  \varepsilon  }\,,
\eeqa
\beqa
[F^6R]_7&=&e^{6 \Phi }c_{6} F_{\alpha  }{}^{\gamma  } F^{\alpha  \beta  } F_{\beta  }{}^{\delta  } F_{\epsilon  }{}^{\mu  } F^{\epsilon  \varepsilon  } F_{\varepsilon  }{}^{\nu  } R_{\gamma  \mu  \delta  \nu  } + e^{6 \Phi }c_{7} F_{\alpha  }{}^{\gamma  } F^{\alpha  \beta  } F_{\beta  }{}^{\delta  } F_{\gamma  }{}^{\epsilon  } F_{\varepsilon  }{}^{\nu  } F^{\varepsilon  \mu  } R_{\delta  \mu  \epsilon  \nu  }  \nn\\&&+ e^{6 \Phi }c_{8} F_{\alpha  \beta  } F^{\alpha  \beta  } F_{\gamma  }{}^{\epsilon  } F^{\gamma  \delta  } F_{\varepsilon  }{}^{\nu  } F^{\varepsilon  \mu  } R_{\delta  \mu  \epsilon  \nu  } + e^{6 \Phi }c_{9} F_{\alpha  }{}^{\gamma  } F^{\alpha  \beta  } F_{\beta  }{}^{\delta  } F_{\gamma  \delta  } F^{\epsilon  \varepsilon  } F^{\mu  \nu  } R_{\epsilon  \mu  \varepsilon  \nu  }  \nn\\&&+ e^{6 \Phi }c_{10} F_{\alpha  \beta  } F^{\alpha  \beta  } F_{\gamma  \delta  } F^{\gamma  \delta  } F^{\epsilon  \varepsilon  } F^{\mu  \nu  } R_{\epsilon  \mu  \varepsilon  \nu  } + e^{6 \Phi }c_{24} F_{\alpha  }{}^{\gamma  } F^{\alpha  \beta  } F_{\beta  }{}^{\delta  } F_{\gamma  }{}^{\epsilon  } F_{\delta  }{}^{\varepsilon  } F^{\mu  \nu  } R_{\mu  \nu  \epsilon  \varepsilon  } \nn\\&& + e^{6 \Phi }c_{25} F_{\alpha  \beta  } F^{\alpha  \beta  } F_{\gamma  }{}^{\epsilon  } F^{\gamma  \delta  } F_{\delta  }{}^{\varepsilon  } F^{\mu  \nu  } R_{\mu  \nu  \epsilon  \varepsilon  }\,,
\eeqa
\beqa
[F^6\Phi'']_4&=&e^{6 \Phi }c_{83} F_{\alpha  }{}^{\gamma  } F_{\beta  }{}^{\delta  } F_{\gamma  }{}^{\epsilon  } F_{\delta  }{}^{\varepsilon  } F_{\epsilon  }{}^{\mu  } F_{\varepsilon  \mu  } \nabla^{\alpha  }\nabla^{\beta  }\Phi  + e^{6 \Phi }c_{84} F_{\alpha  }{}^{\gamma  } F_{\beta  \gamma  } F_{\delta  }{}^{\varepsilon  } F^{\delta  \epsilon  } F_{\epsilon  }{}^{\mu  } F_{\varepsilon  \mu  } \nabla^{\alpha  }\nabla^{\beta  }\Phi   \nn\\&&+ e^{6 \Phi }c_{85} F_{\alpha  }{}^{\gamma  } F_{\beta  }{}^{\delta  } F_{\gamma  }{}^{\epsilon  } F_{\delta  \epsilon  } F_{\varepsilon  \mu  } F^{\varepsilon  \mu  } \nabla^{\alpha  }\nabla^{\beta  }\Phi \nn\\&& + e^{6 \Phi }c_{86} F_{\alpha  }{}^{\gamma  } F_{\beta  \gamma  } F_{\delta  \epsilon  } F^{\delta  \epsilon  } F_{\varepsilon  \mu  } F^{\varepsilon  \mu  } \nabla^{\alpha  }\nabla^{\beta  }\Phi,
\eeqa
\beqa
[F^6\Phi'^2]_1&=&e^{6 \Phi }c_{158} F_{\alpha  }{}^{\gamma  } F_{\beta  \gamma  } F_{\delta  \epsilon  } F^{\delta  \epsilon  } F_{\varepsilon  \mu  } F^{\varepsilon  \mu  } \nabla^{\alpha  }\Phi  \nabla^{\beta  }\Phi\,,
\eeqa
\beqa
[F^4\Phi'^2\Phi'']_2&\!\!\!=\!\!\!&e^{4 \Phi }c_{239} F_{\alpha  \gamma  } F_{\beta  }{}^{\epsilon  } F_{\delta  }{}^{\varepsilon  } F_{\epsilon  \varepsilon  } \nabla^{\alpha  }\Phi  \nabla^{\beta  }\Phi  \nabla^{\gamma  }\nabla^{\delta  }\Phi  \nn\\&&+ e^{4 \Phi }c_{240} F_{\alpha  \gamma  } F_{\beta  \delta  } F_{\epsilon  \varepsilon  } F^{\epsilon  \varepsilon  } \nabla^{\alpha  }\Phi  \nabla^{\beta  }\Phi  \nabla^{\gamma  }\nabla^{\delta  }\Phi,
\eeqa
\beqa
[FR^2F'\Phi']_{15}&=&e^{2 \Phi }c_{72} F_{\beta  }{}^{\delta  } R_{\gamma  }{}^{\epsilon  \varepsilon  \mu  } R_{\varepsilon  \mu  \delta  \epsilon  } \nabla_{\alpha  }F^{\beta  \gamma  } \nabla^{\alpha  }\Phi  + e^{2 \Phi }c_{143} F_{\beta  }{}^{\delta  } R_{\gamma  }{}^{\epsilon  \varepsilon  \mu  } R_{\varepsilon  \mu  \delta  \epsilon  } \nabla^{\alpha  }\Phi  \nabla^{\beta  }F_{\alpha  }{}^{\gamma  }   \nn\\&&+ e^{2 \Phi }c_{144} F_{\beta  \gamma  } R^{\delta  \epsilon  \varepsilon  \mu  } R_{\varepsilon  \mu  \delta  \epsilon  } \nabla^{\alpha  }\Phi  \nabla^{\beta  }F_{\alpha  }{}^{\gamma  } + e^{2 \Phi }c_{145} F^{\delta  \epsilon  } R_{\varepsilon  \mu  \delta  \epsilon  } R^{\varepsilon  \mu  }{}_{\beta  \gamma  } \nabla^{\alpha  }\Phi  \nabla^{\beta  }F_{\alpha  }{}^{\gamma  }  \nn\\&& + e^{2 \Phi }c_{146} F^{\delta  \epsilon  } R_{\varepsilon  \mu  \gamma  \epsilon  } R^{\varepsilon  \mu  }{}_{\beta  \delta  } \nabla^{\alpha  }\Phi  \nabla^{\beta  }F_{\alpha  }{}^{\gamma  } + e^{2 \Phi }c_{147} F^{\epsilon  \varepsilon  } R_{\beta  }{}^{\mu  }{}_{\alpha  \gamma  } R_{\epsilon  \mu  \delta  \varepsilon  } \nabla^{\alpha  }\Phi  \nabla^{\beta  }F^{\gamma  \delta  }   \nn\\&&+ e^{2 \Phi }c_{148} F_{\alpha  }{}^{\epsilon  } R_{\gamma  }{}^{\varepsilon  }{}_{\beta  }{}^{\mu  } R_{\epsilon  \mu  \delta  \varepsilon  } \nabla^{\alpha  }\Phi  \nabla^{\beta  }F^{\gamma  \delta  } + e^{2 \Phi }c_{149} F^{\epsilon  \varepsilon  } R_{\gamma  }{}^{\mu  }{}_{\alpha  \beta  } R_{\epsilon  \mu  \delta  \varepsilon  } \nabla^{\alpha  }\Phi  \nabla^{\beta  }F^{\gamma  \delta  }   \nn\\&&+ e^{2 \Phi }c_{150} F^{\epsilon  \varepsilon  } R_{\gamma  }{}^{\mu  }{}_{\beta  \epsilon  } R_{\varepsilon  \mu  \alpha  \delta  } \nabla^{\alpha  }\Phi  \nabla^{\beta  }F^{\gamma  \delta  } + e^{2 \Phi }c_{151} F^{\epsilon  \varepsilon  } R_{\beta  }{}^{\mu  }{}_{\alpha  \epsilon  } R_{\varepsilon  \mu  \gamma  \delta  } \nabla^{\alpha  }\Phi  \nabla^{\beta  }F^{\gamma  \delta  }   \nn\\&&+ e^{2 \Phi }c_{152} F^{\epsilon  \varepsilon  } R_{\epsilon  }{}^{\mu  }{}_{\alpha  \beta  } R_{\varepsilon  \mu  \gamma  \delta  } \nabla^{\alpha  }\Phi  \nabla^{\beta  }F^{\gamma  \delta  } + e^{2 \Phi }c_{153} F_{\alpha  \gamma  } R_{\beta  }{}^{\epsilon  \varepsilon  \mu  } R_{\varepsilon  \mu  \delta  \epsilon  } \nabla^{\alpha  }\Phi  \nabla^{\beta  }F^{\gamma  \delta  }   \nn\\&&+ e^{2 \Phi }c_{154} F_{\gamma  }{}^{\epsilon  } R_{\varepsilon  \mu  \beta  \epsilon  } R^{\varepsilon  \mu  }{}_{\alpha  \delta  } \nabla^{\alpha  }\Phi  \nabla^{\beta  }F^{\gamma  \delta  } + e^{2 \Phi }c_{155} F_{\beta  }{}^{\epsilon  } R_{\varepsilon  \mu  \gamma  \delta  } R^{\varepsilon  \mu  }{}_{\alpha  \epsilon  } \nabla^{\alpha  }\Phi  \nabla^{\beta  }F^{\gamma  \delta  }   \nn\\&&+ e^{2 \Phi }c_{156} F_{\alpha  }{}^{\epsilon  } R_{\varepsilon  \mu  \gamma  \delta  } R^{\varepsilon  \mu  }{}_{\beta  \epsilon  } \nabla^{\alpha  }\Phi  \nabla^{\beta  }F^{\gamma  \delta  }\,,
\eeqa
\beqa
[F^4\Phi'^4]_5&=&e^{4 \Phi }c_{189} F_{\gamma  }{}^{\epsilon  } F^{\gamma  \delta  } F_{\delta  }{}^{\varepsilon  } F_{\epsilon  \varepsilon  } \nabla_{\alpha  }\Phi  \nabla^{\alpha  }\Phi  \nabla_{\beta  }\Phi  \nabla^{\beta  }\Phi  + e^{4 \Phi }c_{190} F_{\gamma  \delta  } F^{\gamma  \delta  } F_{\epsilon  \varepsilon  } F^{\epsilon  \varepsilon  } \nabla_{\alpha  }\Phi  \nabla^{\alpha  }\Phi  \nabla_{\beta  }\Phi  \nabla^{\beta  }\Phi  \nn\\&&+ e^{4 \Phi }c_{219} F_{\beta  }{}^{\delta  } F_{\gamma  }{}^{\epsilon  } F_{\delta  }{}^{\varepsilon  } F_{\epsilon  \varepsilon  } \nabla_{\alpha  }\Phi  \nabla^{\alpha  }\Phi  \nabla^{\beta  }\Phi  \nabla^{\gamma  }\Phi  + e^{4 \Phi }c_{220} F_{\beta  }{}^{\delta  } F_{\gamma  \delta  } F_{\epsilon  \varepsilon  } F^{\epsilon  \varepsilon  } \nabla_{\alpha  }\Phi  \nabla^{\alpha  }\Phi  \nabla^{\beta  }\Phi  \nabla^{\gamma  }\Phi \nn\\&& + e^{4 \Phi }c_{326} F_{\alpha  }{}^{\epsilon  } F_{\beta  \epsilon  } F_{\gamma  }{}^{\varepsilon  } F_{\delta  \varepsilon  } \nabla^{\alpha  }\Phi  \nabla^{\beta  }\Phi  \nabla^{\gamma  }\Phi  \nabla^{\delta  }\Phi\,,
\eeqa
\beqa
[F^2R\Phi'^4]_4&\!\!\!\!\!=\!\!\!\!\!&e^{2 \Phi }c_{191} F^{\gamma  \delta  } F^{\epsilon  \varepsilon  } R_{\gamma  \epsilon  \delta  \varepsilon  } \nabla_{\alpha  }\Phi  \nabla^{\alpha  }\Phi  \nabla_{\beta  }\Phi  \nabla^{\beta  }\Phi  + e^{2 \Phi }c_{221} F_{\delta  }{}^{\varepsilon  } F^{\delta  \epsilon  } R_{\beta  \epsilon  \gamma  \varepsilon  } \nabla_{\alpha  }\Phi  \nabla^{\alpha  }\Phi  \nabla^{\beta  }\Phi  \nabla^{\gamma  }\Phi \nn\\&& + e^{2 \Phi }c_{222} F_{\beta  }{}^{\delta  } F^{\epsilon  \varepsilon  } R_{\epsilon  \varepsilon  \gamma  \delta  } \nabla_{\alpha  }\Phi  \nabla^{\alpha  }\Phi  \nabla^{\beta  }\Phi  \nabla^{\gamma  }\Phi \nn\\&& + e^{2 \Phi }c_{327} F_{\alpha  }{}^{\epsilon  } F_{\beta  }{}^{\varepsilon  } R_{\gamma  \epsilon  \delta  \varepsilon  } \nabla^{\alpha  }\Phi  \nabla^{\beta  }\Phi  \nabla^{\gamma  }\Phi  \nabla^{\delta  }\Phi, 
\eeqa
\beqa
[FR\Phi'^3F']_{10}&\!\!\!\!=\!\!\!\!&e^{2 \Phi }c_{208} F^{\epsilon  \varepsilon  } R_{\epsilon  \varepsilon  \gamma  \delta  } \nabla_{\alpha  }\Phi  \nabla^{\alpha  }\Phi  \nabla^{\beta  }\Phi  \nabla^{\gamma  }F_{\beta  }{}^{\delta  } + e^{2 \Phi }c_{211} F_{\gamma  }{}^{\varepsilon  } R_{\delta  \varepsilon  \beta  \epsilon  } \nabla_{\alpha  }\Phi  \nabla^{\alpha  }\Phi  \nabla^{\beta  }\Phi  \nabla^{\gamma  }F^{\delta  \epsilon  } \nn\\&&+ e^{2 \Phi }c_{212} F_{\beta  }{}^{\varepsilon  } R_{\delta  \varepsilon  \gamma  \epsilon  } \nabla_{\alpha  }\Phi  \nabla^{\alpha  }\Phi  \nabla^{\beta  }\Phi  \nabla^{\gamma  }F^{\delta  \epsilon  } + e^{2 \Phi }c_{213} F_{\delta  }{}^{\varepsilon  } R_{\epsilon  \varepsilon  \beta  \gamma  } \nabla_{\alpha  }\Phi  \nabla^{\alpha  }\Phi  \nabla^{\beta  }\Phi  \nabla^{\gamma  }F^{\delta  \epsilon  } \nn\\&&+ e^{2 \Phi }c_{216} F^{\epsilon  \varepsilon  } R_{\epsilon  \varepsilon  \gamma  \delta  } \nabla_{\alpha  }F_{\beta  }{}^{\delta  } \nabla^{\alpha  }\Phi  \nabla^{\beta  }\Phi  \nabla^{\gamma  }\Phi  + e^{2 \Phi }c_{218} F_{\delta  }{}^{\varepsilon  } R_{\beta  \varepsilon  \gamma  \epsilon  } \nabla_{\alpha  }F^{\delta  \epsilon  } \nabla^{\alpha  }\Phi  \nabla^{\beta  }\Phi  \nabla^{\gamma  }\Phi \nn\\&& + e^{2 \Phi }c_{269} F_{\delta  }{}^{\varepsilon  } R_{\beta  \varepsilon  \gamma  \epsilon  } \nabla^{\alpha  }\Phi  \nabla^{\beta  }\Phi  \nabla^{\gamma  }\Phi  \nabla^{\delta  }F_{\alpha  }{}^{\epsilon  } + e^{2 \Phi }c_{270} F_{\beta  }{}^{\varepsilon  } R_{\delta  \varepsilon  \gamma  \epsilon  } \nabla^{\alpha  }\Phi  \nabla^{\beta  }\Phi  \nabla^{\gamma  }\Phi  \nabla^{\delta  }F_{\alpha  }{}^{\epsilon  } \nn\\&&+ e^{2 \Phi }c_{271} F_{\beta  }{}^{\varepsilon  } R_{\epsilon  \varepsilon  \gamma  \delta  } \nabla^{\alpha  }\Phi  \nabla^{\beta  }\Phi  \nabla^{\gamma  }\Phi  \nabla^{\delta  }F_{\alpha  }{}^{\epsilon  } \nn\\&&+ e^{2 \Phi }c_{322} F_{\alpha  \epsilon  } R_{\beta  \varepsilon  \gamma  \delta  } \nabla^{\alpha  }\Phi  \nabla^{\beta  }\Phi  \nabla^{\gamma  }\Phi  \nabla^{\delta  }F^{\epsilon  \varepsilon  },
\eeqa
\beqa
[F^4R\Phi'']_9&=&e^{4 \Phi }c_{87} F_{\gamma  }{}^{\epsilon  } F^{\gamma  \delta  } F_{\delta  }{}^{\varepsilon  } F_{\epsilon  }{}^{\mu  } R_{\alpha  \varepsilon  \beta  \mu  } \nabla^{\alpha  }\nabla^{\beta  }\Phi  + e^{4 \Phi }c_{88} F_{\gamma  \delta  } F^{\gamma  \delta  } F_{\epsilon  }{}^{\mu  } F^{\epsilon  \varepsilon  } R_{\alpha  \varepsilon  \beta  \mu  } \nabla^{\alpha  }\nabla^{\beta  }\Phi \nn\\&& + e^{4 \Phi }c_{89} F_{\alpha  }{}^{\gamma  } F_{\beta  }{}^{\delta  } F_{\epsilon  }{}^{\mu  } F^{\epsilon  \varepsilon  } R_{\gamma  \varepsilon  \delta  \mu  } \nabla^{\alpha  }\nabla^{\beta  }\Phi  + e^{4 \Phi }c_{90} F_{\alpha  }{}^{\gamma  } F_{\gamma  }{}^{\delta  } F_{\epsilon  }{}^{\mu  } F^{\epsilon  \varepsilon  } R_{\delta  \varepsilon  \beta  \mu  } \nabla^{\alpha  }\nabla^{\beta  }\Phi \nn\\&& + e^{4 \Phi }c_{91} F_{\alpha  }{}^{\gamma  } F_{\beta  \gamma  } F^{\delta  \epsilon  } F^{\varepsilon  \mu  } R_{\delta  \varepsilon  \epsilon  \mu  } \nabla^{\alpha  }\nabla^{\beta  }\Phi  + e^{4 \Phi }c_{98} F_{\alpha  }{}^{\gamma  } F_{\delta  }{}^{\varepsilon  } F^{\delta  \epsilon  } F_{\epsilon  }{}^{\mu  } R_{\varepsilon  \mu  \beta  \gamma  } \nabla^{\alpha  }\nabla^{\beta  }\Phi \nn\\&& + e^{4 \Phi }c_{99} F_{\alpha  }{}^{\gamma  } F_{\delta  \epsilon  } F^{\delta  \epsilon  } F^{\varepsilon  \mu  } R_{\varepsilon  \mu  \beta  \gamma  } \nabla^{\alpha  }\nabla^{\beta  }\Phi  + e^{4 \Phi }c_{102} F_{\alpha  }{}^{\gamma  } F_{\gamma  }{}^{\delta  } F_{\delta  }{}^{\epsilon  } F^{\varepsilon  \mu  } R_{\varepsilon  \mu  \beta  \epsilon  } \nabla^{\alpha  }\nabla^{\beta  }\Phi  \nn\\&&+ e^{4 \Phi }c_{104} F_{\alpha  }{}^{\gamma  } F_{\beta  }{}^{\delta  } F_{\gamma  }{}^{\epsilon  } F^{\varepsilon  \mu  } R_{\varepsilon  \mu  \delta  \epsilon  } \nabla^{\alpha  }\nabla^{\beta  }\Phi \,,
\eeqa
\beqa
[F^2\Phi'^6]_2&\!\!\!\!\!=\!\!\!\!\!&e^{2 \Phi }c_{224} F_{\delta  \epsilon  } F^{\delta  \epsilon  } \nabla_{\alpha  }\Phi  \nabla^{\alpha  }\Phi  \nabla_{\beta  }\Phi  \nabla^{\beta  }\Phi  \nabla_{\gamma  }\Phi  \nabla^{\gamma  }\Phi \nn\\&&+ e^{2 \Phi }c_{330} F_{\gamma  }{}^{\epsilon  } F_{\delta  \epsilon  } \nabla_{\alpha  }\Phi  \nabla^{\alpha  }\Phi  \nabla_{\beta  }\Phi  \nabla^{\beta  }\Phi  \nabla^{\gamma  }\Phi  \nabla^{\delta  }\Phi,
\eeqa
\beqa
[F^4\Phi''^2]_8&=&e^{4 \Phi }c_{113} F_{\beta  }{}^{\delta  } F_{\gamma  }{}^{\epsilon  } F_{\delta  }{}^{\varepsilon  } F_{\epsilon  \varepsilon  } \nabla_{\alpha  }\nabla^{\gamma  }\Phi  \nabla^{\alpha  }\nabla^{\beta  }\Phi  + e^{4 \Phi }c_{114} F_{\beta  }{}^{\delta  } F_{\gamma  \delta  } F_{\epsilon  \varepsilon  } F^{\epsilon  \varepsilon  } \nabla_{\alpha  }\nabla^{\gamma  }\Phi  \nabla^{\alpha  }\nabla^{\beta  }\Phi \nn\\&& + e^{4 \Phi }c_{137} F_{\gamma  }{}^{\epsilon  } F^{\gamma  \delta  } F_{\delta  }{}^{\varepsilon  } F_{\epsilon  \varepsilon  } \nabla^{\alpha  }\nabla^{\beta  }\Phi  \nabla_{\beta  }\nabla_{\alpha  }\Phi  + e^{4 \Phi }c_{138} F_{\gamma  \delta  } F^{\gamma  \delta  } F_{\epsilon  \varepsilon  } F^{\epsilon  \varepsilon  } \nabla^{\alpha  }\nabla^{\beta  }\Phi  \nabla_{\beta  }\nabla_{\alpha  }\Phi  \nn\\&&+ e^{4 \Phi }c_{226} F_{\alpha  }{}^{\epsilon  } F_{\beta  }{}^{\varepsilon  } F_{\gamma  \epsilon  } F_{\delta  \varepsilon  } \nabla^{\alpha  }\nabla^{\beta  }\Phi  \nabla^{\gamma  }\nabla^{\delta  }\Phi  + e^{4 \Phi }c_{227} F_{\alpha  }{}^{\epsilon  } F_{\beta  \epsilon  } F_{\gamma  }{}^{\varepsilon  } F_{\delta  \varepsilon  } \nabla^{\alpha  }\nabla^{\beta  }\Phi  \nabla^{\gamma  }\nabla^{\delta  }\Phi \nn\\&& + e^{4 \Phi }c_{228} F_{\alpha  \gamma  } F_{\beta  }{}^{\epsilon  } F_{\delta  }{}^{\varepsilon  } F_{\epsilon  \varepsilon  } \nabla^{\alpha  }\nabla^{\beta  }\Phi  \nabla^{\gamma  }\nabla^{\delta  }\Phi \nn\\&& + e^{4 \Phi }c_{229} F_{\alpha  \gamma  } F_{\beta  \delta  } F_{\epsilon  \varepsilon  } F^{\epsilon  \varepsilon  } \nabla^{\alpha  }\nabla^{\beta  }\Phi  \nabla^{\gamma  }\nabla^{\delta  }\Phi ,
\eeqa
\beqa
[F^2R\Phi''^2]_9&=&e^{2 \Phi }c_{115} F_{\delta  }{}^{\varepsilon  } F^{\delta  \epsilon  } R_{\beta  \epsilon  \gamma  \varepsilon  } \nabla_{\alpha  }\nabla^{\gamma  }\Phi  \nabla^{\alpha  }\nabla^{\beta  }\Phi  + e^{2 \Phi }c_{116} F_{\beta  }{}^{\delta  } F^{\epsilon  \varepsilon  } R_{\epsilon  \varepsilon  \gamma  \delta  } \nabla_{\alpha  }\nabla^{\gamma  }\Phi  \nabla^{\alpha  }\nabla^{\beta  }\Phi  \nn\\&&+ e^{2 \Phi }c_{139} F^{\gamma  \delta  } F^{\epsilon  \varepsilon  } R_{\gamma  \epsilon  \delta  \varepsilon  } \nabla^{\alpha  }\nabla^{\beta  }\Phi  \nabla_{\beta  }\nabla_{\alpha  }\Phi  + e^{2 \Phi }c_{230} F_{\epsilon  \varepsilon  } F^{\epsilon  \varepsilon  } R_{\alpha  \gamma  \beta  \delta  } \nabla^{\alpha  }\nabla^{\beta  }\Phi  \nabla^{\gamma  }\nabla^{\delta  }\Phi  \nn\\&&+ e^{2 \Phi }c_{231} F_{\alpha  }{}^{\epsilon  } F_{\gamma  }{}^{\varepsilon  } R_{\beta  \varepsilon  \delta  \epsilon  } \nabla^{\alpha  }\nabla^{\beta  }\Phi  \nabla^{\gamma  }\nabla^{\delta  }\Phi  + e^{2 \Phi }c_{232} F_{\alpha  }{}^{\epsilon  } F_{\beta  }{}^{\varepsilon  } R_{\gamma  \epsilon  \delta  \varepsilon  } \nabla^{\alpha  }\nabla^{\beta  }\Phi  \nabla^{\gamma  }\nabla^{\delta  }\Phi \nn\\&& + e^{2 \Phi }c_{233} F_{\alpha  }{}^{\epsilon  } F_{\epsilon  }{}^{\varepsilon  } R_{\gamma  \varepsilon  \beta  \delta  } \nabla^{\alpha  }\nabla^{\beta  }\Phi  \nabla^{\gamma  }\nabla^{\delta  }\Phi  + e^{2 \Phi }c_{236} F_{\alpha  }{}^{\epsilon  } F_{\gamma  }{}^{\varepsilon  } R_{\epsilon  \varepsilon  \beta  \delta  } \nabla^{\alpha  }\nabla^{\beta  }\Phi  \nabla^{\gamma  }\nabla^{\delta  }\Phi  \nn\\&&+ e^{2 \Phi }c_{237} F_{\alpha  \gamma  } F^{\epsilon  \varepsilon  } R_{\epsilon  \varepsilon  \beta  \delta  } \nabla^{\alpha  }\nabla^{\beta  }\Phi  \nabla^{\gamma  }\nabla^{\delta  }\Phi\,,
\eeqa
\beqa
[F^2F'^2\Phi'']_{21}&=&e^{4 \Phi }c_{136} F_{\gamma  \delta  } F_{\epsilon  \varepsilon  } \nabla_{\alpha  }\nabla^{\varepsilon  }\Phi  \nabla^{\alpha  }F^{\beta  \gamma  } \nabla_{\beta  }F^{\delta  \epsilon  } + e^{4 \Phi }c_{141} F_{\gamma  \delta  } F_{\epsilon  \varepsilon  } \nabla_{\alpha  }F^{\delta  \epsilon  } \nabla^{\alpha  }F^{\beta  \gamma  } \nabla_{\beta  }\nabla^{\varepsilon  }\Phi   \nn\\&&+ e^{4 \Phi }c_{200} F_{\delta  }{}^{\varepsilon  } F_{\epsilon  \varepsilon  } \nabla_{\alpha  }F_{\beta  }{}^{\delta  } \nabla^{\alpha  }F^{\beta  \gamma  } \nabla_{\gamma  }\nabla^{\epsilon  }\Phi  + e^{4 \Phi }c_{201} F_{\delta  }{}^{\varepsilon  } F_{\epsilon  \varepsilon  } \nabla^{\alpha  }F^{\beta  \gamma  } \nabla_{\beta  }F_{\alpha  }{}^{\delta  } \nabla_{\gamma  }\nabla^{\epsilon  }\Phi   \nn\\&&+ e^{4 \Phi }c_{256} F_{\gamma  }{}^{\varepsilon  } F_{\epsilon  \varepsilon  } \nabla^{\alpha  }F^{\beta  \gamma  } \nabla_{\beta  }F^{\delta  \epsilon  } \nabla_{\delta  }\nabla_{\alpha  }\Phi  + e^{4 \Phi }c_{257} F_{\gamma  }{}^{\varepsilon  } F_{\epsilon  \varepsilon  } \nabla_{\alpha  }F^{\delta  \epsilon  } \nabla^{\alpha  }F^{\beta  \gamma  } \nabla_{\delta  }\nabla_{\beta  }\Phi   \nn\\&&+ e^{4 \Phi }c_{259} F_{\epsilon  \varepsilon  } F^{\epsilon  \varepsilon  } \nabla_{\alpha  }F_{\beta  }{}^{\delta  } \nabla^{\alpha  }F^{\beta  \gamma  } \nabla_{\delta  }\nabla_{\gamma  }\Phi  + e^{4 \Phi }c_{260} F_{\epsilon  \varepsilon  } F^{\epsilon  \varepsilon  } \nabla^{\alpha  }F^{\beta  \gamma  } \nabla_{\beta  }F_{\alpha  }{}^{\delta  } \nabla_{\delta  }\nabla_{\gamma  }\Phi   \nn\\&&+ e^{4 \Phi }c_{265} F_{\alpha  \gamma  } F_{\epsilon  \varepsilon  } \nabla^{\alpha  }F^{\beta  \gamma  } \nabla_{\beta  }F^{\delta  \epsilon  } \nabla_{\delta  }\nabla^{\varepsilon  }\Phi  + e^{4 \Phi }c_{289} F_{\gamma  \varepsilon  } F_{\delta  \epsilon  } \nabla_{\alpha  }\nabla^{\varepsilon  }\Phi  \nabla^{\alpha  }F^{\beta  \gamma  } \nabla^{\delta  }F_{\beta  }{}^{\epsilon  } \nn\\&& + e^{4 \Phi }c_{290} F_{\gamma  \delta  } F_{\epsilon  \varepsilon  } \nabla_{\alpha  }\nabla^{\varepsilon  }\Phi  \nabla^{\alpha  }F^{\beta  \gamma  } \nabla^{\delta  }F_{\beta  }{}^{\epsilon  } + e^{4 \Phi }c_{293} F_{\delta  }{}^{\varepsilon  } F_{\epsilon  \varepsilon  } \nabla^{\alpha  }F^{\beta  \gamma  } \nabla_{\gamma  }\nabla_{\alpha  }\Phi  \nabla^{\delta  }F_{\beta  }{}^{\epsilon  } \nn\\&& + e^{4 \Phi }c_{294} F_{\alpha  \delta  } F_{\epsilon  \varepsilon  } \nabla^{\alpha  }F^{\beta  \gamma  } \nabla_{\gamma  }\nabla^{\varepsilon  }\Phi  \nabla^{\delta  }F_{\beta  }{}^{\epsilon  } + e^{4 \Phi }c_{297} F_{\gamma  }{}^{\varepsilon  } F_{\epsilon  \varepsilon  } \nabla^{\alpha  }F^{\beta  \gamma  } \nabla_{\delta  }\nabla_{\alpha  }\Phi  \nabla^{\delta  }F_{\beta  }{}^{\epsilon  } \nn\\&& + e^{4 \Phi }c_{318} F_{\gamma  \delta  } F_{\epsilon  \varepsilon  } \nabla^{\alpha  }F^{\beta  \gamma  } \nabla_{\beta  }\nabla_{\alpha  }\Phi  \nabla^{\delta  }F^{\epsilon  \varepsilon  } + e^{4 \Phi }c_{324} F_{\beta  \epsilon  } F_{\gamma  \varepsilon  } \nabla^{\alpha  }F^{\beta  \gamma  } \nabla_{\delta  }\nabla_{\alpha  }\Phi  \nabla^{\delta  }F^{\epsilon  \varepsilon  } \nn\\&& + e^{4 \Phi }c_{325} F_{\beta  \gamma  } F_{\epsilon  \varepsilon  } \nabla^{\alpha  }F^{\beta  \gamma  } \nabla_{\delta  }\nabla_{\alpha  }\Phi  \nabla^{\delta  }F^{\epsilon  \varepsilon  } + e^{4 \Phi }c_{336} F_{\delta  }{}^{\varepsilon  } F_{\epsilon  \varepsilon  } \nabla^{\alpha  }F^{\beta  \gamma  } \nabla_{\beta  }F_{\alpha  \gamma  } \nabla^{\delta  }\nabla^{\epsilon  }\Phi   \nn\\&&+ e^{4 \Phi }c_{345} F_{\alpha  \delta  } F_{\gamma  \varepsilon  } \nabla^{\alpha  }F^{\beta  \gamma  } \nabla^{\delta  }F^{\epsilon  \varepsilon  } \nabla_{\epsilon  }\nabla_{\beta  }\Phi  + e^{4 \Phi }c_{371} F_{\gamma  \epsilon  } F_{\delta  \varepsilon  } \nabla_{\alpha  }F_{\beta  }{}^{\delta  } \nabla^{\alpha  }F^{\beta  \gamma  } \nabla^{\epsilon  }\nabla^{\varepsilon  }\Phi  \nn\\&& + e^{4 \Phi }c_{372} F_{\gamma  \epsilon  } F_{\delta  \varepsilon  } \nabla^{\alpha  }F^{\beta  \gamma  } \nabla_{\beta  }F_{\alpha  }{}^{\delta  } \nabla^{\epsilon  }\nabla^{\varepsilon  }\Phi \,,
\eeqa
\beqa
[F^3R\Phi'F']_1&=&e^{4 \Phi }c_{142} F_{\beta  }{}^{\delta  } F_{\gamma  }{}^{\epsilon  } F^{\varepsilon  \mu  } R_{\varepsilon  \mu  \delta  \epsilon  } \nabla^{\alpha  }\Phi  \nabla^{\beta  }F_{\alpha  }{}^{\gamma  }\,,
\eeqa
\beqa
[FR\Phi''\Phi'F']_2&=&e^{2 \Phi }c_{157} F_{\beta  }{}^{\varepsilon  } R_{\epsilon  \varepsilon  \gamma  \delta  } \nabla_{\alpha  }\nabla^{\epsilon  }\Phi  \nabla^{\alpha  }\Phi  \nabla^{\beta  }F^{\gamma  \delta  } \nn\\&&+ e^{2 \Phi }c_{337} F_{\delta  }{}^{\varepsilon  } R_{\epsilon  \varepsilon  \beta  \gamma  } \nabla^{\alpha  }\Phi  \nabla^{\beta  }F_{\alpha  }{}^{\gamma  } \nabla^{\delta  }\nabla^{\epsilon  }\Phi,
\eeqa
\beqa
[F^2\Phi''^2\Phi'^2]_4&\!\!\!\!\!=\!\!\!\!\!&e^{2 \Phi }c_{194} F_{\gamma  }{}^{\epsilon  } F_{\delta  \epsilon  } \nabla_{\alpha  }\nabla^{\gamma  }\Phi  \nabla^{\alpha  }\Phi  \nabla_{\beta  }\nabla^{\delta  }\Phi  \nabla^{\beta  }\Phi  + e^{2 \Phi }c_{199} F_{\delta  \epsilon  } F^{\delta  \epsilon  } \nabla_{\alpha  }\nabla^{\gamma  }\Phi  \nabla^{\alpha  }\Phi  \nabla^{\beta  }\Phi  \nabla_{\gamma  }\nabla_{\beta  }\Phi  \nn\\&& + e^{2 \Phi }c_{248} F_{\gamma  }{}^{\epsilon  } F_{\delta  \epsilon  } \nabla^{\alpha  }\Phi  \nabla_{\beta  }\nabla_{\alpha  }\Phi  \nabla^{\beta  }\Phi  \nabla^{\gamma  }\nabla^{\delta  }\Phi  \nn\\&&+ e^{2 \Phi }c_{338} F_{\beta  \delta  } F_{\gamma  \epsilon  } \nabla_{\alpha  }\nabla^{\gamma  }\Phi  \nabla^{\alpha  }\Phi  \nabla^{\beta  }\Phi  \nabla^{\delta  }\nabla^{\epsilon  }\Phi,
\eeqa
\beqa
[F^2\Phi'^2F'^2]_{17}&=&e^{4 \Phi }c_{195} F_{\epsilon  \varepsilon  } F^{\epsilon  \varepsilon  } \nabla_{\alpha  }\Phi  \nabla^{\alpha  }\Phi  \nabla^{\beta  }F^{\gamma  \delta  } \nabla_{\gamma  }F_{\beta  \delta  } + e^{4 \Phi }c_{196} F_{\delta  }{}^{\varepsilon  } F_{\epsilon  \varepsilon  } \nabla_{\alpha  }\Phi  \nabla^{\alpha  }\Phi  \nabla^{\beta  }F^{\gamma  \delta  } \nabla_{\gamma  }F_{\beta  }{}^{\epsilon  }  \nn\\&&+ e^{4 \Phi }c_{197} F_{\delta  }{}^{\varepsilon  } F_{\epsilon  \varepsilon  } \nabla_{\alpha  }F^{\gamma  \delta  } \nabla^{\alpha  }\Phi  \nabla^{\beta  }\Phi  \nabla_{\gamma  }F_{\beta  }{}^{\epsilon  } + e^{4 \Phi }c_{203} F_{\epsilon  \varepsilon  } F^{\epsilon  \varepsilon  } \nabla^{\alpha  }\Phi  \nabla^{\beta  }\Phi  \nabla_{\gamma  }F_{\beta  \delta  } \nabla^{\gamma  }F_{\alpha  }{}^{\delta  }  \nn\\&&+ e^{4 \Phi }c_{204} F_{\delta  }{}^{\varepsilon  } F_{\epsilon  \varepsilon  } \nabla^{\alpha  }\Phi  \nabla^{\beta  }\Phi  \nabla_{\gamma  }F_{\beta  }{}^{\epsilon  } \nabla^{\gamma  }F_{\alpha  }{}^{\delta  } + e^{4 \Phi }c_{205} F_{\beta  \epsilon  } F_{\delta  \varepsilon  } \nabla^{\alpha  }\Phi  \nabla^{\beta  }\Phi  \nabla_{\gamma  }F^{\epsilon  \varepsilon  } \nabla^{\gamma  }F_{\alpha  }{}^{\delta  } \nn\\&& + e^{4 \Phi }c_{250} F_{\epsilon  \varepsilon  } F^{\epsilon  \varepsilon  } \nabla^{\alpha  }\Phi  \nabla^{\beta  }\Phi  \nabla^{\gamma  }F_{\alpha  }{}^{\delta  } \nabla_{\delta  }F_{\beta  \gamma  } + e^{4 \Phi }c_{252} F_{\beta  }{}^{\varepsilon  } F_{\epsilon  \varepsilon  } \nabla^{\alpha  }\Phi  \nabla^{\beta  }\Phi  \nabla^{\gamma  }F_{\alpha  }{}^{\delta  } \nabla_{\delta  }F_{\gamma  }{}^{\epsilon  }  \nn\\&&+ e^{4 \Phi }c_{298} F_{\delta  }{}^{\varepsilon  } F_{\epsilon  \varepsilon  } \nabla_{\alpha  }F_{\beta  }{}^{\gamma  } \nabla^{\alpha  }\Phi  \nabla^{\beta  }\Phi  \nabla^{\delta  }F_{\gamma  }{}^{\epsilon  } + e^{4 \Phi }c_{320} F_{\gamma  \delta  } F_{\epsilon  \varepsilon  } \nabla_{\alpha  }F_{\beta  }{}^{\gamma  } \nabla^{\alpha  }\Phi  \nabla^{\beta  }\Phi  \nabla^{\delta  }F^{\epsilon  \varepsilon  }  \nn\\&&+ e^{4 \Phi }c_{356} F_{\delta  }{}^{\varepsilon  } F_{\epsilon  \varepsilon  } \nabla_{\alpha  }F^{\gamma  \delta  } \nabla^{\alpha  }\Phi  \nabla^{\beta  }\Phi  \nabla^{\epsilon  }F_{\beta  \gamma  } + e^{4 \Phi }c_{358} F_{\gamma  \epsilon  } F_{\delta  \varepsilon  } \nabla_{\alpha  }F^{\gamma  \delta  } \nabla^{\alpha  }\Phi  \nabla^{\beta  }\Phi  \nabla^{\epsilon  }F_{\beta  }{}^{\varepsilon  }  \nn\\&&+ e^{4 \Phi }c_{360} F_{\gamma  \epsilon  } F_{\delta  \varepsilon  } \nabla^{\alpha  }\Phi  \nabla^{\beta  }\Phi  \nabla^{\gamma  }F_{\alpha  }{}^{\delta  } \nabla^{\epsilon  }F_{\beta  }{}^{\varepsilon  } + e^{4 \Phi }c_{361} F_{\gamma  \delta  } F_{\epsilon  \varepsilon  } \nabla^{\alpha  }\Phi  \nabla^{\beta  }\Phi  \nabla^{\gamma  }F_{\alpha  }{}^{\delta  } \nabla^{\epsilon  }F_{\beta  }{}^{\varepsilon  }  \nn\\&&+ e^{4 \Phi }c_{363} F_{\beta  }{}^{\varepsilon  } F_{\epsilon  \varepsilon  } \nabla^{\alpha  }\Phi  \nabla^{\beta  }\Phi  \nabla^{\gamma  }F_{\alpha  }{}^{\delta  } \nabla^{\epsilon  }F_{\gamma  \delta  } + e^{4 \Phi }c_{367} F_{\beta  \epsilon  } F_{\delta  \varepsilon  } \nabla^{\alpha  }\Phi  \nabla^{\beta  }\Phi  \nabla^{\gamma  }F_{\alpha  }{}^{\delta  } \nabla^{\epsilon  }F_{\gamma  }{}^{\varepsilon  }  \nn\\&&+ e^{4 \Phi }c_{368} F_{\beta  \delta  } F_{\epsilon  \varepsilon  } \nabla^{\alpha  }\Phi  \nabla^{\beta  }\Phi  \nabla^{\gamma  }F_{\alpha  }{}^{\delta  } \nabla^{\epsilon  }F_{\gamma  }{}^{\varepsilon  }\,,
\eeqa
\beqa
[FF'\Phi'\Phi''^2]_1&=&e^{2 \Phi }c_{202} F_{\delta  \epsilon  } \nabla^{\alpha  }\Phi  \nabla_{\beta  }\nabla_{\alpha  }\Phi  \nabla^{\beta  }F^{\gamma  \delta  } \nabla_{\gamma  }\nabla^{\epsilon  }\Phi \,,
\eeqa
\beqa
[FF'\Phi'^3\Phi'']_5&\!\!\!\!=\!\!\!\!&e^{2 \Phi }c_{225} F_{\delta  \epsilon  } \nabla_{\alpha  }F_{\beta  }{}^{\delta  } \nabla^{\alpha  }\Phi  \nabla^{\beta  }\Phi  \nabla_{\gamma  }\nabla^{\epsilon  }\Phi  \nabla^{\gamma  }\Phi  + e^{2 \Phi }c_{258} F_{\gamma  \epsilon  } \nabla_{\alpha  }F^{\delta  \epsilon  } \nabla^{\alpha  }\Phi  \nabla^{\beta  }\Phi  \nabla^{\gamma  }\Phi  \nabla_{\delta  }\nabla_{\beta  }\Phi \nn\\&&  + e^{2 \Phi }c_{264} F_{\gamma  \epsilon  } \nabla_{\alpha  }F_{\beta  }{}^{\delta  } \nabla^{\alpha  }\Phi  \nabla^{\beta  }\Phi  \nabla^{\gamma  }\Phi  \nabla_{\delta  }\nabla^{\epsilon  }\Phi  + e^{2 \Phi }c_{272} F_{\delta  \epsilon  } \nabla^{\alpha  }\Phi  \nabla^{\beta  }\Phi  \nabla_{\gamma  }\nabla_{\beta  }\Phi  \nabla^{\gamma  }\Phi  \nabla^{\delta  }F_{\alpha  }{}^{\epsilon  }  \nn\\&&+ e^{2 \Phi }c_{273} F_{\gamma  \epsilon  } \nabla^{\alpha  }\Phi  \nabla^{\beta  }\Phi  \nabla^{\gamma  }\Phi  \nabla_{\delta  }\nabla_{\beta  }\Phi  \nabla^{\delta  }F_{\alpha  }{}^{\epsilon  }\,,
\eeqa
\beqa
[R\Phi'^2\Phi''^2]_2&\!\!\!=\!\!\!&c_{249} R_{\alpha  \delta  \beta  \epsilon  } \nabla^{\alpha  }\Phi  \nabla^{\beta  }\Phi  \nabla_{\gamma  }\nabla^{\epsilon  }\Phi  \nabla^{\gamma  }\nabla^{\delta  }\Phi  +c_{339} R_{\gamma  \delta  \beta  \epsilon  } \nabla_{\alpha  }\nabla^{\gamma  }\Phi  \nabla^{\alpha  }\Phi  \nabla^{\beta  }\Phi  \nabla^{\delta  }\nabla^{\epsilon  }\Phi ,
\eeqa
\beqa
[\Phi'^4F'^2]_4&\!\!\!\!=\!\!\!\!&e^{2 \Phi }c_{251} \nabla_{\alpha  }\Phi  \nabla^{\alpha  }\Phi  \nabla_{\beta  }\Phi  \nabla^{\beta  }\Phi  \nabla^{\gamma  }F^{\delta  \epsilon  } \nabla_{\delta  }F_{\gamma  \epsilon  } + e^{2 \Phi }c_{296} \nabla_{\alpha  }\Phi  \nabla^{\alpha  }\Phi  \nabla^{\beta  }\Phi  \nabla^{\gamma  }\Phi  \nabla_{\delta  }F_{\gamma  \epsilon  } \nabla^{\delta  }F_{\beta  }{}^{\epsilon  } \nn\\&&+ e^{2 \Phi }c_{332} \nabla_{\alpha  }F_{\beta  }{}^{\epsilon  } \nabla^{\alpha  }\Phi  \nabla^{\beta  }\Phi  \nabla_{\gamma  }F_{\delta  \epsilon  } \nabla^{\gamma  }\Phi  \nabla^{\delta  }\Phi \nn\\&& + e^{2 \Phi }c_{343} \nabla_{\alpha  }\Phi  \nabla^{\alpha  }\Phi  \nabla^{\beta  }\Phi  \nabla^{\gamma  }\Phi  \nabla^{\delta  }F_{\beta  }{}^{\epsilon  } \nabla_{\epsilon  }F_{\gamma  \delta  },
\eeqa
\beqa
[\Phi''^4]_2&=&c_{261} \nabla_{\alpha  }\nabla^{\gamma  }\Phi  \nabla^{\alpha  }\nabla^{\beta  }\Phi  \nabla_{\beta  }\nabla^{\delta  }\Phi  \nabla_{\delta  }\nabla_{\gamma  }\Phi  +c_{263} \nabla^{\alpha  }\nabla^{\beta  }\Phi  \nabla_{\beta  }\nabla_{\alpha  }\Phi  \nabla^{\gamma  }\nabla^{\delta  }\Phi  \nabla_{\delta  }\nabla_{\gamma  }\Phi ,
\eeqa
\beqa
[FF'\Phi'^5]_2&=&e^{2 \Phi }c_{299} F_{\delta  \epsilon  } \nabla_{\alpha  }\Phi  \nabla^{\alpha  }\Phi  \nabla_{\beta  }\Phi  \nabla^{\beta  }\Phi  \nabla^{\gamma  }\Phi  \nabla^{\delta  }F_{\gamma  }{}^{\epsilon  }\nn\\&& + e^{2 \Phi }c_{329} F_{\delta  \epsilon  } \nabla_{\alpha  }\Phi  \nabla^{\alpha  }\Phi  \nabla_{\beta  }F_{\gamma  }{}^{\epsilon  } \nabla^{\beta  }\Phi  \nabla^{\gamma  }\Phi  \nabla^{\delta  }\Phi, 
\eeqa
\beqa
[F^2\Phi'^4\Phi'']_2&=&e^{2 \Phi }c_{331} F_{\gamma  }{}^{\epsilon  } F_{\delta  \epsilon  } \nabla^{\alpha  }\Phi  \nabla_{\beta  }\nabla_{\alpha  }\Phi  \nabla^{\beta  }\Phi  \nabla^{\gamma  }\Phi  \nabla^{\delta  }\Phi\nn\\&&  + e^{2 \Phi }c_{340} F_{\beta  \delta  } F_{\gamma  \epsilon  } \nabla_{\alpha  }\Phi  \nabla^{\alpha  }\Phi  \nabla^{\beta  }\Phi  \nabla^{\gamma  }\Phi  \nabla^{\delta  }\nabla^{\epsilon  }\Phi,
\eeqa
\beqa
[R\Phi''^3]_1&=&c_{335} R_{\beta  \delta  \gamma  \epsilon  } \nabla_{\alpha  }\nabla^{\gamma  }\Phi  \nabla^{\alpha  }\nabla^{\beta  }\Phi  \nabla^{\delta  }\nabla^{\epsilon  }\Phi \,,
\eeqa
\beqa
[R\Phi'^4\Phi'']_1&=&c_{341} R_{\beta  \delta  \gamma  \epsilon  } \nabla_{\alpha  }\Phi  \nabla^{\alpha  }\Phi  \nabla^{\beta  }\Phi  \nabla^{\gamma  }\Phi  \nabla^{\delta  }\nabla^{\epsilon  }\Phi\,,
\eeqa
\beqa
[F'^2\Phi''^2]_6&=&e^{2 \Phi }c_{347} \nabla_{\alpha  }F^{\delta  \epsilon  } \nabla^{\alpha  }F^{\beta  \gamma  } \nabla_{\delta  }\nabla_{\beta  }\Phi  \nabla_{\epsilon  }\nabla_{\gamma  }\Phi  + e^{2 \Phi }c_{348} \nabla^{\alpha  }F^{\beta  \gamma  } \nabla_{\delta  }\nabla_{\alpha  }\Phi  \nabla^{\delta  }F_{\beta  }{}^{\epsilon  } \nabla_{\epsilon  }\nabla_{\gamma  }\Phi \nn\\&&+ e^{2 \Phi }c_{350} \nabla_{\alpha  }F_{\beta  }{}^{\delta  } \nabla^{\alpha  }F^{\beta  \gamma  } \nabla_{\gamma  }\nabla^{\epsilon  }\Phi  \nabla_{\epsilon  }\nabla_{\delta  }\Phi  + e^{2 \Phi }c_{351} \nabla^{\alpha  }F^{\beta  \gamma  } \nabla_{\beta  }F_{\alpha  }{}^{\delta  } \nabla_{\gamma  }\nabla^{\epsilon  }\Phi  \nabla_{\epsilon  }\nabla_{\delta  }\Phi \nn\\&& + e^{2 \Phi }c_{353} \nabla^{\alpha  }F^{\beta  \gamma  } \nabla_{\gamma  }\nabla_{\alpha  }\Phi  \nabla^{\delta  }F_{\beta  }{}^{\epsilon  } \nabla_{\epsilon  }\nabla_{\delta  }\Phi  \nn\\&&+ e^{2 \Phi }c_{355} \nabla^{\alpha  }F^{\beta  \gamma  } \nabla_{\beta  }F_{\alpha  \gamma  } \nabla^{\delta  }\nabla^{\epsilon  }\Phi  \nabla_{\epsilon  }\nabla_{\delta  }\Phi,\labell{CCPP}
\eeqa
\beqa
[F'^2\Phi'^2\Phi'']_4&\!\!\!\!\!=\!\!\!\!\!&e^{2 \Phi }c_{349} \nabla_{\alpha  }F^{\gamma  \delta  } \nabla^{\alpha  }\Phi  \nabla^{\beta  }\Phi  \nabla_{\gamma  }F_{\beta  }{}^{\epsilon  } \nabla_{\epsilon  }\nabla_{\delta  }\Phi  + e^{2 \Phi }c_{352} \nabla^{\alpha  }\Phi  \nabla^{\beta  }\Phi  \nabla_{\gamma  }F_{\beta  }{}^{\epsilon  } \nabla^{\gamma  }F_{\alpha  }{}^{\delta  } \nabla_{\epsilon  }\nabla_{\delta  }\Phi\nn\\&&  + e^{2 \Phi }c_{354} \nabla_{\alpha  }F_{\beta  }{}^{\gamma  } \nabla^{\alpha  }\Phi  \nabla^{\beta  }\Phi  \nabla^{\delta  }F_{\gamma  }{}^{\epsilon  } \nabla_{\epsilon  }\nabla_{\delta  }\Phi  \nn\\&&+ e^{2 \Phi }c_{357} \nabla_{\alpha  }F^{\gamma  \delta  } \nabla^{\alpha  }\Phi  \nabla^{\beta  }\Phi  \nabla_{\epsilon  }\nabla_{\delta  }\Phi  \nabla^{\epsilon  }F_{\beta  \gamma  },
\eeqa
\beqa
[FF'^3\Phi']_1&=&e^{4 \Phi }c_{377} F_{\epsilon  \varepsilon  } \nabla^{\alpha  }\Phi  \nabla^{\beta  }F_{\alpha  }{}^{\gamma  } \nabla^{\delta  }F_{\beta  }{}^{\epsilon  } \nabla^{\varepsilon  }F_{\gamma  \delta  }\,,
\eeqa
\beqa
[\Phi'^4\Phi''^2]_2&=&c_{262} \nabla_{\alpha  }\Phi  \nabla^{\alpha  }\Phi  \nabla_{\beta  }\nabla^{\delta  }\Phi  \nabla^{\beta  }\Phi  \nabla^{\gamma  }\Phi  \nabla_{\delta  }\nabla_{\gamma  }\Phi \nn\\&& +c_{334} \nabla^{\alpha  }\Phi  \nabla_{\beta  }\nabla_{\alpha  }\Phi  \nabla^{\beta  }\Phi  \nabla^{\gamma  }\Phi  \nabla_{\delta  }\nabla_{\gamma  }\Phi  \nabla^{\delta  }\Phi,
\eeqa
\beqa
[\Phi'^6\Phi'']_1&=&c_{333} \nabla_{\alpha  }\Phi  \nabla^{\alpha  }\Phi  \nabla_{\beta  }\Phi  \nabla^{\beta  }\Phi  \nabla_{\gamma  }\Phi  \nabla^{\gamma  }\Phi  \nabla_{\delta  }\Phi  \nabla^{\delta  }\Phi\,,
\eeqa
\beqa
[F^8]_5&=&e^{8 \Phi }c_{1} F_{\alpha  }{}^{\gamma  } F^{\alpha  \beta  } F_{\beta  }{}^{\delta  } F_{\gamma  }{}^{\epsilon  } F_{\delta  }{}^{\varepsilon  } F_{\epsilon  }{}^{\mu  } F_{\varepsilon  }{}^{\nu  } F_{\mu  \nu  } + e^{8 \Phi }c_{2} F_{\alpha  \beta  } F^{\alpha  \beta  } F_{\gamma  }{}^{\epsilon  } F^{\gamma  \delta  } F_{\delta  }{}^{\varepsilon  } F_{\epsilon  }{}^{\mu  } F_{\varepsilon  }{}^{\nu  } F_{\mu  \nu  }\nn\\&& + e^{8 \Phi }c_{3} F_{\alpha  }{}^{\gamma  } F^{\alpha  \beta  } F_{\beta  }{}^{\delta  } F_{\gamma  \delta  } F_{\epsilon  }{}^{\mu  } F^{\epsilon  \varepsilon  } F_{\varepsilon  }{}^{\nu  } F_{\mu  \nu  } + e^{8 \Phi }c_{4} F_{\alpha  \beta  } F^{\alpha  \beta  } F_{\gamma  \delta  } F^{\gamma  \delta  } F_{\epsilon  }{}^{\mu  } F^{\epsilon  \varepsilon  } F_{\varepsilon  }{}^{\nu  } F_{\mu  \nu  } \nn\\&&+ e^{8 \Phi }c_{5} F_{\alpha  \beta  } F^{\alpha  \beta  } F_{\gamma  \delta  } F^{\gamma  \delta  } F_{\epsilon  \varepsilon  } F^{\epsilon  \varepsilon  } F_{\mu  \nu  } F^{\mu  \nu  }\,,
\eeqa
\beqa
[F^4F'^2]_{20}&=&e^{6 \Phi }c_{57} F_{\gamma  }{}^{\epsilon  } F_{\delta  }{}^{\varepsilon  } F_{\epsilon  }{}^{\mu  } F_{\varepsilon  \mu  } \nabla_{\alpha  }F_{\beta  }{}^{\delta  } \nabla^{\alpha  }F^{\beta  \gamma  } + e^{6 \Phi }c_{58} F_{\gamma  }{}^{\epsilon  } F_{\delta  \epsilon  } F_{\varepsilon  \mu  } F^{\varepsilon  \mu  } \nabla_{\alpha  }F_{\beta  }{}^{\delta  } \nabla^{\alpha  }F^{\beta  \gamma  } \nn\\&&+ e^{6 \Phi }c_{61} F_{\beta  }{}^{\varepsilon  } F_{\gamma  }{}^{\mu  } F_{\delta  \varepsilon  } F_{\epsilon  \mu  } \nabla_{\alpha  }F^{\delta  \epsilon  } \nabla^{\alpha  }F^{\beta  \gamma  } + e^{6 \Phi }c_{62} F_{\beta  \delta  } F_{\gamma  }{}^{\varepsilon  } F_{\epsilon  }{}^{\mu  } F_{\varepsilon  \mu  } \nabla_{\alpha  }F^{\delta  \epsilon  } \nabla^{\alpha  }F^{\beta  \gamma  }\nn\\&& + e^{6 \Phi }c_{63} F_{\beta  \gamma  } F_{\delta  }{}^{\varepsilon  } F_{\epsilon  }{}^{\mu  } F_{\varepsilon  \mu  } \nabla_{\alpha  }F^{\delta  \epsilon  } \nabla^{\alpha  }F^{\beta  \gamma  } + e^{6 \Phi }c_{64} F_{\beta  \delta  } F_{\gamma  \epsilon  } F_{\varepsilon  \mu  } F^{\varepsilon  \mu  } \nabla_{\alpha  }F^{\delta  \epsilon  } \nabla^{\alpha  }F^{\beta  \gamma  } \nn\\&&+ e^{6 \Phi }c_{65} F_{\beta  \gamma  } F_{\delta  \epsilon  } F_{\varepsilon  \mu  } F^{\varepsilon  \mu  } \nabla_{\alpha  }F^{\delta  \epsilon  } \nabla^{\alpha  }F^{\beta  \gamma  } + e^{6 \Phi }c_{118} F_{\delta  }{}^{\varepsilon  } F^{\delta  \epsilon  } F_{\epsilon  }{}^{\mu  } F_{\varepsilon  \mu  } \nabla^{\alpha  }F^{\beta  \gamma  } \nabla_{\beta  }F_{\alpha  \gamma  }\nn\\&& + e^{6 \Phi }c_{119} F_{\delta  \epsilon  } F^{\delta  \epsilon  } F_{\varepsilon  \mu  } F^{\varepsilon  \mu  } \nabla^{\alpha  }F^{\beta  \gamma  } \nabla_{\beta  }F_{\alpha  \gamma  } + e^{6 \Phi }c_{122} F_{\gamma  }{}^{\epsilon  } F_{\delta  }{}^{\varepsilon  } F_{\epsilon  }{}^{\mu  } F_{\varepsilon  \mu  } \nabla^{\alpha  }F^{\beta  \gamma  } \nabla_{\beta  }F_{\alpha  }{}^{\delta  } \nn\\&&+ e^{6 \Phi }c_{123} F_{\gamma  }{}^{\epsilon  } F_{\delta  \epsilon  } F_{\varepsilon  \mu  } F^{\varepsilon  \mu  } \nabla^{\alpha  }F^{\beta  \gamma  } \nabla_{\beta  }F_{\alpha  }{}^{\delta  } + e^{6 \Phi }c_{128} F_{\alpha  \delta  } F_{\gamma  }{}^{\varepsilon  } F_{\epsilon  }{}^{\mu  } F_{\varepsilon  \mu  } \nabla^{\alpha  }F^{\beta  \gamma  } \nabla_{\beta  }F^{\delta  \epsilon  }\nn\\&& + e^{6 \Phi }c_{274} F_{\alpha  }{}^{\varepsilon  } F_{\gamma  }{}^{\mu  } F_{\delta  \varepsilon  } F_{\epsilon  \mu  } \nabla^{\alpha  }F^{\beta  \gamma  } \nabla^{\delta  }F_{\beta  }{}^{\epsilon  } + e^{6 \Phi }c_{275} F_{\alpha  }{}^{\varepsilon  } F_{\gamma  \varepsilon  } F_{\delta  }{}^{\mu  } F_{\epsilon  \mu  } \nabla^{\alpha  }F^{\beta  \gamma  } \nabla^{\delta  }F_{\beta  }{}^{\epsilon  } \nn\\&&+ e^{6 \Phi }c_{276} F_{\alpha  \delta  } F_{\gamma  }{}^{\varepsilon  } F_{\epsilon  }{}^{\mu  } F_{\varepsilon  \mu  } \nabla^{\alpha  }F^{\beta  \gamma  } \nabla^{\delta  }F_{\beta  }{}^{\epsilon  } + e^{6 \Phi }c_{277} F_{\alpha  \delta  } F_{\gamma  \epsilon  } F_{\varepsilon  \mu  } F^{\varepsilon  \mu  } \nabla^{\alpha  }F^{\beta  \gamma  } \nabla^{\delta  }F_{\beta  }{}^{\epsilon  } \nn\\&&+ e^{6 \Phi }c_{300} F_{\alpha  \epsilon  } F_{\beta  \delta  } F_{\gamma  }{}^{\mu  } F_{\varepsilon  \mu  } \nabla^{\alpha  }F^{\beta  \gamma  } \nabla^{\delta  }F^{\epsilon  \varepsilon  } + e^{6 \Phi }c_{301} F_{\alpha  \delta  } F_{\beta  \epsilon  } F_{\gamma  }{}^{\mu  } F_{\varepsilon  \mu  } \nabla^{\alpha  }F^{\beta  \gamma  } \nabla^{\delta  }F^{\epsilon  \varepsilon  }\nn\\&& + e^{6 \Phi }c_{302} F_{\alpha  \beta  } F_{\gamma  }{}^{\mu  } F_{\delta  \epsilon  } F_{\varepsilon  \mu  } \nabla^{\alpha  }F^{\beta  \gamma  } \nabla^{\delta  }F^{\epsilon  \varepsilon  }\nn\\&& + e^{6 \Phi }c_{303} F_{\alpha  \beta  } F_{\gamma  \epsilon  } F_{\delta  }{}^{\mu  } F_{\varepsilon  \mu  } \nabla^{\alpha  }F^{\beta  \gamma  } \nabla^{\delta  }F^{\epsilon  \varepsilon  },
\eeqa
where the coupling constants $c_1, \dots, c_{377}$ are determined in this work through dimensional reduction of 11-dimensional gravity at order $\ell_p^6$. Note that the above basis represents one particular scheme, in which the 377 independent couplings omit certain structures like $[F^3F'\Phi'\Phi'']$, $[F^5F'\Phi']$, $[F^2\Phi''^3]$, $[\Phi'^8]$, $[\Phi'^2\Phi''^3]$, and $[R\Phi'^6]$. These structures may have independent couplings in other schemes, while some of the structures in \reef{T55} may, in turn, have no coupling. The basis in \reef{T55} is a specific one we have chosen to express the KK reduction couplings.

\end{document}